\documentclass[final,3p]{elsarticle}
\usepackage{amsmath}
\usepackage{amssymb}
\usepackage{amsfonts}
\usepackage{amsthm}
\usepackage{color}
\usepackage[T1]{fontenc}
\usepackage{bm}
\usepackage{tikz}
\usetikzlibrary{matrix,calc}
\usepackage{tabulary}
\usepackage{tabularx}
\usepackage{booktabs}   
\usepackage{multirow}   
\usepackage{array}
\usepackage{graphicx}
\usepackage{pgfplots}
\usepackage{placeins}
\usepackage{xfrac}
\usepgfplotslibrary{groupplots}
\usepackage{tcolorbox}
\usepackage{algorithm}
\usepackage[noend]{algpseudocode}
\usepackage{standalone}
\usepackage{hyperref}
\usepackage[capitalise]{cleveref}
\usepackage{parskip}
\makeatletter
\renewcommand{\paragraph}{%
  \@startsection{paragraph}{4}{\z@}%
  {3.25ex \@plus1ex \@minus.2ex}%
  {-0.6em}%
  {\normalfont\normalsize\bfseries\paragraphdot}%
}
\newcommand{\paragraphdot}[1]{#1.\ }
\makeatother
\definecolor{clr1}{RGB}{0,84,159}       
\definecolor{clr2}{RGB}{161,16,53}      
\definecolor{clr3}{RGB}{0,97,101}       
\definecolor{clr4}{RGB}{246,168,0}      
\definecolor{clr5}{RGB}{87,171,39}      
\definecolor{clr6}{RGB}{156,158,159}    
\definecolor{clr7}{RGB}{100,101,103}    
\definecolor{clr8}{RGB}{122,111,172}    
\definecolor{clr9}{RGB}{0,152,161}      
\definecolor{clr10}{RGB}{227,0,102}     
\definecolor{rwth1}{RGB}{0,84,159}      
\definecolor{rwth2}{RGB}{142,186,229}   
\definecolor{rwth3}{RGB}{0,97,101}      
\definecolor{rwth4}{RGB}{0,152,161}     
\definecolor{rwth5}{RGB}{87,171,39}     
\definecolor{rwth6}{RGB}{189,205,0}     
\definecolor{rwth7}{RGB}{255,237,0}     
\definecolor{rwth8}{RGB}{246,168,0}     
\definecolor{rwth9}{RGB}{227,0,102}     
\definecolor{rwth10}{RGB}{204,7,30}     
\definecolor{rwth11}{RGB}{161,16,53}    
\definecolor{rwth12}{RGB}{97,33,88}     
\definecolor{rwth13}{RGB}{122,111,172}  
\definecolor{loss1}{RGB}{0,114,178}   
\definecolor{loss2}{RGB}{213,94,0}    
\definecolor{loss3}{RGB}{0,158,115}   
\definecolor{loss4}{RGB}{204,121,167} 
\definecolor{loss5}{RGB}{230,159,0}   
\definecolor{ekA}{RGB}{5,48,97}      
\definecolor{ekB}{RGB}{67,147,195}   
\definecolor{ekC}{RGB}{161,230,242}
\definecolor{ekD}{RGB}{244,165,130}  
\definecolor{ekE}{RGB}{178,24,43}    
\newcommand{\RN}[1]{\uppercase\expandafter{\romannumeral#1}}
\usepackage{lineno}
\modulolinenumbers[100]
\journal{}
\newcommand{\beq}{\begin{equation}}
\newcommand{\eeq}{\end{equation}}

\def\tens  #1{\mbox{\boldmath{\scriptsize{$#1$}}}{}}
\def\ten   #1{\mbox{\boldmath $#1$}{}}
\begin{document} 
\begin{frontmatter}
\title{A Convex Route to Thermoelasticity: Learning Internal Energy and Dissipation\tnoteref{t1}}
\tnotetext[t1]{
Earlier versions of this preprint appeared on arXiv under the title
``A Convex Route to Thermomechanics: Learning Internal Energy and Dissipation''.
The title has been updated to match the published journal article:
\emph{A Convex Route to Thermoelasticity: Learning Internal Energy and Dissipation},
\href{https://doi.org/10.1016/j.cma.2026.119082}{https://doi.org/10.1016/j.cma.2026.119082}.
}
\cortext[cor1]{Corresponding author}
\author[add1]{Hagen Holthusen\corref{cor1}} 
\ead{hagen.holthusen@fau.de}
\author[add1]{Paul Steinmann}
\ead{paul.steinmann@fau.de}
\author[add1,add2]{Ellen Kuhl}
\ead{ekuhl@stanford.edu}
\address[add1]{Institute of Applied Mechanics, University of Erlangen-Nuremberg, Egerlandstra{\ss}e 5, 91058 Erlangen, Germany}
\address[add2]{Department of Mechanical Engineering, Stanford University, United States}
\begin{abstract}
We present a physics-based neural network framework for the discovery of constitutive models in fully coupled thermomechanics. In contrast to classical formulations based on the Helmholtz energy, we adopt the internal energy and a dissipation potential as primary constitutive functions, expressed in terms of deformation and entropy. This choice avoids the need to enforce mixed convexity--concavity conditions and facilitates a consistent incorporation of thermodynamic principles. 
In this contribution, we focus on materials without preferred directions or internal variables.

While the formulation is posed in terms of entropy, the temperature is treated as the independent observable, and the entropy is inferred internally through the constitutive relation, enabling thermodynamically consistent modeling without requiring entropy data.

Thermodynamic admissibility of the networks is guaranteed by construction. The internal energy and dissipation potential are represented by input convex neural networks, ensuring convexity and compliance with the second law. Objectivity, material symmetry, and normalization are embedded directly into the architecture through invariant-based representations and zero-anchored formulations.

We demonstrate the performance of the proposed framework on synthetic and experimental datasets, including purely thermal problems and fully coupled thermomechanical responses of soft tissues and filled rubbers. The results show that the learned models accurately capture the underlying constitutive behavior. All code, data, and trained models are made publicly available via \href{https://doi.org/10.5281/zenodo.19248596}{Zenodo.org}.
\end{abstract}
\end{frontmatter}
\section{Introduction}

Mechanics and thermodynamics are intrinsically intertwined. Whenever materials deform, they store energy, exchange heat, and dissipate part of the supplied power through irreversible mechanisms \cite{ColemanNoll1963,LuPister1975,Ziegler2012}. 
This interplay has been studied extensively in classical continuum thermomechanics. Early thermoelastic formulations already highlighted the coupling between deformation and temperature \cite{Biot1958}, while later developments introduced internal-variable frameworks to describe irreversible processes \cite{ColemanGurtin1967,Halphen1975}. 
Variational approaches further unified energy storage and dissipation within a consistent thermodynamic setting \cite{Yang2006,Stainier2013,Teichtmeister2022}. 
Despite these advances, constructing constitutive models that remain thermodynamically consistent while capturing complex coupled responses across loading paths remains a challenging task.

Recent progress in physics-based machine learning aims to address this challenge by combining data-driven flexibility with physical structure. 
Within this context, two complementary research directions have emerged, which differ primarily in how physical principles are incorporated into the learning process.

The first direction focuses on constitutive modeling based on \textit{strongly-enforcing} network architectures. 
Here, neural networks are designed to represent material behavior directly, while thermodynamic admissibility is ensured by construction through architectural constraints. 
This idea has been established in Constitutive Artificial Neural Networks for automated model discovery \cite{Linka2021,Linka2023} and Physics-Augmented Neural Network formulations \cite{Klein2022}.
An alternative network architecture relies on the concept of Kolmogorov-Arnold Networks and is already applied for constitutive discovery \cite{ABDOLAZIZI2025106212,THAKOLKARAN2025118089,ji2026ickan}.
Subsequent developments improve robustness and flexibility, for example through enhanced constraint handling and parametrized model classes \cite{ChenGuilleminot2022,Klein2023Parametric,Linden2023}. 
These approaches have been successfully applied to increasingly complex inelastic materials \cite{Holthusen2024,FLASCHEL2025106103}, for instance, viscoelasticity \cite{ABDOLAZIZI2024112704,Kalina2025,Holthusen2026generalized}, plasticity \cite{Jadoon2025plasticity,Boes2026}, strain-induced crystallization \cite{Friedrichs2026crystal}, growth and remodeling \cite{Holthusen2025growth}, and non-convex anisotropic inelasticity \cite{holthusen2026complement}. 
A hierarchy of thermodynamically consistent learning frameworks is recently presented in \cite{jones2026hierarchy}.  
In the context of multiphysical extensions, \cite{Kalina2024Magneto,roth2025datadriven} introduce magneto-elastic multiscale formulations. 
Overall, this line of work provides a powerful framework for learning constitutive relations that are consistent by construction.

Complementary to this, the second direction is rooted in partial differential equations and relies on \textit{weakly-enforcing} network architectures, most prominently Physics-Informed Neural Networks (PINNs) \cite{Raissi2019}. 
In this setting, the governing equations are embedded into the loss function, enabling the solution of forward and inverse boundary value problems without explicit discretization schemes. 
Over the past years, PINNs have evolved into a versatile tool for scientific machine learning \cite{Karniadakis2021,Cuomo2022}, with numerous applications in thermal modeling. 
These include heat conduction in heterogeneous and anisotropic media \cite{Tian2021HeatPINN,Zhang2022MPINN,Xing2023HeatPINN}, as well as coupled transport phenomena such as porous and conjugate heat transfer \cite{Zhu2023PorousHeatPINN,Lee2025ConjugateHeatPINN}. 
Such developments demonstrate the capability of PINN-based approaches to handle strongly coupled field problems.

At the interface of these two directions, first studies addressing coupled thermomechanical behavior have emerged. 
PINN-based approaches have been applied to thermoelasticity, wave propagation, and poroelasticity \cite{Yang2024ThermoelasticPINN,Sabour2024MGT,Roy2024PoroelasticPINN,TANG2024104574}, while classical continuum formulations continue to provide the theoretical foundation for non-isothermal inelasticity at finite strains \cite{Casey1998,Casey2011,Lubarda2004,Junker2014}. 
More recently, gradient-enhanced and phase-field models have been developed to describe coupled damage and failure processes \cite{Felder2022,Lamm2024,Dittmann2020,Ambati2015}. 
However, a unified data-driven framework that combines constitutive learning with fully coupled thermomechanical consistency is still lacking.

From a data-driven constitutive perspective, existing approaches typically rely on specific thermodynamic potentials. 
Early work often prescribes parts of the temperature dependence and focuses on learning the remaining constitutive response \cite{Zlati2023}. 
More recent developments incorporate thermodynamic structure directly into the network architecture, most notably through formulations based on the Helmholtz energy \cite{Fuhg2024}. 
To the authors' knowledge, the latter represents the first strongly-enforcing network architecture for thermomechanics. 
While these approaches ensure consistency for thermoelastic processes, they do not address fully coupled thermomechanical behavior with dissipation in a general setting.

\paragraph{Research gap}
Recent progress in physics-based neural networks for thermomechanics has primarily focused on thermoelastic constitutive modeling. 
These approaches typically rely on Helmholtz energy formulations with convexity-like constraints in deformation and concavity in temperature. 
While this provides a rigorous framework for thermo-hyperelasticity, the extension to fully coupled thermomechanical processes with dissipation remains largely unexplored. 
In particular, a general framework that incorporates irreversible thermal effects, embeds thermomechanical principles by construction, and remains compatible with the observable temperature field is still an open challenge.

\paragraph{Aim of the study}
The aim of this work is to develop a physics-based neural network framework for the discovery of constitutive models in fully coupled thermomechanics. 
To this end, we formulate the constitutive behavior in terms of the internal energy and a dissipation potential, while treating temperature as the independent observable and entropy as an auxiliary variable inferred through the constitutive relation. 
This choice avoids the mixed convexity--concavity requirements of Helmholtz-based formulations and enables a thermodynamically consistent treatment of dissipative thermomechanical processes. 
Building on this foundation, we construct neural network architectures that satisfy thermodynamic admissibility and fundamental material principles by design.

\paragraph{Outline}
The remainder of this contribution is organized as follows. 
In \cref{sec:theory}, the thermomechanical framework is introduced, including the internal energy formulation and the underlying thermodynamic structure. 
The neural network architecture and the embedding of thermodynamic constraints are presented in \cref{sec:neural_network}. 
Subsequently, a series of numerical examples is investigated: a purely thermal problem (\cref{sec:heat_diffusion}), temperature-dependent mechanical behavior of experimental datasets (\cref{sec:porcine_tissue,sec:carbon_filler}), and a fully coupled structural setting (\cref{sec:structural_example}). 
The results are analyzed and discussed in \cref{sec:discussion}, followed by concluding remarks.
\section{Theoretical foundations}
\label{sec:theory}
We aim to design a neural network architecture that satisfies the governing principles of thermomechanics \emph{a priori}. 
For this purpose, we consider the continuum mechanical formulation of a fully coupled thermomechanical problem in the reference configuration. 
We first state the dissipation requirement and introduce constitutive potentials that ensure thermodynamic admissibility. 
We then summarize the balance equations for the coupled mechanical and thermal fields. 
Building on this foundation, we discuss material principles that are mandatory for admissible constitutive modeling, with a focus on isotropic materials. 
Thereafter, we outline additional material constraints that are not strictly required by the balance laws and the second law, but are highly beneficial when constructing robust and reliable network architectures. 
We conclude with the discretized weak forms of the balance equations.

\paragraph{Thermodynamic potentials}

In solid mechanics, the Helmholtz energy is commonly employed because it represents the mechanically available energy and admits the natural variables deformation gradient $\ten{F}$ and absolute temperature $T>0$. 
However, thermodynamic stability requires concavity of the Helmholtz energy with respect to $T$, together with mechanical stability conditions that are typically expressed through convexity-type requirements in appropriate deformation measures. 
Imposing these mixed curvature properties directly within neural network parametrizations is non-trivial.

We therefore adopt the internal energy $e$ and describe the material state in terms of the pair $(\ten{F},s)$, where $s$ denotes the referential entropy density. 
The constitutive structure is characterized by an energetic potential and a dissipation potential,
\begin{equation}
  e = e(\ten{F}, s), 
  \qquad
  \phi = \phi(\dot{\ten{F}}, \dot{s}, \ten{g}; \ten{F}, s),
  \label{eq:state_potentials}
\end{equation}
where the semicolon indicates that $(\ten{F},s)$ act solely as parameters of the dissipation potential. 
The dissipation potential $\phi$ accounts for irreversible processes in the sense of generalized standard materials \cite{Halphen1975,Ziegler2012}. 
The referential thermal gradient is defined as
\begin{equation}
  \ten{g} := -\frac{1}{T}\,\mathrm{Grad}(T) = -\mathrm{Grad}(\ln(T)),
  \label{eq:heat_gradient}
\end{equation}
whose contribution to $\phi$ is consistent with the concept of a conduction potential proposed in \cite{Biot1958}.

It is important to emphasize that, although the core formulation is expressed in terms of $(\ten{F},s)$, entropy is not measured directly in experiments. 
Instead, the temperature field $T$ is accessible. 
Through the constitutive relation linking $T$ and $s$, we infer the corresponding entropy from the prescribed deformation and temperature fields. 
Hence, while the model is formulated in terms of the internal energy and entropy, it remains fully compatible with measurable quantities\footnote{The relations induced by Legendre transformations between the internal energy and the Helmholtz energy, as well as the transformation of the dissipation potential with respect to the entropy rate $\dot{s}$ and its conjugate variable, are closely connected to variational formulations of thermomechanics. The interested reader is referred to \cite{Yang2006} for a detailed discussion.}.

\paragraph{Clausius--Duhem inequality}

To construct a constitutive framework that is thermodynamically admissible by design, we start from the second law of thermodynamics in its local form. 
In the reference configuration, the Clausius--Duhem inequality reads
\begin{equation}
  \ten{P}:\dot{\ten{F}} - \dot{e} + T\,\dot{s} + \ten{q}\cdot\ten{g} \ge 0,
  \label{eq:clausius_duhem}
\end{equation}
where $\ten{P}$ denotes the Piola stress tensor and $\ten{q}$ the referential heat flux. 
Using the constitutive potentials in \cref{eq:state_potentials}, we postulate the state laws
\begin{equation}
  \ten{P} - \frac{\partial e}{\partial \ten{F}} = \frac{\partial \phi}{\partial \dot{\ten{F}}},
  \qquad
  T - \frac{\partial e}{\partial s} = \frac{\partial \phi}{\partial \dot{s}},
  \qquad
  \ten{q} = \frac{\partial \phi}{\partial \ten{g}},
  \label{eq:state_laws}
\end{equation}
which define stress, temperature, and heat flux directly in terms of the energetic and dissipative potentials. 
In particular, the second relation establishes the link between the measurable temperature field and the entropy, which will play a central role in our discovery framework.

\paragraph{Balance equations}

While the Clausius--Duhem inequality constrains the local constitutive response, the evolution of the mechanical and thermal fields is governed by the corresponding balance laws. 
In the reference configuration and neglecting inertia for clarity, the balance of linear momentum is given by
\begin{equation}
  \mathrm{Div}\,\ten{P} + \ten{f} = \ten{0},
  \label{eq:balance_momentum}
\end{equation}
where $\ten{f}$ denotes the referential body force. 
The balance of internal energy reads
\begin{equation}
  \dot{e} = \ten{P}:\dot{\ten{F}} - \mathrm{Div}\,\ten{q} + r,
  \label{eq:balance_energy}
\end{equation}
with referential heat source $r$. 
Together with suitable Dirichlet and Neumann boundary conditions for the deformation and temperature fields, these equations define the fully coupled thermomechanical boundary value problem.
\subsection{Material principles}
\label{sec:principles}

We now discuss fundamental material principles that constitute mandatory requirements for admissible constitutive modeling. 
In the following, we restrict our attention to isotropic materials, that is, materials without preferred internal directions.

\paragraph{Objectivity}

The principle of objectivity requires invariance of the constitutive response under superposed rigid body motions. 
For every time-dependent rotation tensor $\ten{Q}(t) \in \mathrm{SO}(3)$, the internal energy and the dissipation potential must satisfy
\begin{equation}
\begin{aligned}
  e(\ten{F},s) 
  &= e(\ten{Q}\ten{F},s), \\
  \phi(\dot{\ten{F}}, \dot{s}, \ten{g}; \ten{F}, s)
  &= \phi(\dot{\overline{\ten{Q}\ten{F}}}, \dot{s}, \ten{g}; \ten{Q}\ten{F}, s),
\end{aligned}
\label{eq:objectivity_potentials}
\end{equation}
where $\dot{\overline{\ten{Q}\ten{F}}}$ denotes the material time derivative of $\ten{Q}\ten{F}$.
Note that $\ten{g}$ is a referential quantity.
In addition, all constitutively determined quantities must transform consistently. 
In particular, the Piola stress tensor and the referential heat flux are required to satisfy
\begin{equation}
\begin{aligned}
  \ten{Q}\ten{P}(\ten{F},s,\dot{\ten{F}}, \dot{s}, \ten{g})
  &= \ten{P}(\ten{Q}\ten{F},s,\dot{\overline{\ten{Q}\ten{F}}}, \dot{s}, \ten{g}), \\
  \ten{q}(\ten{F},s,\dot{\ten{F}}, \dot{s}, \ten{g})
  &= \ten{q}(\ten{Q}\ten{F},s,\dot{\overline{\ten{Q}\ten{F}}}, \dot{s}, \ten{g}).
\end{aligned}
\label{eq:objectivity_forces}
\end{equation}
Since the absolute temperature $T$ is a scalar field, it is invariant under superposed rigid body motions.

\paragraph{Material symmetry}

The principle of material symmetry states that the material response must be invariant under superposed time-independent rotations of the reference configuration that belong to the symmetry group of the material. 
For isotropic materials, this symmetry group coincides with the full special orthogonal group, that is $\ten{Q} \in \mathrm{SO}(3)$.
Accordingly, the constitutive potentials are required to satisfy
\begin{equation}
\begin{aligned}
  e(\ten{F},s) 
  &= e(\ten{F}\ten{Q},s), \\
  \phi(\dot{\ten{F}}, \dot{s}, \ten{g}; \ten{F}, s)
  &= \phi(\dot{\ten{F}}\ten{Q}, \dot{s}, \ten{Q}^{T}\ten{g}; \ten{F}\ten{Q}, s),
\end{aligned}
\label{eq:symmetry_potentials}
\end{equation}
and the associated constitutive quantities must fulfill
\begin{equation}
\begin{aligned}
  \ten{P}(\ten{F},s,\dot{\ten{F}}, \dot{s}, \ten{g})\ten{Q}
  &= \ten{P}(\ten{F}\ten{Q},s,\dot{\ten{F}}\ten{Q}, \dot{s}, \ten{Q}^{T}\ten{g}), \\
  \ten{Q}^{T}\ten{q}(\ten{F},s,\dot{\ten{F}}, \dot{s}, \ten{g})
  &= \ten{q}(\ten{F}\ten{Q},s,\dot{\ten{F}}\ten{Q}, \dot{s}, \ten{Q}^{T}\ten{g}).
\end{aligned}
\label{eq:symmetry_forces}
\end{equation}
The absolute temperature again satisfies this requirement trivially, since it is a scalar quantity.

Combining objectivity and material symmetry for isotropic materials, the internal energy and the dissipation potential can be regarded as \textit{scalar-valued isotropic functions},
\begin{equation}
  e(\ten{F},s) = e(\ten{C},s),
  \qquad
  \phi(\dot{\ten{F}}, \dot{s}, \ten{g}; \ten{F}, s)
  =
  \phi(\dot{\ten{C}}, \dot{s}, \ten{g}; \ten{C}, s),
  \label{eq:isotropic_function}
\end{equation}
depending on the right Cauchy--Green tensor $\ten{C}=\ten{F}^{T}\ten{F}$ and its rate. 
With this representation and the state laws \eqref{eq:state_laws}, the transformation properties in \cref{eq:objectivity_forces,eq:symmetry_forces} are satisfied.
\subsection{Material constraints}
\label{sec:constraints}
The following constraints are not strictly required by the balance laws and the Clausius--Duhem inequality. 
However, they are frequently imposed in order to improve stability, identifiability, and extrapolation properties of learned constitutive models. 
In this sense, they serve as guiding principles for the design of our physics-based neural network architecture.

\paragraph{Standard dissipative solids}

So far, the dissipation potential allows us to account for dissipative effects related to both deformation and thermal processes. 
The former is typically associated with fluid-like or viscous material behavior, whereas the latter becomes relevant for pronounced temperature gradients or rapid thermal processes, for instance under strong transient heating.\newline
In the present work, we restrict ourselves to the class of standard dissipative solids. 
Accordingly, we neglect the pair $(\dot{\ten{F}},\dot{s})$ in the argument list of $\phi$, while still allowing the potential to be parameterized by $(\ten{F},s)$. 
Under this assumption, the state laws \eqref{eq:state_laws} reduce to
\begin{equation}
  \ten{P} = \frac{\partial e}{\partial \ten{F}},
  \qquad
  T = \frac{\partial e}{\partial s} > 0,
  \label{eq:reduced_state_laws}
\end{equation}
which implies that the internal energy is \textit{monotonically increasing} with respect to the entropy.

\paragraph{Normalization}

When the material is in its rest state, it is common to impose normalization conditions on the potentials as well as on the constitutively dependent quantities. 
The rest state is characterized by a deformation gradient equal to the identity tensor, a reference temperature $T_0>0$, and the corresponding reference entropy $s_0$. 
While the reference temperature must be strictly positive, the reference entropy is less constrained. 
A common choice in continuum mechanics is to set $s_0=0$ at $T=T_0$, which we adopt here\footnote{For instance, a standard choice for the caloric part of the Helmholtz energy $\psi(T)=c_{T0}[ T-T_0-T\ln(\tfrac{T}{T_0}) ]$, differentiation with respect to $T$ yields an entropy satisfying $s_0=0$ at $T=T_0$. Here, $c_{T0}$ denotes the heat capacity.}.\newline
The normalization conditions for the potentials therefore read
\begin{equation}
  e(\ten{I},0) = 0,
  \qquad
  \phi(\ten{0};\ten{F},s) = 0.
\end{equation}
The latter condition expresses that the dissipation potential vanishes in the absence of its driving force, independently of the parametrization.
For the stress and temperature, we analogously obtain
\begin{equation}
  \ten{P}(\ten{I},0)
  =
  \left.
  \frac{\partial e}{\partial \ten{F}}
  \right|_{\substack{\tens{F}=\tens{I} \\ s=0}}
  =
  \ten{0},
  \qquad
  T(\ten{I},0)
  =
  \left.
  \frac{\partial e}{\partial s}
  \right|_{\substack{\tens{F}=\tens{I} \\ s=0}}
  =
  T_0.
\label{eq:normalization}
\end{equation}

\paragraph{Convexity of the internal energy}

The reduced temperature relation in \cref{eq:reduced_state_laws} provides a constitutive link between entropy and temperature. 
Up to this point, however, this relation is not necessarily uniquely invertible. 
While each entropy value yields a unique temperature, a given temperature could, in principle, correspond to multiple entropy values. 
Such behavior may arise, for example, in the presence of phase transformations, which are not considered in the present work.\newline
We therefore additionally require
\begin{equation}
  \frac{\partial^2 e}{\partial s^2} > 0,
  \label{eq:convexity_temperature}
\end{equation}
that is, \textit{strict convexity} of the internal energy with respect to the entropy.\newline
Similarly, we impose convexity-related conditions with respect to $\ten{F}$. 
Since convexity in $\ten{F}$ is generally too restrictive in finite elasticity, we adopt the concept of polyconvexity \cite{ball_convexity_1976}. 
Together with coercivity, polyconvexity provides a sufficient condition for the existence of minimizers \cite{ciarlet_mathematical_1988}. 
In the thermomechanical setting, the internal energy is said to be polyconvex if it admits a representation
\begin{equation}
  e = W(\ten{F}, \mathrm{cof}\ten{F}, J; s),
\end{equation}
which is convex in $(\ten{F},\mathrm{cof}\ten{F},J)$ for fixed entropy $s$. 
Here, $\mathrm{cof}\ten{F}$ denotes the cofactor of the deformation gradient and $J$ its determinant.

A subtle but important point is that the energy is required to be \textit{jointly convex} in $(\ten{F},\mathrm{cof}\ten{F},J)$ and convex in $s$, whereas between $s$ and the set $(\ten{F},\mathrm{cof}\ten{F},J)$ only \textit{separate convexity} is imposed. 
Consequently, for the extended set of variables $(\ten{F},\mathrm{cof}\ten{F},J,s)$ the full Hessian is not required to be positive (semi-)definite.

\paragraph{Convexity of the dissipation potential}

Since we restrict ourselves to standard dissipative solids, the derivatives of $\phi$ with respect to $\dot{\ten{F}}$ and $\dot{s}$ vanish identically. 
For the remaining contribution $\ten{q}\cdot\ten{g}$ in the dissipation inequality \eqref{eq:clausius_duhem}, appropriate structural restrictions on the dissipation potential are required.
To this end, we recall a classical result from convex analysis \cite{Rockafellar1970}. 
If the potential satisfies
\begin{equation}
  \phi(\ten{0}; \ten{F}, s) = 0,
  \qquad
  \phi(\ten{g}; \ten{F}, s) \ge 0,
  \qquad
  \phi(\ten{0}; \ten{F}, s)
  \ge
  \phi(\ten{g}; \ten{F}, s)
  - \partial_{\tens{g}}\phi \cdot \ten{g},
  \label{eq:constraints_potential}
\end{equation}
then the Clausius--Duhem inequality is satisfied \cite{Germain1998}. 
Here, $\partial_{\tens{g}}\phi$ denotes the subgradient of $\phi$ with respect to $\ten{g}$.
While \cref{eq:constraints_potential} essentially enforces convexity of the potential, convexity is only a sufficient, not a necessary, condition. 
More general formulations based on monotone potentials and their incorporation into physics-based neural networks can be found in \cite{holthusen2026complement}. 
In the present work, however, we restrict ourselves to convex dissipation potentials.
\subsection{Weak forms and their linearization}
\label{sec:weak_forms}

The internal energy and the dissipation potential are represented by physics-based neural networks.
For the identification of the constitutive behavior, we adopt an unsupervised learning strategy.
Accordingly, the neural network representations of the internal energy and the dissipation potential are embedded into a temporal and spatial discretization of the balance laws \cref{eq:balance_momentum,eq:balance_energy}. 
This results in a weak formulation of the coupled thermomechanical boundary value problem, which forms the basis of the numerical implementation.

\paragraph{Principle of virtual work}

For a given displacement field $\ten{u}$ and the temperature field $T$, the internal and external virtual work must be in equilibrium,
\begin{equation}
  w_{\mathrm{tot}} := w_{\mathrm{int}} - w_{\mathrm{ext}} \overset{!}{=} 0.
\end{equation}
The internal virtual work is given by
\begin{equation}
  w_{\mathrm{int}} 
  :=
  \int_\mathcal{B}
    \ten{P} : \mathrm{Grad}(\delta\ten{u})\,
  \mathrm{d}V
  +
    \int_\mathcal{B}
      \left[
    T\,\dot{s}\,\delta T
    -
    \ten{q}\cdot\mathrm{Grad}(\delta T)
  \right]
  \mathrm{d}V,
\end{equation}
where $\delta\ten{u}$ and $\delta T$ denote admissible test functions for the mechanical and thermal fields, respectively. 
The first term corresponds to the virtual work of the stresses, while the remaining terms represent the weak form of the thermal balance, expressed in terms of the entropy rate and the heat flux.
The external virtual work reads
\begin{equation}
  w_{\mathrm{ext}}
  :=
  \int_\mathcal{B}
  \ten{f}\cdot\delta\ten{u}\,\mathrm{d}V
  +
  \int_{\partial\mathcal{B}_t}
  \ten{t}\cdot\delta\ten{u}\,\mathrm{d}A
  +
  \int_\mathcal{B}
  r\,\delta T\,\mathrm{d}V
  +
  \int_{\partial\mathcal{B}_q}
  q\,\delta T\,\mathrm{d}A,
\end{equation}
where $\ten{t}$ denotes the prescribed traction on the Neumann boundary $\partial\mathcal{B}_t$ and $q$ refers to the prescribed heat flux on $\partial\mathcal{B}_q$.
On the Dirichlet boundaries, the essential boundary conditions
\begin{equation}
  \ten{u} = \tilde{\ten{u}},
  \qquad
  T = \tilde{T}
\end{equation}
are imposed.
On the Neumann boundaries, the natural boundary conditions are given by
\begin{equation}
  \ten{P}\cdot\ten{n} = \ten{t},
  \qquad
  \ten{q}\cdot\ten{n} = -q,
\end{equation}
where $\ten{n}$ denotes the outward unit normal vector in the reference configuration.

\paragraph{Spatial and temporal discretization}

In space, we adopt an isoparametric finite element discretization. The displacement and temperature fields are approximated as
\begin{equation}
  \ten{u}^h(\ten{X})
  =
  \sum_{a=1}^{n_{\mathrm{node}}}
  N_a(\ten{X})\,\ten{u}_a,
  \qquad
  T^h(\ten{X})
  =
  \sum_{a=1}^{n_{\mathrm{node}}}
  N_a(\ten{X})\,T_a,
\end{equation}
where $N_a$ denote the shape functions and $\ten{u}_a$, $T_a$ the corresponding nodal degrees of freedom. The same interpolation functions are employed for the geometrical mapping.
For the temporal discretization, we use an implicit backward Euler scheme for the entropy rate,
\begin{equation}
  \dot{s}^{n+1}
  =
  \frac{s^{n+1} - s^{n}}{\Delta t},
\end{equation}
with time increment $\Delta t$. All constitutive quantities are evaluated at time level $n+1$, leading to a fully coupled nonlinear system at each time step.
The entropy is treated as an auxiliary variable and is determined implicitly from the state law~\eqref{eq:reduced_state_laws}
\begin{equation}
  T^{n+\alpha}
  -
  \frac{\partial e^{n+\alpha}}{\partial s^{n+\alpha}}
  =
  0,
  \qquad
  \alpha \in \{0,1\},
\label{eq:local_discretized_state_law}
\end{equation}
which is solved locally at each quadrature point and time level.
Insertion of the discretized fields into the weak form yields a nonlinear algebraic system in residual form,
\begin{equation}
  \ten{r}
  =
  \begin{bmatrix}
    \ten{r}_{\tens{u}}(\ten{u}^h, T^h) \\
    \Delta t \,\ten{r}_{T}(\ten{u}^h, T^h)
  \end{bmatrix}
  =
  \ten{0},
  \label{eq:weak_residual}
\end{equation}
where the global residual vector collects the mechanical and thermal contributions. 
The thermal residual is scaled by the time increment $\Delta t$ to improve numerical conditioning. 
In analogy to the procedure of pseudo potentials \cite{Korelc2016}, the individual residual vectors are obtained by differentiating the discretized total virtual work with respect to the corresponding global vectors $(\bullet)_{\mathrm{glo}}$ of nodal trial values,
\begin{equation}
  \ten{r}_{\tens{u}}
  =
  \frac{\partial w^h_{\mathrm{tot}}}{\partial \delta \ten{u}_{\mathrm{glo}}^h},
  \qquad
  \ten{r}_{T}
  =
  \frac{\partial w^h_{\mathrm{tot}}}{\partial \delta T_{\mathrm{glo}}^h}.
\end{equation}
Since the trial values enter the weak form linearly, these derivatives are independent of their specific representation and can be evaluated efficiently using algorithmic differentiation.
The global residual in \cref{eq:weak_residual} plays a central role in the discovery procedure, as it serves as the physics-based loss term enforcing the balance laws at the discrete level.
\section{Physics-based neural network architecture}
\label{sec:neural_network}

With the theoretical foundations of the previous section, we are now enabled to design a physics-based neural network architecture that satisfies physics \textit{a priori}.
To this end, we will learn both the internal energy $e$ and the dissipation potential $\phi$ by means of physics-constraint neural networks.
As we have seen, convexity for both potentials is an essential ingredient.
To this end, we will employ Input Convex Neural Networks (ICNNs) \cite{Amos2017} to construct generally convex networks.
We will address how to incorporate the material principles discussed in \cref{sec:principles} and constraints in \cref{sec:constraints} into the overall architecture

\subsection{Zero-Anchored Neural Networks}

In the following, we introduce the general architecture of the neural networks employed in this contribution.
In \cref{sec:neural_representation}, we discuss how these architectures are tailored to the requirements of joint and separate convexity for the internal energy, the convex dissipation potential being parameterized by \((\ten{F}, s)\), and the prediction of entropy for stabilizing the training process.

\paragraph{Fully Input Convex Neural Network}

To begin with, we comment on fully Input Convex Neural Networks (FICNNs).
Their construction relies on three fundamental properties of convex functions:
the sum of convex functions is convex,
a non-negative scaling of a convex function remains convex,
and the composition $f \circ g$ is convex if $f$ is convex and monotonically non-decreasing and $g$ is convex.\newline
Based on these principles, the architecture reads
\begin{equation}
  \begin{aligned}
    \mathbf{x}_{1}
    &=
    \sigma_{0}\left(
      \mathbf{V}_{0}\,\mathbf{x}_{0}
      +
      \mathbf{b}_{0}
    \right)
    -
    \sigma_{0}\left(
      \mathbf{b}_{0}
    \right), \\
    \mathbf{x}_{\ell+1}
    &=
    \sigma_{\ell}\left(
      \mathbf{W}_{\ell}\,\mathbf{x}_{\ell}
      +
      \mathbf{V}_{\ell}\,\mathbf{x}_{0}
      +
      \mathbf{b}_{\ell}
    \right)
    -
    \sigma_{\ell}\left(
      \mathbf{b}_{\ell}
    \right),
  \end{aligned}
  \label{eq:FICNN}
\end{equation}
where the activation functions $\sigma_{\ell}$ are required to be convex and monotonically non-decreasing.
The weight matrices in the recurrent branch satisfy
$\mathbf{W}_{\ell} \in \mathbb{R}_{\ge 0}$,
while initially $\mathbf{V}_{\ell} \in \mathbb{R}$.
Additionally, there are no constraints imposed on the biases.
The subtraction of $\sigma_{\ell}(\mathbf{b}_{\ell})$ is in line with \cite{holthusen2026complement} and ensures that the network is zero-anchored, i.e., the output vanishes for vanishing input without affecting convexity.\newline
In our setting, convex functions are often provided as inputs to the ICNN rather than the raw arguments.
To preserve convexity under such compositions, we additionally restrict
$\mathbf{V}_{\ell} \in \mathbb{R}_{\ge 0}$.

\paragraph{Partially Input Convex Neural Networks}

While the internal energy depends exclusively on arguments for which convexity is required, the dissipation potential may be parameterized by $\ten{F}$ and $s$.
For this reason, we employ a partially Input Convex Neural Network (PICNN) \cite{Amos2017}, slightly adapted following \cite{holthusen2026complement},
\begin{equation}
  \begin{aligned}
    \mathbf{x}_{1}^c
    &=
    \sigma_{0}\!\left(
      \mathbf{V}_{0}^c\!\left[
        \mathbf{x}_{0}^c
        \odot
        \left[
          \mathbf{V}_{0}^{cp}\,\mathbf{x}_{0}^p
          +
          \mathbf{b}_{0}^{cp}
        \right]_+
      \right]
      +
      \mathbf{U}_{0}^{cp}\,\mathbf{x}_{0}^p
      +
      \mathbf{b}_{0}^c
    \right)
    -
    \sigma_{0}\!\left(
      \mathbf{U}_{0}^{cp}\,\mathbf{x}_{0}^p
      +
      \mathbf{b}_{0}^c
    \right), \\
    \mathbf{x}_{\ell+1}^c
    &=
    \sigma_{\ell}\!\left(
      \mathbf{W}_{\ell}^c\!\left[
        \mathbf{x}_{\ell}^c
        \odot
        \left[
          \mathbf{W}_{\ell}^{cp}\mathbf{x}_{\ell}^p
          +
          \mathbf{b}_{\ell}^{cp}
        \right]_+
      \right]
      +
      \mathbf{V}_{\ell}^c\!\left[
        \mathbf{x}_{0}^c
        \odot
        \left[
          \mathbf{V}_{\ell}^{cp}\mathbf{x}_{\ell}^p
          +
          \mathbf{c}_{\ell}^{cp}
        \right]_+
      \right]
      +
      \mathbf{U}_{\ell}^{cp}\,\mathbf{x}_{\ell}^p
      +
      \mathbf{b}_{\ell}^c
    \right)
    -
    \sigma_{\ell}\!\left(
      \mathbf{U}_{\ell}^{cp}\,\mathbf{x}_{\ell}^p
      +
      \mathbf{b}_{\ell}^c
    \right), \\
    \mathbf{x}_{\ell+1}^p
    &=
    f_{\ell}\!\left(
      \mathbf{W}_{\ell}^p\,\mathbf{x}_{\ell}^p
      +
      \mathbf{b}_{\ell}^p
    \right)
    -
    f_{\ell}\!\left(
      \mathbf{b}_{\ell}^p
    \right),
  \end{aligned}
  \label{eq:PICNN}
\end{equation}
where $[\bullet]_+$ denotes the \texttt{ReLU} activation.
All weights associated with the convex branch $(\bullet)^c$ are constrained to be non-negative in order to preserve convexity. 
In contrast, neither the weights nor the activation functions $f_{\ell}$ in the parameterization branch $(\bullet)^p$, nor the weights appearing in the coupling terms $(\bullet)^{cp}$, are subject to sign constraints.

Finally, we emphasize a crucial difference compared to neural networks employing ICNNs for elastic \cite{Linden2023,Klein2022} and inelastic materials \cite{holthusen2026complement,Kalina2025}.
In the present setting, strict convexity with respect to the entropy is required.
Although monotonicity is preserved under composition with increasing convex functions, multiplication by zero would reduce a strictly convex function to a merely convex one.
To prevent such degeneracy, all weights associated with the convex branch are constrained to be strictly positive.

\paragraph{Auxiliary Multilayer Perceptron}

Although the entropy can be obtained by iteratively solving \cref{eq:local_discretized_state_law}, this approach may be inefficient during training. 
The reason is that automatic differentiation would need to propagate gradients through all iterations of the solver unless implicit differentiation is employed.
As an alternative, we approximate the entropy using an auxiliary multilayer perceptron (MLP)
\begin{equation}
    \mathbf{x}_{\ell+1}
    =
    f_{\ell}\!\left(
      \mathbf{W}_{\ell}\,\mathbf{x}_{\ell}
      +
      \mathbf{b}_{\ell}
    \right)
    -
    f_{\ell}\!\left(
      \mathbf{b}_{\ell}
    \right),
    \label{eq:auxMLP}
\end{equation}
which is anchored at the origin in the same manner as the previously introduced networks. 
No additional constraints are imposed on the activation functions, weights, or biases.
After training, the iterative solver of \cref{eq:local_discretized_state_law} is reintroduced during inference.

\paragraph{Loss function}

The training objective consists of a physics-based loss, a regularization term, and an auxiliary loss,
\begin{equation}
  \mathcal{L}
  =
  \underbrace{
    \lambda_{\mathrm{D}}\,
    \mathcal{L}_{\mathrm{D}}
    +
    \lambda_{\mathrm{N}}\,
    \mathcal{L}_{\mathrm{N}}
  }_{\mathcal{L}_{\mathrm{phys}}}
  +
  \lambda_{\mathrm{R}}\,
  \mathcal{L}_{\mathrm{R}}
  +
  \lambda_{\mathrm{A}}\,
  \mathcal{L}_{\mathrm{aux}}.
\end{equation}

The physics loss $\mathcal{L}_{\mathrm{phys}}$ enforces the discrete balance laws and is constructed from the global residual vector introduced in \cref{eq:weak_residual}. 
It is decomposed into contributions associated with Dirichlet boundaries and Neumann boundaries (including free nodes), where $\lambda_{\mathrm{D}}$ and $\lambda_{\mathrm{N}}$ are scaling parameters.\newline
Each contribution is defined as the mean squared error of the corresponding residual components. 
Denoting by $\ten{r}_{\mathrm{D}}$ and $\ten{r}_{\mathrm{N}}$ the residual vectors associated with Dirichlet and Neumann boundaries, respectively, we define
\begin{equation}
  \mathcal{L}_{\mathrm{D}}
  =
  \mathrm{MSE}(\underline{\ten{r}}_{\mathrm{D}}),
  \qquad
  \mathcal{L}_{\mathrm{N}}
  =
  \mathrm{MSE}(\underline{\ten{r}}_{\mathrm{N}}).
\label{eq:L_phys_normalized}
\end{equation}
To ensure a consistent scaling of the residual contributions, the external loads entering the weak form are normalized. 
In particular, the prescribed mechanical tractions and body forces, as well as the thermal fluxes and heat sources, are scaled such that their respective maximum absolute values are equal to one. 
This normalization, indicated by $(\underline{\bullet})$, is performed separately for the mechanical and thermal problems based on their individual maximal absolute load values. 
Consequently, the mechanical and thermal contributions are scaled independently, preventing an artificial imbalance between the two fields.\newline
The regularization term \(\mathcal{L}_{\mathrm{R}}\) is introduced to enhance the stability of the training procedure and to control the complexity of the neural networks.
In particular, sparsity-promoting regularization can be employed to drive insignificant parameters towards zero, which is beneficial from both a mechanical interpretation and a model reduction perspective.
To this end, the regularization term is formulated as a penalty on the parameter tensors of selected subnetworks.
Considering the parameter tensors $\mathbf{W}_i$, $\mathbf{V}_i$, $\mathbf{U}_i$, and $\mathbf{b}_i$ associated with subnetwork $i$, a combination of \(L^1\)- and \(L^2\)-type contributions is employed to balance sparsity and smoothness, i.e.,
\begin{equation}
  \mathcal{L}_{\mathrm{R}}
  =
  \sum_{i \in \mathcal{J}_{\mathrm{reg}}}
  \sum_{\mathbf{P} \in \{\mathbf{W}_i, \mathbf{V}_i, \mathbf{U}_i, \mathbf{b}_i\}}
  \left[
    \lambda_{i}^{(1)} \, \lVert \mathbf{P} \rVert_{1}
    +
    \lambda_{i}^{(2)} \, \lVert \mathbf{P} \rVert_{F}^2
  \right],
\end{equation}
where \(\lVert \mathbf{P} \rVert_{1} = \sum_k |P_k|\) denotes the entry-wise \(L^1\)-norm and \(\lVert \mathbf{P} \rVert_{F}\) the Frobenius norm.
Here, \(\mathcal{J}_{\mathrm{reg}}\) denotes the set of subnetworks subject to regularization, and \(\lambda_{j}^{(1)}\), \(\lambda_{j}^{(2)}\) are the corresponding regularization coefficients.\newline
Finally, the auxiliary loss enforces consistency between the thermodynamic state law \cref{eq:reduced_state_laws} and the entropy predicted by the auxiliary MLP introduced in \cref{eq:auxMLP}. 
It is defined as
\begin{equation}
  \mathcal{L}_{\mathrm{aux}}
  =
  \mathrm{MSE}\!\left(
    T^{n+1}
    -
    \frac{\partial e^{n+1}}{\partial s^{n+1}}
  \right),
\end{equation}
and is evaluated at each quadrature point in the domain.

\subsection{Neural network representation of internal energy, dissipation potential, and entropy}
\label{sec:neural_representation}

\begin{figure}[!ht]
\centering
\vspace*{-2.5cm}
\begin{tikzpicture}
\node[anchor=center, inner sep=0] (img) at (0,0)
    {\includegraphics[width=0.8\textwidth]{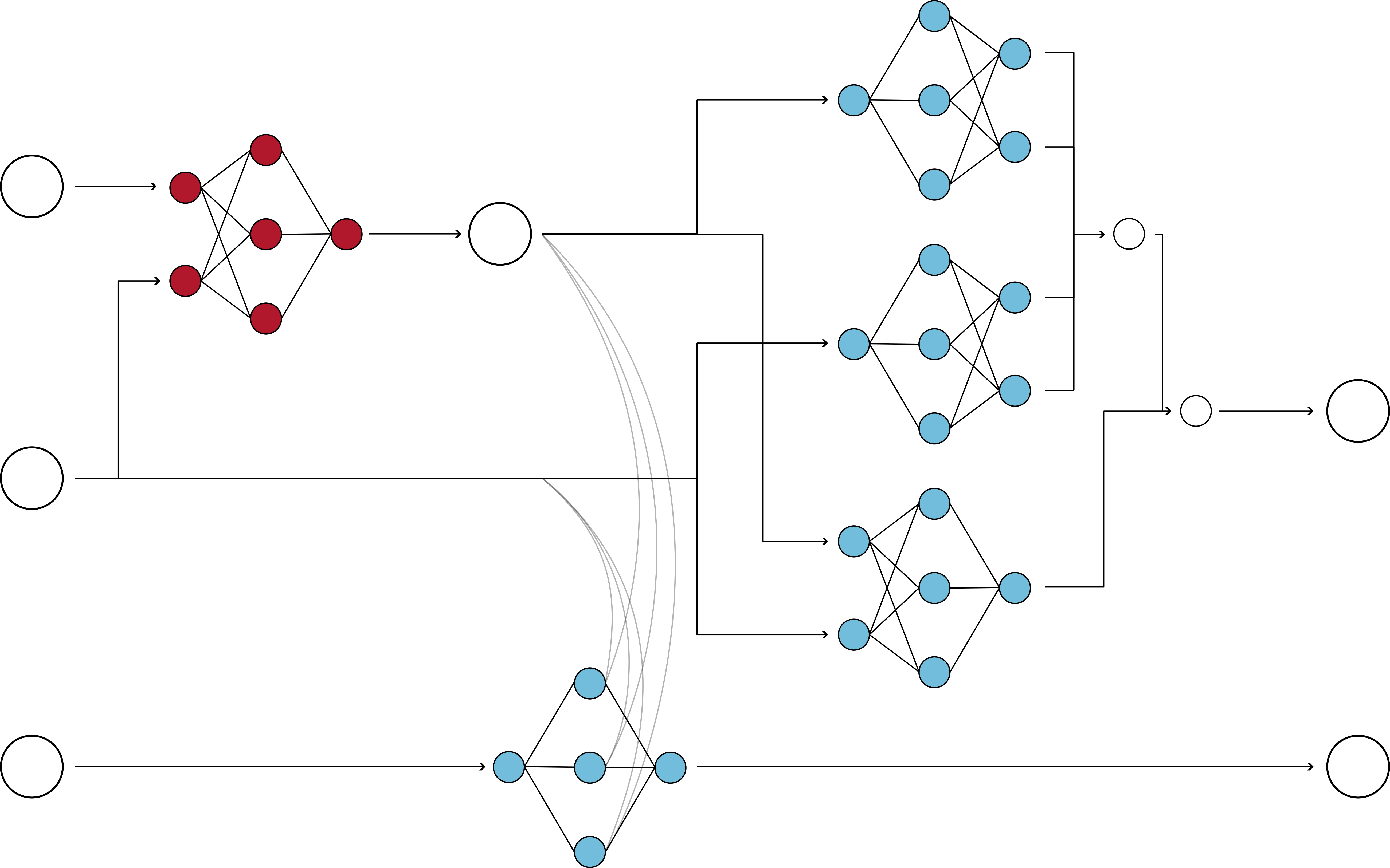}};

\node at (-7.0,-0.45) {$\ten{F}$};
\node at (-7.0,+2.62) {$T$};
\node at (-7.0,-3.51) {$\ten{g}$};

\node at (-2.07,+2.1) {$s$};

\node at (+4.59,+2.1) {$\ast$};
\node at (+5.29,0.24) {\tiny $+$};

\node at (+7.0,0.24) {$\ten{P}$};
\node at (+7.0,-3.51) {$\ten{q}$};

\node[anchor=west] at (+0.7,-0.5) {\small $\mathrm{FICNN}_{\tens{F},s}$};
\node[anchor=west] at (+0.7,+2.05) {\small $\mathrm{FICNN}_{\tens{F}}$};
\node[anchor=west] at (+0.7,+4.6) {\small $\mathrm{FICNN}_{s}$};

\node[anchor=west] at (-2.8,-2.50) {\small $\mathrm{PICNN}_{\tens{g}}$};

\node[anchor=west] at (-5.9,+3.2) {\small $\mathrm{MLP}_{s}$};

\node[] at (+6.0,+0.55) {$\frac{\partial}{\partial\tens{F}}$};
\node[] at (+6.0,-3.2) {$\frac{\partial}{\partial\tens{g}}$};


\end{tikzpicture}
\caption{Overall network architecture consisting of the subnetworks $\mathrm{MLP}_{s}$, $\mathrm{FICNN}_{s}$, $\mathrm{FICNN}_{\tens{F}}$, $\mathrm{FICNN}_{\tens{F},s}$, and $\mathrm{PICNN}_{\tens{g}}$, together with their interactions.
In a first step, the entropy $s$ is predicted by $\mathrm{MLP}_{s}$.
Subsequently, $s$ and the deformation gradient $\ten{F}$ are provided as inputs to the respective subnetworks, from which the first Piola--Kirchhoff stress $\ten{P}$ is obtained via differentiation.
In parallel, the referential heat gradient $\ten{g}$ is processed by $\mathrm{PICNN}_{\tens{g}}$, which is additionally parameterized by $(\ten{F},s)$, yielding the referential heat flux $\ten{q}$ through differentiation.}
\label{fig:overall_network}
\end{figure}
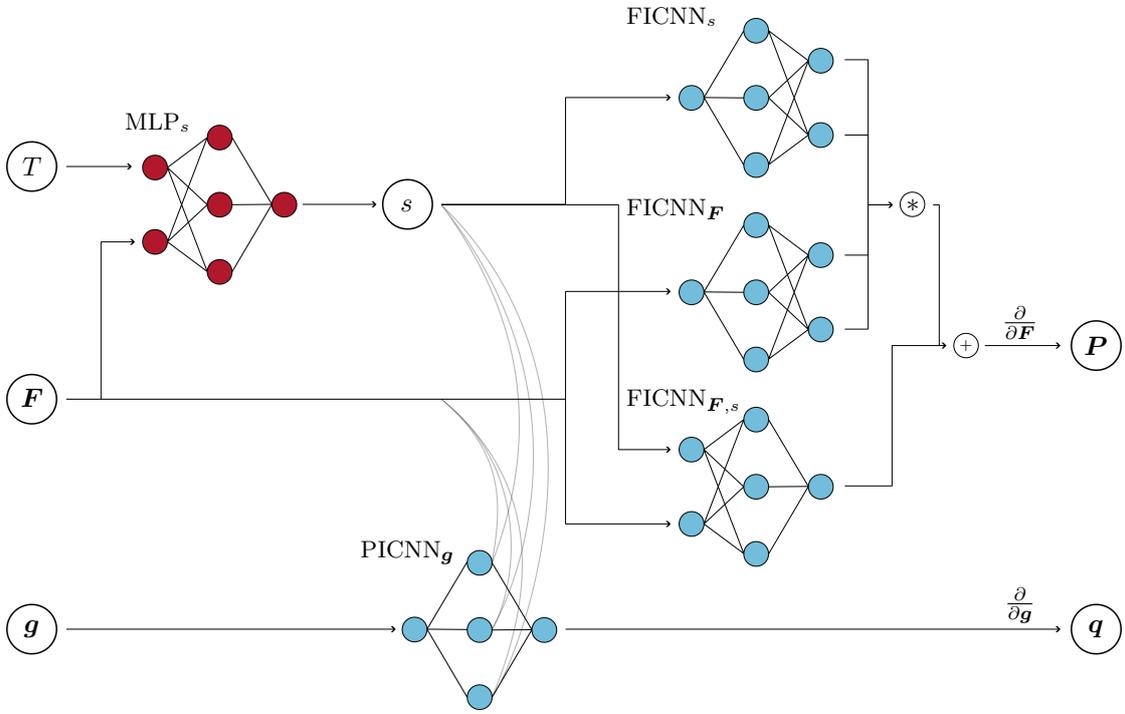

After we introduced the general neural network architectures, we now detail their specific design for the internal energy and the dissipation potential, which satisfies the material principles and constraints outlined in \cref{sec:principles,sec:constraints}. Furthermore, we describe the design for the prediction of entropy.
An overview of the network architecture and the interactions between the subnetworks is provided in \cref{fig:overall_network}.

\paragraph{Internal energy}

We begin by specifying the material principles imposed on the internal energy. 
As discussed above, the internal energy is formulated as a scalar-valued isotropic function and is therefore expressed in terms of the principal invariants \cite{Rivlin1955,Spencer1971,Smith1971}
\begin{equation}
  I_1 := \mathrm{tr}(\ten{C}),
  \qquad
  I_2 := \mathrm{tr}(\mathrm{cof}\ten{C}),
  \qquad
  J := \det(\ten{F}),
\end{equation}
which ensure objectivity and material symmetry and provide a polyconvex set of arguments. 
All inputs are shifted with respect to the rest configuration $\ten{F}=\ten{I}$,
\begin{equation}
  \bar I_1 := I_1 - 3,
  \qquad
  \bar I_2 := I_2 - 3,
  \qquad
  \bar J := J - 1,
  \qquad
  \bar s := s - s_0,
\end{equation}
such that the reference state corresponds to zero input. 
Based on these ingredients, we postulate the following decomposition of the internal energy,
\begin{equation}
  e(\ten{F},s)
  =
  n_T
  \left[
    \hat e(\ten{F},s)
    +
    e_{\mathrm{gr}}(\ten{F})
    -
    n_{\tens{P}}\,\bar J
  \right],
\end{equation}
where the individual contributions are introduced in the following.

To construct a convex representation, we introduce an auxiliary internal energy $\hat{e}$ based on fully input-convex neural networks (FICNNs), cf.~\cref{eq:FICNN},
\begin{equation}
\begin{aligned}
  \hat{e}(\ten{F},s)
  =
  \mathrm{FICNN}_{\tens{F},s}
  \!\left(
    [\,\bar I_1,\;\bar I_2,\;\bar J,\;-\bar J,\;\bar s\,]
  \right)
  +
  \left\{\mathrm{FICNN}_{\tens{F}}
  \!\left(
    [\,\bar I_1,\;\bar I_2,\;\bar J,\;-\bar J\,]
  \right)\right\}_+
  \;\ast\;
  \left\{\mathrm{FICNN}_{s}(\bar s) \right\}_+
  - n\, \left\{0 \right\}_+^2,
\end{aligned}
\end{equation}
where $\mathrm{FICNN}_{\tens{F},s}$ is scalar-valued, whereas $\mathrm{FICNN}_{\tens{F}}$ and $\mathrm{FICNN}_{s}$ produce feature vectors of equal dimension. 
The \texttt{Softplus} activation function is denoted by $\{\bullet\}_+$, and the operator $\ast$ denotes the elementwise product followed by summation over the feature dimension, such that the resulting contribution is scalar-valued. 
The parameter $n$ corresponds to the number of output features, and the subtraction term ensures proper normalization in the trivial case $\mathrm{FICNN}_{\tens{F}}=\mathrm{FICNN}_{s}=\ten{0}$.
The above decomposition reflects the partially separable convex structure of the internal energy: the first term captures joint convexity in deformation and entropy, while the second term enriches the representation through additional separable convex contributions.

Since polyconvexity alone is not sufficient to guarantee the existence of minimizers, a suitable coercive growth behavior of the energy is additionally required. 
To promote this property, we augment the internal energy by a small growth contribution,
\begin{equation}
\begin{split}
  e_{\mathrm{gr}}(\ten{F})
  &=
  \epsilon_{\mathrm{gr}}
  \left[
  \left[
    I_1 - 3 - \ln(\mathrm{det}(\ten{C}))
  \right]
    +
  \left[
    I_2 - 3 - \ln(\mathrm{det}(\mathrm{cof}\ten{C}))
  \right]
    +
  \left[
    J - 1 - \ln(J)
  \right]
  \right] \\
  &=
  \epsilon_{\mathrm{gr}}
  \left[
    \bar I_1
    +
    \bar I_2
    +
    \bar J
    -
    7 \ln(J)
  \right],
\end{split}
\end{equation}
where $\epsilon_{\mathrm{gr}} > 0$ is chosen sufficiently small. 
This term enhances the growth of the energy under large distortional deformations and prevents volumetric collapse as $J \to 0$. 
Notably, $e_{\mathrm{gr}}$ satisfies the normalization conditions of both energy and stress.

We next enforce the normalization condition of the Piola stress tensor stated in \cref{eq:normalization}. 
Following \cite{Linden2023}, we introduce the stress normalization factor
\begin{equation}
  n_{\tens{P}}
  =
  \left.
  \left(
    2\,\frac{\partial \hat e}{\partial I_1}
    +
    4\,\frac{\partial \hat e}{\partial I_2}
    +
    \frac{\partial \hat e}{\partial J}
  \right)
  \right|_{\substack{\tens{F}=\tens{I} \\ s=0}},
\end{equation}
which corresponds to the stress scaling at the reference configuration. 
In addition, temperature normalization is imposed via
\begin{equation}
  n_T
  =
  T_0
  \left[
    \left.
    \frac{\partial \hat e}{\partial s}
    \right|_{\substack{\tens{F}=\tens{I} \\ s=0}}
  \right]^{-1}.
\end{equation}
Since the auxiliary energy is constructed to be monotonically increasing with respect to entropy, the derivative $\partial \hat e / \partial s$ is strictly positive, ensuring that this normalization is well-defined. 
Moreover, due to the adopted ICNN architecture, the normalization condition of the energy is satisfied by construction.

\paragraph{Dissipation potential}

Lastly, we represent the dissipation potential by a neural network. 
Analogous to the internal energy, it is formulated as a scalar-valued isotropic function. 
Following \cite{Smith1971}, we introduce the invariants
\begin{equation}
  I_4 := \ten{g} \cdot \ten{g}, 
  \qquad 
  I_5 := \ten{g} \cdot \ten{C} \cdot \ten{g}, 
  \qquad 
  I_6 := \ten{g} \cdot \mathrm{cof}\ten{C} \cdot \ten{g}.
\end{equation}
Further, we reformulate these quantities in terms of tensor-induced norms. 
For a symmetric positive definite tensor $\ten{A}$, we define
\begin{equation}
  \| \ten{v} \|_{\tens{A}}
  :=
  \sqrt{\ten{v} \cdot \ten{A} \cdot \ten{v}},
\end{equation}
which satisfies absolute homogeneity and the triangle inequality and is therefore convex in $\ten{v}$.
Using this construction, we introduce the modified invariants
\begin{equation}
  \bar I_4 := \|\ten{g}\|_{\tens{I}}, 
  \qquad 
  \bar I_5 := \|\ten{g}\|_{\tens{C}}, 
  \qquad 
  \bar I_6 := \|\ten{g}\|_{\mathrm{cof}\tens{C}},
  \label{eq:mod_invars_potential}
\end{equation}
which preserve isotropy and ensure convexity with respect to $\ten{g}$.
The dissipation potential is then represented by a partially Input Convex Neural Network (PICNN), see \cref{eq:PICNN},
\begin{equation}
  \phi(\ten{g};\ten{F},s)
  =
  \mathrm{PICNN}_{\tens{g}}
  \left(
    [\,\bar I_4,\;\bar I_5,\;\bar I_6\,],\,
    [\,\bar J,\;\bar s\,]
  \right),
\end{equation}
which is convex with respect to $\ten{g}$ while being parameterized by $\bar J$ and $\bar s$.
Additional scaling by invariants such as $\bar I_1$ or $\bar I_2$ is omitted, since the dependence on $\ten{F}$ and its cofactor is already captured through $\bar I_5$ and $\bar I_6$.

Due to the adopted architecture, the dissipation potential satisfies $\phi(\ten{0};\ten{F},s)=0$ by construction. 
Non-negativity is ensured by applying an appropriate non-negative activation function to the scalar output neuron.
Noteworthy, a shifted activation function must be omitted for the output layer; otherwise, also non-negative output functions such as \texttt{ReLU} may produce negative values.
Consequently, the architecture guarantees non-negative dissipation by design, cf. \cref{eq:constraints_potential}.

\paragraph{Auxiliary entropy network}

Lastly, we specify the auxiliary multilayer perceptron (MLP) in \cref{eq:auxMLP}, which is employed during training to predict the entropy $s$. 
To this end, we introduce the temperature shift with respect to the reference state
\begin{equation}
  \bar{T} := T - T_0,
\end{equation}
which, together with the shifted deformation invariants, serves as input to the network. 
Accordingly, the entropy is approximated as
\begin{equation}
  s = \mathrm{MLP}_s\!\left([\,\bar{I}_1,\;\bar{I}_2,\;\bar{J},\;\bar{T}\,]\right).
\end{equation}
Since the entropy in the reference configuration is chosen as $s_0 = 0$, the zero-anchored architecture of the MLP is consistent with this normalization.
After training, the auxiliary MLP is replaced by the iterative solver enforcing the implicit state law for the entropy in \cref{eq:reduced_state_laws}.

\section{Training and Testing}

In the following, we examine the proposed physics-based network architecture for learning constitutive behavior in thermomechanics.
To this end, we study four different test cases: two based on experimental data, presented in \cref{sec:porcine_tissue,sec:carbon_filler}, and two based on synthetic data, presented in \cref{sec:heat_diffusion,sec:structural_example}.
The first three examples address behavior at the material-point level, or over a small spatial domain in the case of transient heat diffusion.
In the final example, we investigate the ability of the architecture to learn from full-field data and assess its accuracy in predicting nodal reactions for an unseen boundary value problem.

In this study, we do not address the measurement of physical fields, such as displacement, temperature, and their associated reactions, and instead assume full access to all fields.
Although this is clearly an idealized setting, the training of physics-based neural networks and the associated acquisition of experimental data constitute a substantial research topic in their own right.

The influence of the displacement field on the constitutive behavior arises primarily through its spatial gradient and time rate.
This contrasts with the temperature field, which enters the formulation indirectly through both the internal energy and the dissipation potential via the entropy.
In our experience, and consistent with common practice in the machine learning community, this tends to bias the training toward the entropy/temperature contribution, whose magnitude is generally larger than that of the spatial and temporal gradients.
Therefore, in line with the normalization procedure of the loss introduced in \cref{eq:L_phys_normalized}, we track the maximum temperature during training and normalize the entire temperature field by this value.
As a consequence, the temperature gradients are scaled accordingly.
During testing, this normalization value must be taken into account when evaluating unseen problems.

In all examples, we employ the relative activity of the individual subnetworks as an evaluation metric.
To this end, we quantify the activity of each subnetwork by aggregating the norms of its parameters (weights and biases).
More specifically, for each parameter tensor $\mathbf{W}_i$, $\mathbf{V}_i$, $\mathbf{U}_i$, and $\mathbf{b}_i$, an activity measure is computed as its Frobenius norm, and subsequently summed over all parameters belonging to the respective subnetwork $i$.
Denoting the resulting activity of subnetwork $i$ by $A_i$, the relative activity is defined as
\begin{equation}
  \tilde{A}_i = \frac{A_i}{\sum_j A_j},
  \label{eq:activity}
\end{equation}
which represents the normalized contribution of each subnetwork to the overall model activity.

The physics-based network is implemented in \texttt{JAX} using the \texttt{Flax} package.

Unless stated otherwise, all hyperparameters (training, constraints, regularization) and network architectures are kept identical across the examples.
Their specific values are summarized in \ref{app:hyperparameters}.
Furthermore, the synthetic data are generated using the constitutive models presented in \ref{app:constitutive}.
Notably, these models are formulated in terms of the Helmholtz energy and a temperature-based dissipation potential.
Hence, we do not explicitly prescribe within the network architecture the constitutive model used to generate the data; rather, during training, the network must learn the intrinsic relation implied by the Legendre transformation.
For completeness, a brief summary of the constitutive equations in terms of the Helmholtz energy is provided in \ref{app:balance_energy_psi}.

\subsection{Synthetic data: Heat diffusion}
\label{sec:heat_diffusion}

To begin with, we consider the discovery of the constitutive behavior of a purely transient heat problem, i.e., a rigid heat conductor.
Due to its transient nature, both the internal energy associated with the heat capacity and the dissipation potential governing the heat flux are non-zero.
As shown in \ref{app:constitutive}, we assume Fourier's law for the heat flux.
From the perspective of the dissipation potential (here equivalent to a conduction potential), Fourier's law induces a specific parametrization with respect to the temperature or, equivalently, the entropy.
Consequently, the network is required to learn this parametrization from the data.

The boundary value problem is loosely inspired by experimental setups for the measurement of \emph{U}-values in civil engineering applications \cite{Anderson1986}.
While heat fluxes and temperatures are reported in the literature, typically only the fluxes at the interior and exterior boundaries are measured.
However, for the identification of transient behavior, access to the temperature field across the wall is required.
Otherwise, one is restricted to assuming a constant temperature gradient, which is not consistent with transient heat conduction.

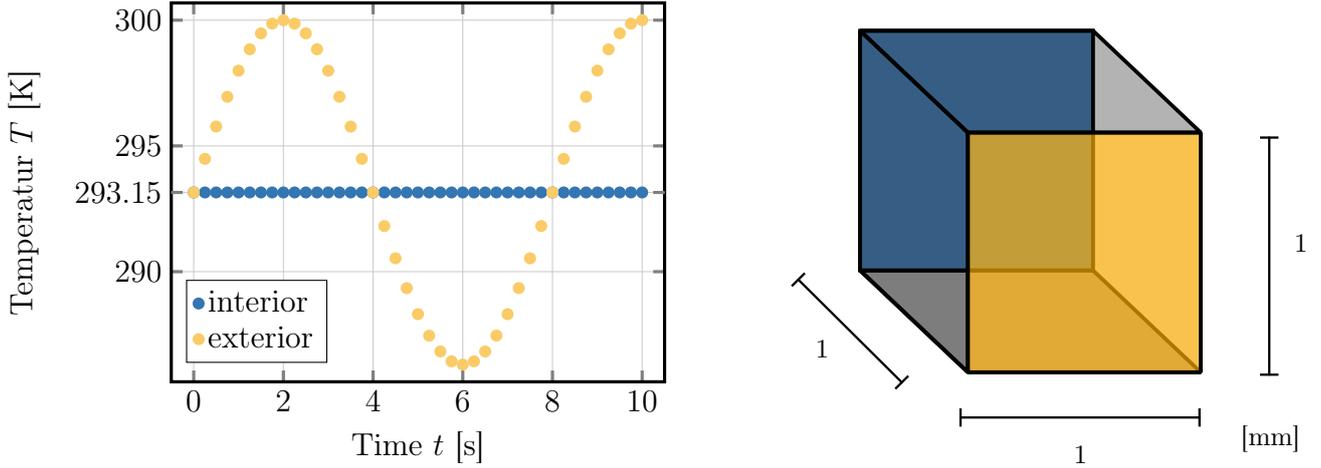
\begin{figure}[!t]
\centering
\begin{tikzpicture}
\begin{axis}[
    width=0.44\textwidth,
    height=0.36\textwidth,
    xlabel={Time $t$ [s]},
    ylabel={Temperatur $T$ [K]},
    ylabel style={yshift=20pt},
    grid=major,
    major grid style={gray!40},
    axis line style={very thick},
    tick style={very thick},
    tick label style={font=\large},
    label style={font=\large},
    legend style={
        at={(0.03,0.05)},
        anchor=south west,
        draw=black,
        fill=white,
        font=\large
    },
    legend cell align=left,
    enlargelimits=0.05,
    extra y ticks={293.15},
    extra y tick labels={293.15},
    extra y tick style={
        grid=major
    },
]

    \addplot[
      only marks,
      mark=*,
      mark size=1.5pt,
      mark options={line width=1.5pt},
      clr1!80,
    ] table[x=t, y=T_0, col sep=space]
    {./figures/ArtificialHeat/thermal_results.txt};
    \addlegendentry{interior}

    \addplot[
      only marks,
      mark=*,
      mark size=1.5pt,
      mark options={line width=1.5pt},
      clr4!60,
    ] table[x=t, y=T_1, col sep=space]
    {./figures/ArtificialHeat/thermal_results.txt};
     \addlegendentry{exterior}

\end{axis}
\end{tikzpicture}
\hspace*{1.4cm}
\begin{tikzpicture}

\node[anchor=center, inner sep=0] (img) at (0,0)
    {\includegraphics[width=0.35\textwidth]{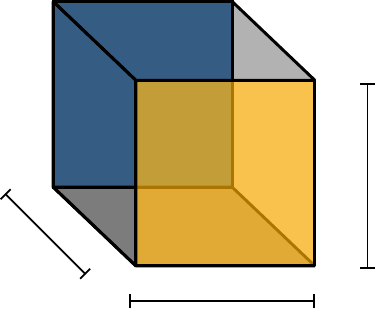}};

\node at (-2.8,-1.6) {$1$};
\node at (0.6,-3.0) {$1$};
\node at (3.5,-0.2) {$1$};
\node at (3.1,-2.8) {[mm]};

\end{tikzpicture}
\caption{Boundary value problem for the transient heat diffusion example. A cubic specimen with edge length $1\,\mathrm{mm}$ is subjected to a time-dependent temperature at the exterior boundary and a constant temperature at the interior boundary. The prescribed loading is given by
$
T(t)=\left[300-293.15\right]\sin\!\left(\frac{\pi}{4}t\right)+293.15,
$
while the initial temperature is $T_{\mathrm{init}}=293.15\,\mathrm{K}$.
The domain is discretized using $4\times 4\times 4$ elements with edge length $0.25\,\mathrm{mm}$.}
\label{fig:sketch_heat_diffusion}
\end{figure}

\begin{figure}[!t]
\centering
\begin{tikzpicture}
\begin{axis}[
    width=0.44\textwidth,
    height=0.36\textwidth,
    xlabel={Position [mm]},
    ylabel={Temperatur $T$ [K]},
    ylabel style={yshift=20pt},
    grid=major,
    major grid style={gray!40},
    axis line style={very thick},
    tick style={very thick},
    tick label style={font=\large},
    label style={font=\large},
    xtick={0,0.25,0.5,0.75,1},
xticklabels={
    \shortstack{0\\\small interior},
    0.25,
    0.5,
    0.75,
    \shortstack{1\\\small exterior}
},
    legend style={
        at={(1.03,1.00)},
        anchor=north west,
        draw=black,
        fill=white,
        font=\large
    },
    legend cell align=left,
    enlargelimits=0.05,
    extra y ticks={293.15},
    extra y tick labels={293.15},
    extra y tick style={
        grid=major
    },
]

    \addplot[
        smooth,
    line width = 1.5pt,
      ekE,
    ] table[x=x_step15, y=T_step15, col sep=space]
    {./graphs/temperature_over_x_multiple_times.txt};
    \addlegendentry{$t=2$ s}

    \addplot[
        smooth,
    line width = 1.5pt,
      ekA,
    ] table[x=x_step31, y=T_step31, col sep=space]
    {./graphs/temperature_over_x_multiple_times.txt};
     \addlegendentry{$t=4$ s}

    \addplot[
      only marks,
      mark=*,
      mark size=1.5pt,
      mark options={line width=1.5pt},
      clr4!60,
    ] coordinates {(1,293.15)};

    \addplot[
      only marks,
      mark=*,
      mark size=1.5pt,
      mark options={line width=1.5pt},
      clr1!80,
    ] coordinates {(0,293.15)};

\end{axis}
\end{tikzpicture}
\caption{
Temperature distribution at two exemplary time steps along the wall thickness direction for the transient heat diffusion problem. Although the temperatures at the interior and exterior boundaries are identical, a nonlinear temperature profile emerges within the wall as a result of transient effects. The edge length of one finite element is $0.25$ mm.
}
\label{fig:heat_diffusion_T_over_x}
\end{figure}
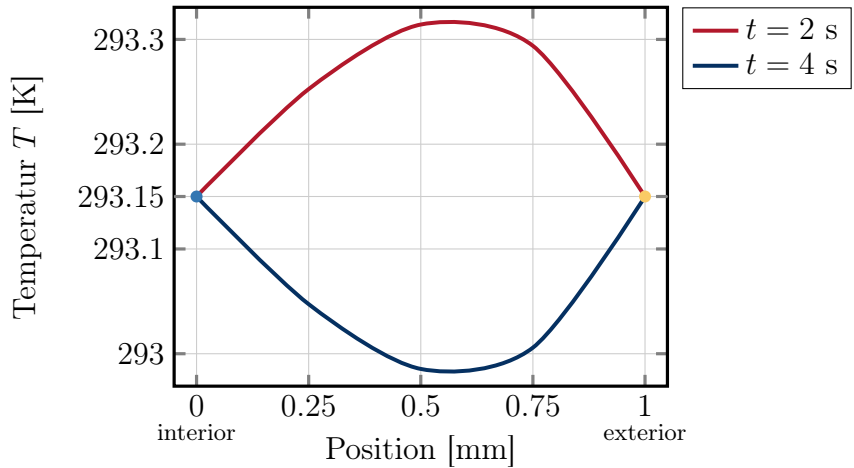

\cref{fig:sketch_heat_diffusion} illustrates the considered boundary value problem.
In accordance with realistic building conditions, the exterior temperature follows a sinusoidal variation over time, whereas the interior temperature is assumed to remain constant, cf.~\cite{Anderson1986}.
The domain is discretized using 64 hexahedral elements with an edge length of \(0.25\,\mathrm{mm}\).
\cref{fig:heat_diffusion_T_over_x} shows the temperature distribution along the wall thickness direction (from interior to exterior) at two distinct time instances \(t = 2\,\mathrm{s}\) and \(t = 4\,\mathrm{s}\).
The resulting temperature profile is nonlinear due to transient effects, which cannot be captured when using a single element.

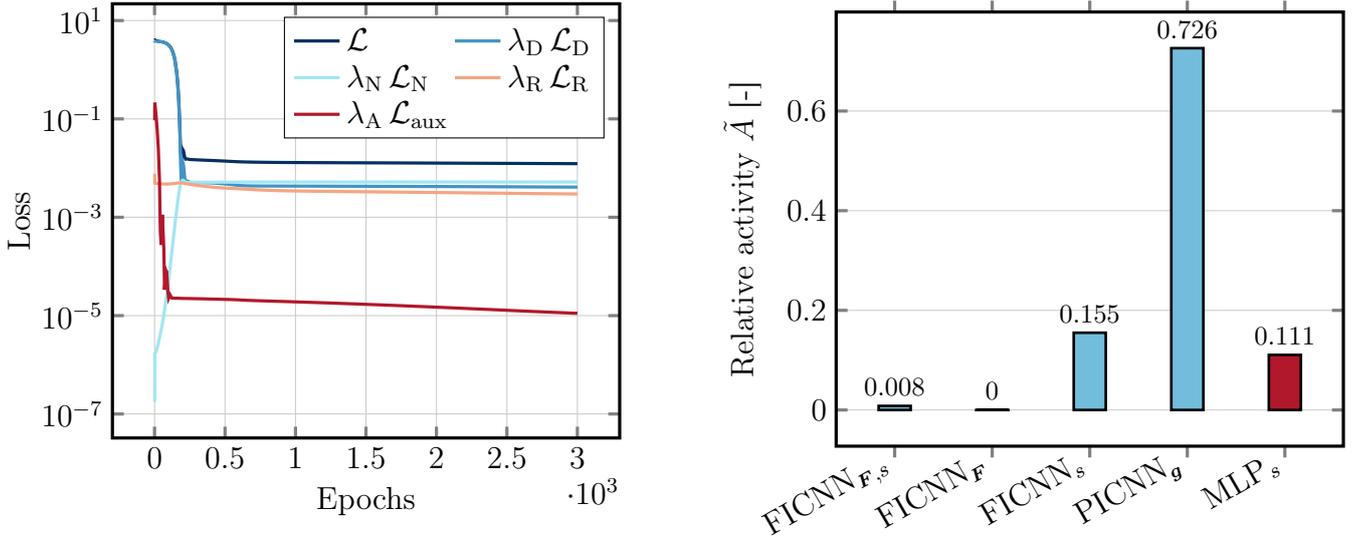
\begin{figure}[!t]
\centering
\begin{tikzpicture}[baseline=(current bounding box.north)]
\begin{axis}[
width=0.45\textwidth,
height=0.40\textwidth,
xlabel={Epochs},
ylabel={Loss},
grid=major,
ymode=log,
major grid style={gray!40},
axis line style={very thick},
tick style={very thick},
tick label style={font=\large},
label style={font=\large},
scaled x ticks=manual:{}{\pgfmathparse{#1/1000}},
xtick scale label code/.code={$\,\cdot 10^{3}$},
xticklabel style={
    /pgf/number format/fixed,
    /pgf/number format/precision=1
},
legend style={
    at={(0.97,0.97)},
    anchor=north east,
    draw=black,
    fill=white,
    font=\large,
},
legend columns=2,
legend cell align=left,
enlargelimits=0.1,
]

\addplot[
    ekA, solid, line width=1.2pt
] table[
    col sep=comma,
    x=epoch,
    y=loss
]{graphs/losses/reduced_loss_output_losses_Heat/loss_reduced.csv};

\addlegendentry{$\mathcal{L}$}

\addplot[
    ekB, line width=1.2pt
] table[
    col sep=comma,
    x=epoch,
    y=loss_force_DC
]{graphs/losses/reduced_loss_output_losses_Heat/loss_force_DC_reduced.csv};

\addlegendentry{$\lambda_{\mathrm{D}}\,\mathcal{L}_{\mathrm{D}}$}

\addplot[
    ekC, line width=1.2pt
] table[
    col sep=comma,
    x=epoch,
    y=loss_force_AC
]{graphs/losses/reduced_loss_output_losses_Heat/loss_force_AC_reduced.csv};

\addlegendentry{$\lambda_{\mathrm{N}}\,\mathcal{L}_{\mathrm{N}}$}

\addplot[
    ekD, line width=1.2pt
] table[
    col sep=comma,
    x=epoch,
    y=Reg
]{graphs/losses/reduced_loss_output_losses_Heat/Reg_reduced.csv};

\addlegendentry{$\lambda_{\mathrm{R}}\,\mathcal{L}_{\mathrm{R}}$}

\addplot[
    ekE, line width=1.2pt
] table[
    col sep=comma,
    x=epoch,
    y=Mismatch
]{graphs/losses/reduced_loss_output_losses_Heat/Mismatch_reduced.csv};

\addlegendentry{$\lambda_{\mathrm{A}}\,\mathcal{L}_{\mathrm{aux}}$}

\end{axis}

\end{tikzpicture}
\hspace*{1cm}
\begin{tikzpicture}[baseline=(current bounding box.north)]
\begin{axis}[
width=0.45\textwidth,
height=0.40\textwidth,
major grid style={gray!40},
axis line style={very thick},
tick style={very thick},
tick label style={font=\large},
label style={font=\large},
ymajorgrids=true,
x tick label style={rotate=30, anchor=east},
ybar,
bar width=12pt,
enlarge x limits=0.15,
ylabel={Relative activity $\tilde{A}$ [-]},
symbolic x coords={Em,Ef,Es,Di,Aux},
xtick={Em,Ef,Es,Di,Aux},
xticklabels={
    $\mathrm{FICNN}_{\tens{F},s}$,
    $\mathrm{FICNN}_{\tens{F}}$,
    $\mathrm{FICNN}_{s}$,
    $\mathrm{PICNN}_{\tens{g}}$,
    $\mathrm{MLP}_{s}$
},
nodes near coords,
nodes near coords style={
    /pgf/number format/.cd,
    fixed,
    precision=3
},
nodes near coords align={vertical},
]
\addplot[draw=black,fill=ekB!50!ekC,line width=1.0pt,bar shift=0pt] coordinates {(Em,8.199485684405e-03)};
\addplot[draw=black,fill=ekB!50!ekC,line width=1.0pt,bar shift=0pt] coordinates {(Ef,2.719317444385e-05)};
\addplot[draw=black,fill=ekB!50!ekC,line width=1.0pt,bar shift=0pt] coordinates {(Es,1.550436708103e-01)};
\addplot[draw=black,fill=ekB!50!ekC,line width=1.0pt,bar shift=0pt] coordinates {(Di,7.261554999301e-01)};
\addplot[draw=black,fill=ekE,line width=1.0pt,bar shift=0pt]        coordinates {(Aux,1.105741504008e-01)};
\end{axis}
\end{tikzpicture}
\caption{Left: Training history of the physics-based neural network for the transient heat diffusion problem. Shown are the total loss as well as its individual contributions, including the physics-based residual losses, the loss of the auxiliary network, and the regularization term, over the course of training iterations.
Right: Relative activity according to \cref{eq:activity} of the individual subnetworks for the transient heat diffusion problem. The diagram shows the normalized contributions of internal energy, dissipation, and auxiliary MLP to the overall activity. It can be observed that the thermal response is dominated by dissipative effects, while the remaining contributions are of significantly smaller magnitude.}
\label{fig:loss_heat_diffusion}
\end{figure}

For training, the weighting of the auxiliary loss is set to $\lambda_{\mathrm{A}} = 10^3$, and the model is trained for \(3{,}000\) epochs.
The evolution of all individual loss terms, as well as the total loss, is shown in \cref{fig:loss_heat_diffusion} (left).
As observed, the loss decreases rapidly during the initial phase of training up to approximately epoch \(250\), followed by a slower convergence regime.

\cref{fig:loss_heat_diffusion} (right) shows the relative activity of the individual subnetworks corresponding to the model with the lowest loss.
The purely mechanical network $\mathrm{FICNN}_{\tens{F}}$ is fully inactive, as expected for a purely thermal problem.
Nevertheless, due to the use of the \texttt{Softplus} activation function, the subnetwork $\mathrm{FICNN}_{s}$ still contributes to the overall energetic response.
The activity of the coupled subnetwork $\mathrm{FICNN}_{\tens{F},s}$ is close to zero, indicating that the energetic behavior is sufficiently captured by entropy alone.
It is plausible that stronger regularization would further suppress this contribution.
In contrast, the dissipation network clearly dominates the overall activity, which is consistent with the fact that the heat flux—and thus the governing physics—is directly encoded in the dissipation potential via the entropy dependence implied by Fourier's law.

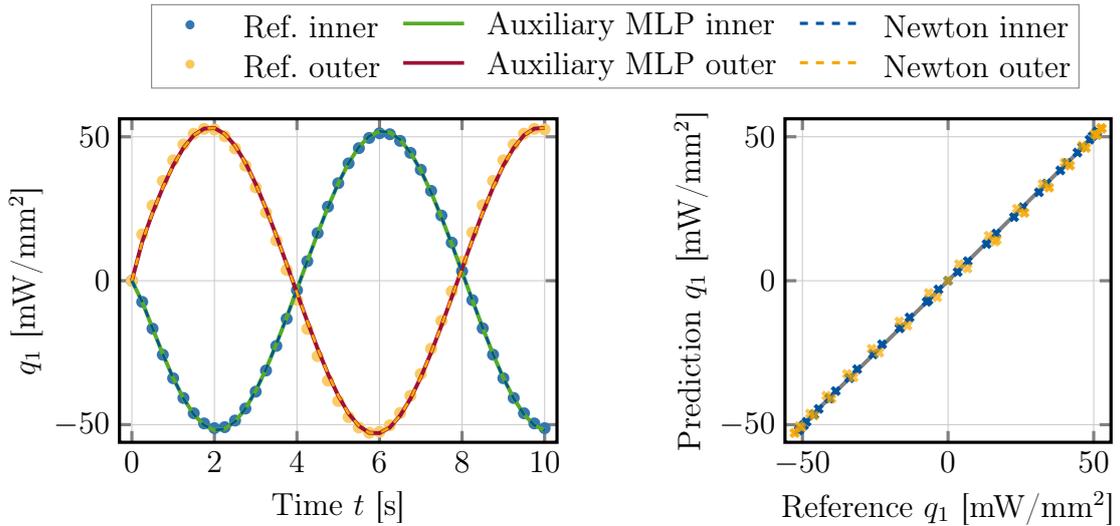
\begin{figure}[!t]
\centering
\begin{tikzpicture}

  \begin{groupplot}[
    group style={
      group size=2 by 1,
      horizontal sep=3cm
    },
    width=0.40\textwidth,
    height=0.32\textwidth,
    grid=major,
    major grid style={gray!40},
    axis line style={very thick},
    tick style={very thick},
    tick label style={font=\large},
    label style={font=\large},
    enlargelimits=0.03,
    clip mode=individual
  ]

    \nextgroupplot[
      xlabel={Time $t$ [s]},
      ylabel={$q_1$ [mW/mm\textsuperscript{2}]},
    ]

    \addplot[
      only marks,
      mark=*,
      mark size=1.5pt,
      mark options={line width=1.5pt},
      clr1!80,
    ] table[x=t, y=FT_0, col sep=space]
    {./figures/ArtificialHeat/thermal_results.txt};

    \addplot[
      only marks,
      mark=*,
      mark size=1.5pt,
      mark options={line width=1.5pt},
      clr4!60,
    ] table[x=t, y=FT_1, col sep=space]
    {./figures/ArtificialHeat/thermal_results.txt};

    \addplot[
      line width=1.5pt,
      clr5
    ] table[x=t, y=FT_NN_0, col sep=space]
    {./figures/ArtificialHeat/thermal_results.txt};

    \addplot[
      line width=1.5pt,
      clr2
    ] table[x=t, y=FT_NN_1, col sep=space]
    {./figures/ArtificialHeat/thermal_results.txt};

    \addplot[
      line width=1.0pt,
      dashed,
      clr1
    ] table[x=t, y=FT_NN_Newton_0, col sep=space]
    {./figures/ArtificialHeat/thermal_results.txt};

    \addplot[
      line width=1.0pt,
      dashed,
      clr4
    ] table[x=t, y=FT_NN_Newton_1, col sep=space]
    {./figures/ArtificialHeat/thermal_results.txt};

    \nextgroupplot[
      width=0.32\textwidth,
      xlabel={Reference $q_1$ [mW/mm\textsuperscript{2}]},
      ylabel={Prediction $q_1$ [mW/mm\textsuperscript{2}]},
    ]

    \addplot[
      gray,
      line width=1.5pt,
      domain=-50:50,
      samples=2
    ] {x};

    \addplot[
      only marks,
      mark=x,
      mark size=2pt,
      mark options={line width=1.5pt},
      clr1
    ] table[x=FT_0, y=FT_NN_Newton_0, col sep=space]
    {./figures/ArtificialHeat/thermal_results.txt};

    \addplot[
      only marks,
      mark=x,
      mark size=2pt,
      opacity=0.7,
      mark options={line width=1.5pt},
      clr4
    ] table[x=FT_1, y=FT_NN_Newton_1, col sep=space]
    {./figures/ArtificialHeat/thermal_results.txt};

  \end{groupplot}

  \coordinate (legendtop) at ($(group c1r1.north west)!0.5!(group c2r1.north east)+(0,4mm)$);

\def\legw{8mm}

\matrix[
  matrix of nodes,
  anchor=south,
  at={(legendtop)},
  draw=gray,
  inner sep=3pt,
  row sep=1.5mm,
  column sep=3mm,
  nodes={
    anchor=west,
    inner sep=0pt,
    outer sep=0pt,
    font=\large
  }
]{
\makebox[\legw][c]{\tikz[baseline=-0.6ex]\filldraw[clr1!80] (0,0) circle (1.5pt);} &
{Ref.\ inner} &
\makebox[\legw][c]{\tikz[baseline=-0.6ex]\draw[clr5, line width=1.4pt] (0,0) -- (\legw,0);} &
{Auxiliary MLP inner} &
\makebox[\legw][c]{\tikz[baseline=-0.6ex]\draw[clr1, dashed, line width=1.2pt] (0,0) -- (\legw,0);} &
{Newton inner} \\

\makebox[\legw][c]{\tikz[baseline=-0.6ex]\filldraw[clr4!65] (0,0) circle (1.5pt);} &
{Ref.\ outer} &
\makebox[\legw][c]{\tikz[baseline=-0.6ex]\draw[clr2, line width=1.4pt] (0,0) -- (\legw,0);} &
{Auxiliary MLP outer} &
\makebox[\legw][c]{\tikz[baseline=-0.6ex]\draw[clr4, dashed, line width=1.2pt] (0,0) -- (\legw,0);} &
{Newton outer} \\
  };

\end{tikzpicture}
\caption{Comparison of the heat flux component $q_1$ for the transient heat-conduction problem. Left: Temporal evolution of the heat flux at selected inner and outer locations, comparing the reference solution with the predictions of the auxiliary MLP and the Newton solver. Right: Parity plot of predicted versus reference heat flux values.}
\label{fig:results_heat_diffusion}
\end{figure}

The results for the discovered constitutive behavior of the heat flux are shown in \cref{fig:results_heat_diffusion}.
Due to the transient nature of the problem, the reference solution exhibits non-zero heat flux at $t=4\,\mathrm{s}$ and $t=8\,\mathrm{s}$, even in the absence of an instantaneous temperature gradient between interior and exterior boundary values.
The predictions obtained from the model using the auxiliary MLP for entropy estimation and from the reinstated formulation using an iterative solver are nearly identical.
This demonstrates that the auxiliary MLP does not introduce a noticeable approximation error.
For the iterative approach, a Newton--Raphson scheme is employed.
Furthermore, the parity plots of the heat flux component in the direction of heat flow show good agreement with the reference solution at both the interior and exterior boundaries.

\subsection{Experimental data: Porcine Tissue}
\label{sec:porcine_tissue}

Next, we investigate the ability of the proposed approach to discover temperature-dependent mechanical behavior from experimentally measured data of porcine tissue \cite{Zhang2018}.
The reported stress--stretch data for uniaxial tension, given in terms of the Piola stress tensor and the deformation gradient, are summarized in \cref{tab:data_porcine_tissue}.
This dataset was also analyzed in \cite{Fuhg2024}, where a physics-based neural network architecture for thermomechanics based on the Helmholtz energy was proposed.
In their approach, incompressible material behavior is assumed and enforced directly within the network architecture.

\begin{table}[!ht]
\centering
\caption{Piola stress $P_{11}$ as a function of the axial stretch $F_{11}$ for incompressible uniaxial loading at different temperatures for the porcine tissue \cite{Zhang2018}. The table summarizes the thermo-mechanical response used to generate the training data, illustrating the temperature dependence of the constitutive behavior.}
\label{tab:data_porcine_tissue}
\small
\renewcommand{\arraystretch}{0.9}
\resizebox{\textwidth}{!}{
\begin{tabular}{c|c||c|c||c|c||c|c||c|c||c|c}
\hline
\multicolumn{2}{c||}{$T=310.15$\,K} &
\multicolumn{2}{c||}{$T=318.15$\,K} &
\multicolumn{2}{c||}{$T=323.15$\,K} &
\multicolumn{2}{c||}{$T=333.15$\,K} &
\multicolumn{2}{c||}{$T=343.15$\,K} &
\multicolumn{2}{c}{$T=353.15$\,K} \\
\hline
$F_{11}$ [-] & $P_{11}$ [MPa] &
$F_{11}$ [-] & $P_{11}$ [MPa] &
$F_{11}$ [-] & $P_{11}$ [MPa] &
$F_{11}$ [-] & $P_{11}$ [MPa] &
$F_{11}$ [-] & $P_{11}$ [MPa] &
$F_{11}$ [-] & $P_{11}$ [MPa] \\
\hline
1.0 & 0.0000 & 1.0 & 0.0000 & 1.0 & 0.0000 & 1.0 & 0.0000 & 1.0 & 0.0000 & 1.0 & 0.0000 \\
1.2 & 0.0217 & 1.2 & 0.0146 & 1.2 & 0.0146 & 1.2 & 0.0108 & 1.2 & 0.0108 & 1.2 & 0.0108 \\
1.4 & 0.0436 & 1.4 & 0.0314 & 1.4 & 0.0314 & 1.4 & 0.0221 & 1.4 & 0.0221 & 1.4 & 0.0221 \\
1.5 & 0.0787 & 1.5 & 0.0527 & 1.5 & 0.0527 & 1.5 & 0.0320 & 1.5 & 0.0320 & 1.5 & 0.0320 \\
1.6 & 0.1756 & 1.6 & 0.1069 & 1.6 & 0.0875 & 1.6 & 0.0469 & 1.6 & 0.0469 & 1.6 & 0.0469 \\
1.7 & 0.3406 & 1.7 & 0.2271 & 1.7 & 0.1700 & 1.7 & 0.0718 & 1.7 & 0.0671 & 1.7 & 0.0594 \\
1.8 & 0.5556 & 1.8 & 0.3850 & 1.8 & 0.2900 & 1.8 & 0.1144 & 1.8 & 0.0928 & 1.8 & 0.0778 \\
\hline
\end{tabular}
}
\end{table}

Before proceeding with the identification, we briefly discuss two modeling inconsistencies present in the dataset.
First, the reported data contain multiple states in which the deformation gradient is equal to the identity tensor and the stress tensor vanishes, while corresponding to different temperatures.
At first glance, this may appear reasonable; however, it is inconsistent with thermomechanical theory when thermal expansion is taken into account.
In the presence of thermal expansion, a stress-free state at varying temperatures generally requires a temperature-dependent rest configuration.
A common interpretation is therefore to introduce an individual reference temperature $T_0$ for each experiment, such that the initial state satisfies $\ten{F}=\ten{I}$ and $\ten{P}=\ten{0}$ at $T=T_0$.
In contrast, within our framework, the reference temperature—and consequently the reference entropy $s_0$—is treated as a material parameter.
As a result, multiple stress-free states at different temperatures cannot be represented without explicitly accounting for thermally induced deformation.
The missing information in the dataset is thus the deformation associated with thermal expansion, even if its magnitude may be small.

Second, the assumption of incompressibility appears questionable in a thermomechanical setting.
An isotropic material undergoing thermal loading exhibits isotropic thermal expansion, which inherently leads to volumetric changes.
Hence, the material response is intrinsically compressible, even in the absence of mechanical loading.

Both issues could, in principle, be addressed by adapting the network architecture accordingly.
For example, the multiplicative structure of the Helmholtz energy employed in \cite{Fuhg2024} provides a convenient mechanism to incorporate such effects.
However, in the present work, we deliberately avoid tailoring the architecture to enforce incompressibility or multiple reference configurations, as this would reduce the generality of the approach.
Instead, we restrict the loading conditions to be consistent with the available data, i.e., incompressible uniaxial deformation with prescribed $F_{11}$, and train the model solely on the corresponding stress component $P_{11}$.
As a consequence, non-zero transverse stresses $P_{22}$ and $P_{33}$ arise implicitly, but are not considered further in this study.

Since the reference temperature is not provided in the dataset, we set the normalization factor to $n_T = 1$ and allow the network to implicitly identify the reference temperature during training.
It is worth noting that normalization of stress and temperature is not a fundamental requirement, although it is commonly employed to improve numerical conditioning.
In our experiments, omitting normalization led to reduced stability when using the auxiliary MLP for entropy prediction.
Therefore, we employ the iterative solver also during training, which increases the computational cost but improves robustness.
In line with \cite{Fuhg2024}, the model is trained for $30{,}000$ epochs.

\begin{figure}[!t]
\centering
\begin{tikzpicture}[baseline=(current bounding box.north)]
\begin{axis}[
    width=0.45\textwidth,
    height=0.40\textwidth,
    xlabel={Epochs},
    ylabel={Loss},
    grid=major,
    ymode=log,
    major grid style={gray!40},
    axis line style={very thick},
    tick style={very thick},
    tick label style={font=\large},
    label style={font=\large},
    legend style={
        at={(0.97,0.97)},
        anchor=north east,
        draw=black,
        fill=white,
        font=\large,
    },
    legend cell align=left,
    enlargelimits=0.1,
]

\addplot[
    ekA, line width=1.2pt
] table[
    col sep=comma,
    x=epoch,
    y=loss
]{graphs/losses/reduced_loss_output_losses_Pig_Tissue/loss_reduced.csv};

\addlegendentry{$\mathcal{L}$}

\addplot[
    ekB, line width=1.2pt
] table[
    col sep=comma,
    x=epoch,
    y=loss_force_DC
]{graphs/losses/reduced_loss_output_losses_Pig_Tissue/loss_force_DC_reduced.csv};

\addlegendentry{$\lambda_{\mathrm{D}}\,\mathcal{L}_{\mathrm{D}}$}

%

\addplot[
    ekD, line width=1.2pt
] table[
    col sep=comma,
    x=epoch,
    y=Reg
]{graphs/losses/reduced_loss_output_losses_Pig_Tissue/Reg_reduced.csv};

\addlegendentry{$\lambda_{\mathrm{R}}\,\mathcal{L}_{\mathrm{R}}$}

%

\end{axis}

\end{tikzpicture}
\hspace*{1cm}
\begin{tikzpicture}[baseline=(current bounding box.north)]
\begin{axis}[
width=0.45\textwidth,
height=0.40\textwidth,
major grid style={gray!40},
axis line style={very thick},
tick style={very thick},
tick label style={font=\large},
label style={font=\large},
ymajorgrids=true,
x tick label style={rotate=30, anchor=east},
ybar,
bar width=12pt,
enlarge x limits=0.15,
ylabel={Relative activity $\tilde{A}$ [-]},
symbolic x coords={Em,Ef,Es,Di,Aux},
xtick={Em,Ef,Es,Di,Aux},
xticklabels={
    $\mathrm{FICNN}_{\tens{F},s}$,
    $\mathrm{FICNN}_{\tens{F}}$,
    $\mathrm{FICNN}_{s}$,
    $\mathrm{PICNN}_{\tens{g}}$,
    $\mathrm{MLP}_{s}$
},
nodes near coords style={
    /pgf/number format/.cd,
    fixed,
    precision=3
},
nodes near coords align={vertical},
visualization depends on={y \as \rawy},
nodes near coords={
    \ifdim\rawy pt=0pt
        $\times$
    \else
        \pgfmathprintnumber[fixed,precision=3]{\rawy}
    \fi
},
]
\addplot[draw=black,fill=ekB!50!ekC,line width=1.0pt,bar shift=0pt] coordinates {(Em,3.454922037515e-01)};
\addplot[draw=black,fill=ekB!50!ekC,line width=1.0pt,bar shift=0pt] coordinates {(Ef,3.340355572643e-01)};
\addplot[draw=black,fill=ekB!50!ekC,line width=1.0pt,bar shift=0pt] coordinates {(Es,3.204722389842e-01)};
\addplot[draw=black,fill=ekB!50!ekC,line width=1.0pt,bar shift=0pt] coordinates {(Di,0.0)};
\addplot[draw=black,fill=ekE,line width=1.0pt,bar shift=0pt]        coordinates {(Aux,0.0)};
\end{axis}
\end{tikzpicture}
\caption{Left: Training history of the physics-based neural network for the temperature dependent porcine tissue dataset \cite{Zhang2018}. Shown are the total loss as well as its individual contributions, including the physics-based residual losses and the regularization term, over the course of training iterations.
Right: Relative activity according to \cref{eq:activity} of the individual subnetworks for the porcine tissue. The diagram shows the normalized contributions of internal energy. No transient thermal effects are present. Training is performed using an iterative solver instead of the auxiliary network.}
\label{fig:loss_porcine_tissue}
\end{figure}
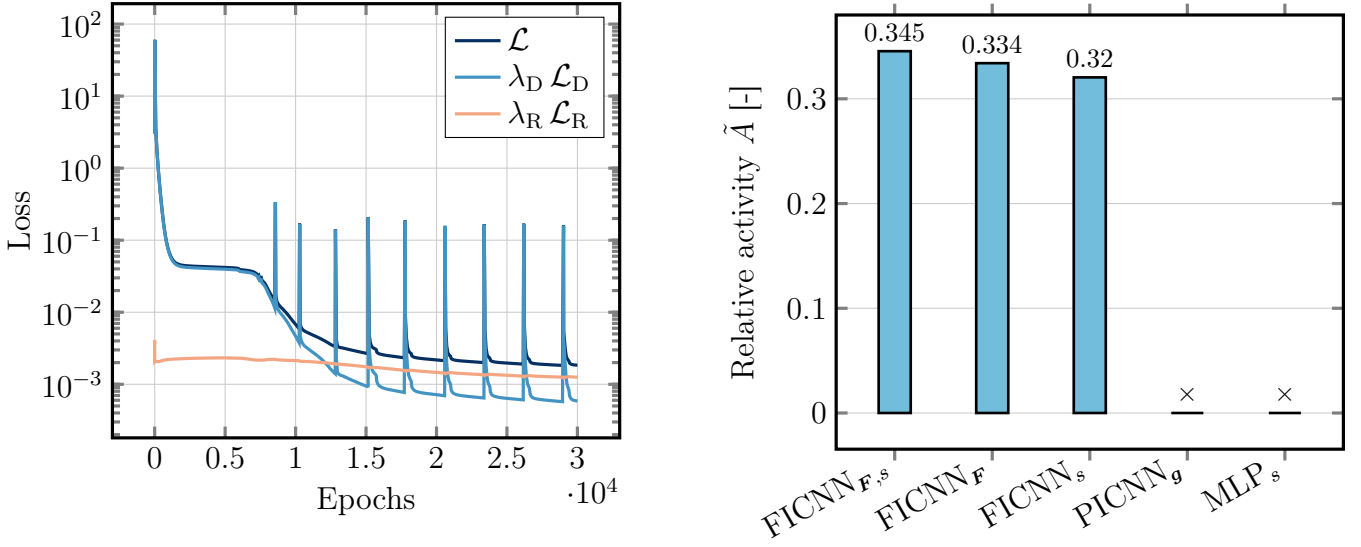

\cref{fig:loss_porcine_tissue} (left) illustrates the evolution of the individual loss terms.
Although the loss exhibits intermittent spikes to higher values, an overall decreasing trend is clearly visible, indicating stable convergence in a broader sense.

The corresponding relative activity of the subnetworks, evaluated at the parameter state with the lowest loss, is shown in \cref{fig:loss_porcine_tissue} (right).
In contrast to the previous example, all constitutive subnetworks exhibit non-negligible activity.
This suggests that a separation based on convexity with respect to deformation and entropy alone may not be sufficient to represent the observed material behavior.
Furthermore, the regularization loss remains consistently active throughout training, indicating that the model complexity is not significantly suppressed and that regularization does not dominate the learning process.

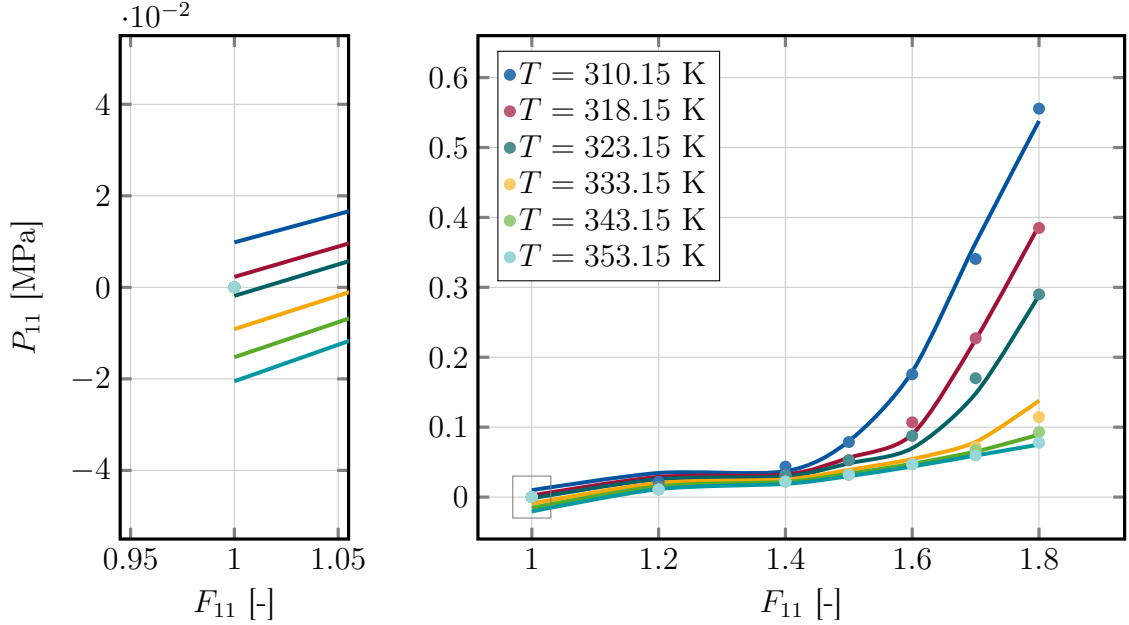
\begin{figure}[!ht]
\centering
\begin{tikzpicture}
\begin{axis}[
    width=0.25\textwidth,
    height=0.45\textwidth,
    xlabel={$F_{11}$ [-]},
    ylabel={$P_{11}$ [MPa]},
    xmin=0.95, xmax=1.05,
    ymin=-0.05, ymax=0.05,
    grid=major,
    major grid style={gray!40},
    axis line style={very thick},
    tick style={very thick},
    tick label style={font=\large},
    label style={font=\large},
    legend style={
        at={(0.03,0.97)},
        anchor=north west,
        draw=black,
        fill=white,
        font=\large
    },
    legend cell align=left,
    enlargelimits=0.05,
]

\addplot[
    only marks,
    mark=*,
    mark size=1.5pt,
    mark options={
        line width=1.5pt
    },
    clr1!80,
] coordinates {
(1.0,0.0000)
(1.2,0.0217)
(1.4,0.0436)
(1.5,0.0787)
(1.6,0.1756)
(1.7,0.3406)
(1.8,0.5556)
};

\addplot[
    only marks,
    mark=*,
    mark size=1.5pt,
    mark options={
        line width=1.5pt
    },
    clr2!70,
] coordinates {
(1.0,0.0000)
(1.2,0.0146)
(1.4,0.0314)
(1.5,0.0527)
(1.6,0.1069)
(1.7,0.2271)
(1.8,0.3850)
};

\addplot[
    only marks,
    mark=*,
    mark size=1.5pt,
    mark options={
        line width=1.5pt
    },
    clr3!70,
] coordinates {
(1.0,0.0000)
(1.2,0.0146)
(1.4,0.0314)
(1.5,0.0527)
(1.6,0.0875)
(1.7,0.1700)
(1.8,0.2900)
};

\addplot[
    only marks,
    mark=*,
    mark size=1.5pt,
    mark options={
        line width=1.5pt
    },
    clr4!60,
] coordinates {
(1.0,0.0000)
(1.2,0.0108)
(1.4,0.0221)
(1.5,0.0320)
(1.6,0.0469)
(1.7,0.0718)
(1.8,0.1144)
};

\addplot[
    only marks,
    mark=*,
    mark size=1.5pt,
    mark options={
        line width=1.5pt
    },
    clr5!60,
] coordinates {
(1.0,0.0000)
(1.2,0.0108)
(1.4,0.0221)
(1.5,0.0320)
(1.6,0.0469)
(1.7,0.0671)
(1.8,0.0928)
};

\addplot[
    only marks,
    mark=*,
    mark size=1.5pt,
    mark options={
        line width=1.5pt
    },
    clr9!40,
] coordinates {
(1.0,0.0000)
(1.2,0.0108)
(1.4,0.0221)
(1.5,0.0320)
(1.6,0.0469)
(1.7,0.0594)
(1.8,0.0778)
};


\addplot[
    line width=1.5pt,
    clr1
] table[x expr={1 + \thisrow{ux_right_mean}}, y=Fx_right_sum, col sep=space]
{./figures/PigTissue/Pig_Tissue_00_F_and_u.dat};

\addplot[
    line width=1.5pt,
    clr2
] table[x expr={1 + \thisrow{ux_right_mean}}, y=Fx_right_sum, col sep=space]
{./figures/PigTissue/Pig_Tissue_01_F_and_u.dat};

\addplot[
    line width=1.5pt,
    clr3
] table[x expr={1 + \thisrow{ux_right_mean}}, y=Fx_right_sum, col sep=space]
{./figures/PigTissue/Pig_Tissue_02_F_and_u.dat};

\addplot[
    line width=1.5pt,
    clr4
] table[x expr={1 + \thisrow{ux_right_mean}}, y=Fx_right_sum, col sep=space]
{./figures/PigTissue/Pig_Tissue_03_F_and_u.dat};

\addplot[
    line width=1.5pt,
    clr5
] table[x expr={1 + \thisrow{ux_right_mean}}, y=Fx_right_sum, col sep=space]
{./figures/PigTissue/Pig_Tissue_04_F_and_u.dat};

\addplot[
    line width=1.5pt,
    clr9
] table[x expr={1 + \thisrow{ux_right_mean}}, y=Fx_right_sum, col sep=space]
{./figures/PigTissue/Pig_Tissue_05_F_and_u.dat};

\end{axis}
\end{tikzpicture}
\hspace*{10pt}
\begin{tikzpicture}
\begin{axis}[
    width=0.55\textwidth,
    height=0.45\textwidth,
    xlabel={$F_{11}$ [-]},
    xmin=1.0, xmax=1.85,
    ymin=0.0, ymax=0.60,
    grid=major,
    major grid style={gray!40},
    axis line style={very thick},
    tick style={very thick},
    tick label style={font=\large},
    label style={font=\large},
    legend style={
        at={(0.03,0.97)},
        anchor=north west,
        draw=black,
        fill=white,
        font=\large
    },
    legend cell align=left,
    enlargelimits=0.1,
]

\addplot[
    only marks,
    mark=*,
    mark size=1.5pt,
    mark options={
        line width=1.5pt
    },
    clr1!80,
] coordinates {
(1.0,0.0000)
(1.2,0.0217)
(1.4,0.0436)
(1.5,0.0787)
(1.6,0.1756)
(1.7,0.3406)
(1.8,0.5556)
};
\addlegendentry{$T=310.15$ K}

\addplot[
    only marks,
    mark=*,
    mark size=1.5pt,
    mark options={
        line width=1.5pt
    },
    clr2!70,
] coordinates {
(1.0,0.0000)
(1.2,0.0146)
(1.4,0.0314)
(1.5,0.0527)
(1.6,0.1069)
(1.7,0.2271)
(1.8,0.3850)
};
\addlegendentry{$T=318.15$ K}

\addplot[
    only marks,
    mark=*,
    mark size=1.5pt,
    mark options={
        line width=1.5pt
    },
    clr3!70,
] coordinates {
(1.0,0.0000)
(1.2,0.0146)
(1.4,0.0314)
(1.5,0.0527)
(1.6,0.0875)
(1.7,0.1700)
(1.8,0.2900)
};
\addlegendentry{$T=323.15$ K}

\addplot[
    only marks,
    mark=*,
    mark size=1.5pt,
    mark options={
        line width=1.5pt
    },
    clr4!60,
] coordinates {
(1.0,0.0000)
(1.2,0.0108)
(1.4,0.0221)
(1.5,0.0320)
(1.6,0.0469)
(1.7,0.0718)
(1.8,0.1144)
};
\addlegendentry{$T=333.15$ K}

\addplot[
    only marks,
    mark=*,
    mark size=1.5pt,
    mark options={
        line width=1.5pt
    },
    clr5!60,
] coordinates {
(1.0,0.0000)
(1.2,0.0108)
(1.4,0.0221)
(1.5,0.0320)
(1.6,0.0469)
(1.7,0.0671)
(1.8,0.0928)
};
\addlegendentry{$T=343.15$ K}

\addplot[
    only marks,
    mark=*,
    mark size=1.5pt,
    mark options={
        line width=1.5pt
    },
    clr9!40,
] coordinates {
(1.0,0.0000)
(1.2,0.0108)
(1.4,0.0221)
(1.5,0.0320)
(1.6,0.0469)
(1.7,0.0594)
(1.8,0.0778)
};
\addlegendentry{$T=353.15$ K}

\addplot[
    smooth,
    line width=1.5pt,
    clr1
] table[x expr={1 + \thisrow{ux_right_mean}}, y=Fx_right_sum, col sep=space]
{./figures/PigTissue/Pig_Tissue_00_F_and_u.dat};

\addplot[
    smooth,
    line width=1.5pt,
    clr2
] table[x expr={1 + \thisrow{ux_right_mean}}, y=Fx_right_sum, col sep=space]
{./figures/PigTissue/Pig_Tissue_01_F_and_u.dat};

\addplot[
    smooth,
    line width=1.5pt,
    clr3
] table[x expr={1 + \thisrow{ux_right_mean}}, y=Fx_right_sum, col sep=space]
{./figures/PigTissue/Pig_Tissue_02_F_and_u.dat};

\addplot[
    smooth,
    line width=1.5pt,
    clr4
] table[x expr={1 + \thisrow{ux_right_mean}}, y=Fx_right_sum, col sep=space]
{./figures/PigTissue/Pig_Tissue_03_F_and_u.dat};

\addplot[
    smooth,
    line width=1.5pt,
    clr5
] table[x expr={1 + \thisrow{ux_right_mean}}, y=Fx_right_sum, col sep=space]
{./figures/PigTissue/Pig_Tissue_04_F_and_u.dat};

\addplot[
    smooth,
    line width=1.5pt,
    clr9
] table[x expr={1 + \thisrow{ux_right_mean}}, y=Fx_right_sum, col sep=space]
{./figures/PigTissue/Pig_Tissue_05_F_and_u.dat};

\draw[gray, thin] (axis cs:0.97,-0.03) rectangle (axis cs:1.03,0.03);

\end{axis}
\end{tikzpicture}
\caption{Comparison of the discovered model and the reference data for porcine tissue under incompressible uniaxial loading at different temperatures. The Piola stress $P_{11}$ is plotted as a function of the axial stretch $F_{11}$. The reference data are shown as discrete points, while the continuous curves represent the response of the discovered model. Left: Enlarged view near $F_{11}=1$. Right: Full stretch range, demonstrating that the discovered model accurately captures the strongly nonlinear and temperature-dependent constitutive response.}
\label{fig:results_porcine_tissue}
\end{figure}

The discovered neural constitutive response is shown in \cref{fig:results_porcine_tissue}.
The model qualitatively reproduces the experimental data, capturing both the temperature-dependent mechanical response and the softening behavior with increasing temperature.
Moreover, the predicted responses are of comparable magnitude to those obtained in \cite{Fuhg2024}.

Interestingly, despite not explicitly enforcing a normalization of the internal energy, the trained network learns a representation whose derivative with respect to entropy yields a reference temperature that is consistent with the near stress-free states across the considered temperature range.
This indicates that the network is capable of implicitly identifying a thermodynamically consistent energy structure.

Finally, we emphasize again that, although the results are qualitatively comparable to those reported in \cite{Fuhg2024}, the underlying modeling assumptions differ significantly.
Therefore, a direct quantitative comparison is not appropriate, and this example should be interpreted as a proof of concept rather than a benchmark comparison.

\subsection{Experimental data: Carbon-filled black rubber}
\label{sec:carbon_filler}

Our final experimental study considers the dataset reported in \cite{Fu2021} for carbon-filled rubber.
We extract fifteen data points from the dataset, which are summarized in \cref{tab:data_rubber}.
This dataset was also analyzed in \cite{Fuhg2024} using a physics-based neural network formulation based on the Helmholtz energy, employing once again the modeling assumption of incompressibility.
Accordingly, we follow a similar procedure as described in the previous example in \cref{sec:porcine_tissue}.

\begin{table}[!ht]
\centering
\caption{Piola stress $P_{11}$ as a function of the axial stretch $F_{11}=\lambda$ for incompressible uniaxial loading at different temperatures for the carbon-filled black rubber \cite{Fu2021}. The table summarizes the thermo-mechanical response used to generate the training data, illustrating the temperature dependence of the constitutive behavior.}
\label{tab:data_rubber}
\small
\renewcommand{\arraystretch}{0.9}
\resizebox{\textwidth}{!}{
\begin{tabular}{c|c||c|c||c|c||c|c||c|c||c|c||c|c||c|c}
\hline
\multicolumn{2}{c||}{$T=283$\,K} & \multicolumn{2}{c||}{$T=303$\,K} & \multicolumn{2}{c||}{$T=323$\,K} & \multicolumn{2}{c||}{$T=343$\,K} & \multicolumn{2}{c||}{$T=363$\,K} & \multicolumn{2}{c||}{$T=373$\,K} & \multicolumn{2}{c||}{$T=383$\,K} & \multicolumn{2}{c}{$T=393$\,K} \\
\hline
$F_{11}$ [-] & $P_{11}$ [MPa] & $F_{11}$ [-] & $P_{11}$ [MPa] & $F_{11}$ [-] & $P_{11}$ [MPa] & $F_{11}$ [-] & $P_{11}$ [MPa] & $F_{11}$ [-] & $P_{11}$ [MPa] & $F_{11}$ [-] & $P_{11}$ [MPa] & $F_{11}$ [-] & $P_{11}$ [MPa] & $F_{11}$ [-] & $P_{11}$ [MPa] \\
\hline
1.00000 & 0.000000 & 1.00000 & 0.000000 & 1.00000 & 0.000000 & 1.00000 & 0.000000 & 1.00000 & 0.000000 & 1.00000 & 0.000000 & 1.00000 & 0.000000 & 1.00000 & 0.000000 \\
1.06605 & 0.450927 & 1.07206 & 0.398578 & 1.06455 & 0.351218 & 1.05925 & 0.302547 & 1.06152 & 0.285128 & 1.07430 & 0.397746 & 1.04399 & 0.238756 & 1.06077 & 0.318900 \\
1.11608 & 0.638094 & 1.12397 & 0.601937 & 1.13816 & 0.617895 & 1.11824 & 0.581138 & 1.11104 & 0.480744 & 1.13038 & 0.598524 & 1.10160 & 0.509104 & 1.10742 & 0.530547 \\
1.17084 & 0.836345 & 1.19056 & 0.852403 & 1.21525 & 0.849735 & 1.17495 & 0.765233 & 1.15913 & 0.622078 & 1.18448 & 0.781625 & 1.15635 & 0.746744 & 1.15566 & 0.724084 \\
1.22446 & 1.003285 & 1.23686 & 0.971640 & 1.27205 & 1.014738 & 1.22774 & 0.899099 & 1.19393 & 0.738158 & 1.22118 & 0.874767 & 1.19527 & 0.863940 & 1.19929 & 0.861042 \\
1.27520 & 1.144868 & 1.29581 & 1.140401 & 1.33427 & 1.207609 & 1.26509 & 0.992175 & 1.23815 & 0.848395 & 1.26319 & 0.986624 & 1.23880 & 0.998858 & 1.24398 & 0.987548 \\
1.31610 & 1.271632 & 1.34446 & 1.284534 & 1.38272 & 1.364871 & 1.31732 & 1.142058 & 1.28148 & 0.951697 & 1.29581 & 1.078572 & 1.28960 & 1.125183 & 1.28524 & 1.108225 \\
1.36402 & 1.435798 & 1.39946 & 1.488493 & 1.42278 & 1.520404 & 1.36520 & 1.304146 & 1.32340 & 1.062818 & 1.33547 & 1.179858 & 1.33547 & 1.273180 & 1.31793 & 1.209076 \\
1.41088 & 1.665726 & 1.44406 & 1.744582 & 1.44906 & 1.689422 & 1.42109 & 1.528480 & 1.36048 & 1.177797 & 1.37632 & 1.306539 & 1.36755 & 1.373510 & 1.35812 & 1.350267 \\
1.44961 & 1.928275 & 1.47214 & 1.959235 & 1.47487 & 1.871115 & 1.46120 & 1.772867 & 1.40003 & 1.297129 & 1.40575 & 1.399503 & 1.40175 & 1.492410 & 1.39659 & 1.472431 \\
1.48409 & 2.297364 & 1.49862 & 2.281025 & 1.49969 & 2.101319 & 1.48787 & 1.998354 & 1.43794 & 1.442473 & 1.44462 & 1.546719 & 1.43794 & 1.652962 & 1.43458 & 1.625800 \\
1.51142 & 2.638657 & 1.52043 & 2.511789 & 1.52464 & 2.323838 & 1.51355 & 2.205599 & 1.46723 & 1.565359 & 1.47214 & 1.687119 & 1.47650 & 1.832872 & 1.47214 & 1.789919 \\
1.53723 & 2.999732 & 1.54088 & 2.825084 & 1.54140 & 2.570015 & 1.53618 & 2.433870 & 1.50290 & 1.771054 & 1.50930 & 1.881512 & 1.50717 & 2.002299 & 1.50504 & 1.987391 \\
1.55437 & 3.310288 & 1.55592 & 3.072414 & 1.55798 & 2.794081 & 1.55540 & 2.649906 & 1.53671 & 1.992777 & 1.53618 & 2.062994 & 1.54140 & 2.188844 & 1.53932 & 2.191808 \\
1.56927 & 3.647574 & 1.57183 & 3.375458 & 1.57183 & 3.018656 & 1.57132 & 2.883669 & 1.56927 & 2.246406 & 1.57029 & 2.301634 & 1.57081 & 2.397227 & 1.57336 & 2.370705 \\
\hline
\end{tabular}
}
\end{table}

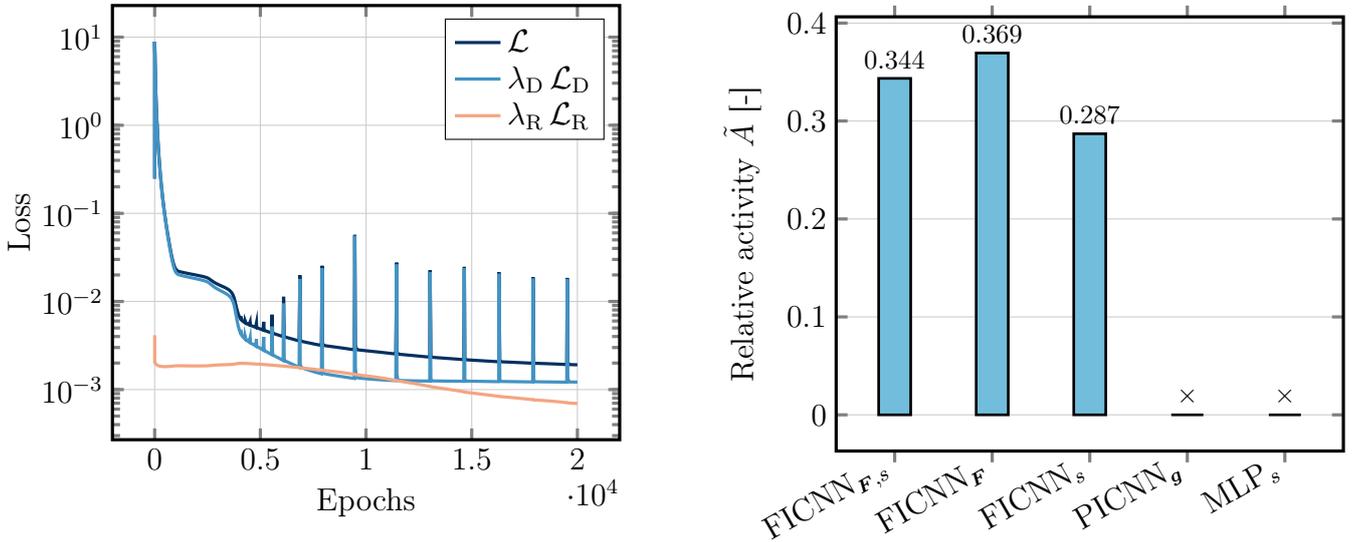
\begin{figure}[!ht]
\centering
\begin{tikzpicture}[baseline=(current bounding box.north)]
\begin{axis}[
    width=0.45\textwidth,
    height=0.40\textwidth,
    xlabel={Epochs},
    ylabel={Loss},
    grid=major,
    ymode=log,
    major grid style={gray!40},
    axis line style={very thick},
    tick style={very thick},
    tick label style={font=\large},
    label style={font=\large},
    legend style={
        at={(0.97,0.97)},
        anchor=north east,
        draw=black,
        fill=white,
        font=\large,
    },
    legend cell align=left,
    enlargelimits=0.1,
]

\addplot[
    ekA, line width=1.2pt
] table[
    col sep=comma,
    x=epoch,
    y=loss
]{graphs/losses/reduced_loss_output_losses_Carbon_Filler/loss_reduced.csv};

\addlegendentry{$\mathcal{L}$}

\addplot[
    ekB, line width=1.2pt
] table[
    col sep=comma,
    x=epoch,
    y=loss_force_DC
]{graphs/losses/reduced_loss_output_losses_Carbon_Filler/loss_force_DC_reduced.csv};

\addlegendentry{$\lambda_{\mathrm{D}}\,\mathcal{L}_{\mathrm{D}}$}

%

\addplot[
    ekD, line width=1.2pt
] table[
    col sep=comma,
    x=epoch,
    y=Reg
]{graphs/losses/reduced_loss_output_losses_Carbon_Filler/Reg_reduced.csv};

\addlegendentry{$\lambda_{\mathrm{R}}\,\mathcal{L}_{\mathrm{R}}$}

%

\end{axis}

\end{tikzpicture}
\hspace*{1cm}
\begin{tikzpicture}[baseline=(current bounding box.north)]
\begin{axis}[
width=0.45\textwidth,
height=0.40\textwidth,
major grid style={gray!40},
axis line style={very thick},
tick style={very thick},
tick label style={font=\large},
label style={font=\large},
ymajorgrids=true,
x tick label style={rotate=30, anchor=east},
ybar,
bar width=12pt,
enlarge x limits=0.15,
ylabel={Relative activity $\tilde{A}$ [-]},
symbolic x coords={Em,Ef,Es,Di,Aux},
xtick={Em,Ef,Es,Di,Aux},
xticklabels={
    $\mathrm{FICNN}_{\tens{F},s}$,
    $\mathrm{FICNN}_{\tens{F}}$,
    $\mathrm{FICNN}_{s}$,
    $\mathrm{PICNN}_{\tens{g}}$,
    $\mathrm{MLP}_{s}$
},
nodes near coords style={
    /pgf/number format/.cd,
    fixed,
    precision=3
},
nodes near coords align={vertical},
visualization depends on={y \as \rawy},
nodes near coords={
    \ifdim\rawy pt=0pt
        $\times$
    \else
        \pgfmathprintnumber[fixed,precision=3]{\rawy}
    \fi
},
]
\addplot[draw=black,fill=ekB!50!ekC,line width=1.0pt,bar shift=0pt] coordinates {(Em,3.435558143838e-01)};
\addplot[draw=black,fill=ekB!50!ekC,line width=1.0pt,bar shift=0pt] coordinates {(Ef,3.694342518909e-01)};
\addplot[draw=black,fill=ekB!50!ekC,line width=1.0pt,bar shift=0pt] coordinates {(Es,2.870099337253e-01)};
\addplot[draw=black,fill=ekB!50!ekC,line width=1.0pt,bar shift=0pt] coordinates {(Di,0.0)};
\addplot[draw=black,fill=ekE,line width=1.0pt,bar shift=0pt]        coordinates {(Aux,0.0)};
\end{axis}
\end{tikzpicture}
\caption{Left: Training history of the physics-based neural network for the temperature dependent carbon-filled black rubber dataset \cite{Fu2021}. Shown are the total loss as well as its individual contributions, including the physics-based residual losses and the regularization term, over the course of training iterations.
Right: Relative activity according to \cref{eq:activity} of the individual subnetworks for the carbon-filled black rubber. The diagram shows the normalized contributions of internal energy. No transient thermal effects are present. Training is performed using an iterative solver instead of the auxiliary network.}
\label{fig:loss_rubber}
\end{figure}

\cref{fig:loss_rubber} (left) illustrates the evolution of both the individual and total loss terms during training.
As in the previous example, occasional spikes are observed; however, the overall trend remains clearly decreasing, indicating convergence of the training process.
The corresponding relative activity of the subnetworks is shown in \cref{fig:loss_rubber} (right).
In this case, the contributions of the individual constitutive subnetworks are more evenly distributed, suggesting that multiple mechanisms are required to represent the material behavior.

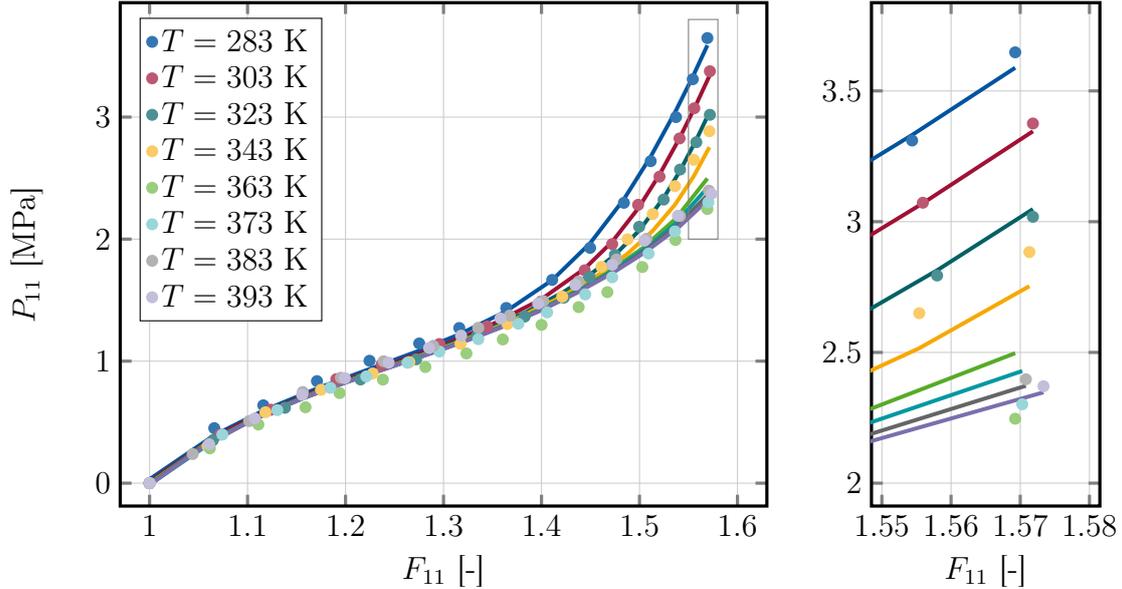
\begin{figure}[!ht]
\centering
\begin{tikzpicture}
\begin{axis}[
    width=0.55\textwidth,
    height=0.45\textwidth,
    xlabel={$F_{11}$ [-]},
    ylabel={$P_{11}$ [MPa]},
    xmin=1.00, xmax=1.60,
    ymin=0.00, ymax=3.75,
    grid=major,
    major grid style={gray!40},
    axis line style={very thick},
    tick style={very thick},
    tick label style={font=\large},
    label style={font=\large},
    legend style={
        at={(0.03,0.97)},
        anchor=north west,
        draw=black,
        fill=white,
        font=\large
    },
    legend cell align=left,
    enlargelimits=0.05,
]

\addplot[
    only marks,
    mark=*,
    mark size=1.5pt,
    mark options={
        line width=1.5pt
    },
    clr1!80,
] coordinates {
(1.00000,0.000000)
(1.06605,0.450927)
(1.11608,0.638094)
(1.17084,0.836345)
(1.22446,1.003285)
(1.27520,1.144868)
(1.31610,1.271632)
(1.36402,1.435798)
(1.41088,1.665726)
(1.44961,1.928275)
(1.48409,2.297364)
(1.51142,2.638657)
(1.53723,2.999732)
(1.55437,3.310288)
(1.56927,3.647574)
};
\addlegendentry{$T=283$ K}

\addplot[
    only marks,
    mark=*,
    mark size=1.5pt,
    mark options={
        line width=1.5pt
    },
    clr2!70,
] coordinates {
(1.00000,0.000000)
(1.07206,0.398578)
(1.12397,0.601937)
(1.19056,0.852403)
(1.23686,0.971640)
(1.29581,1.140401)
(1.34446,1.284534)
(1.39946,1.488493)
(1.44406,1.744582)
(1.47214,1.959235)
(1.49862,2.281025)
(1.52043,2.511789)
(1.54088,2.825084)
(1.55592,3.072414)
(1.57183,3.375458)
};
\addlegendentry{$T=303$ K}

\addplot[
    only marks,
    mark=*,
    mark size=1.5pt,
    mark options={
        line width=1.5pt
    },
    clr3!70,
] coordinates {
(1.00000,0.000000)
(1.06455,0.351218)
(1.13816,0.617895)
(1.21525,0.849735)
(1.27205,1.014738)
(1.33427,1.207609)
(1.38272,1.364871)
(1.42278,1.520404)
(1.44906,1.689422)
(1.47487,1.871115)
(1.49969,2.101319)
(1.52464,2.323838)
(1.54140,2.570015)
(1.55798,2.794081)
(1.57183,3.018656)
};
\addlegendentry{$T=323$ K}

\addplot[
    only marks,
    mark=*,
    mark size=1.5pt,
    mark options={
        line width=1.5pt
    },
    clr4!60,
] coordinates {
(1.00000,0.000000)
(1.05925,0.302547)
(1.11824,0.581138)
(1.17495,0.765233)
(1.22774,0.899099)
(1.26509,0.992175)
(1.31732,1.142058)
(1.36520,1.304146)
(1.42109,1.528480)
(1.46120,1.772867)
(1.48787,1.998354)
(1.51355,2.205599)
(1.53618,2.433870)
(1.55540,2.649906)
(1.57132,2.883669)
};
\addlegendentry{$T=343$ K}

\addplot[
    only marks,
    mark=*,
    mark size=1.5pt,
    mark options={
        line width=1.5pt
    },
    clr5!60,
] coordinates {
(1.00000,0.000000)
(1.06152,0.285128)
(1.11104,0.480744)
(1.15913,0.622078)
(1.19393,0.738158)
(1.23815,0.848395)
(1.28148,0.951697)
(1.32340,1.062818)
(1.36048,1.177797)
(1.40003,1.297129)
(1.43794,1.442473)
(1.46723,1.565359)
(1.50290,1.771054)
(1.53671,1.992777)
(1.56927,2.246406)
};
\addlegendentry{$T=363$ K}

\addplot[
    only marks,
    mark=*,
    mark size=1.5pt,
    mark options={
        line width=1.5pt
    },
    clr9!40,
] coordinates {
(1.00000,0.000000)
(1.07430,0.397746)
(1.13038,0.598524)
(1.18448,0.781625)
(1.22118,0.874767)
(1.26319,0.986624)
(1.29581,1.078572)
(1.33547,1.179858)
(1.37632,1.306539)
(1.40575,1.399503)
(1.44462,1.546719)
(1.47214,1.687119)
(1.50930,1.881512)
(1.53618,2.062994)
(1.57029,2.301634)
};
\addlegendentry{$T=373$ K}

\addplot[
    only marks,
    mark=*,
    mark size=1.5pt,
    mark options={
        line width=1.5pt
    },
    clr7!50,
] coordinates {
(1.00000,0.000000)
(1.04399,0.238756)
(1.10160,0.509104)
(1.15635,0.746744)
(1.19527,0.863940)
(1.23880,0.998858)
(1.28960,1.125183)
(1.33547,1.273180)
(1.36755,1.373510)
(1.40175,1.492410)
(1.43794,1.652962)
(1.47650,1.832872)
(1.50717,2.002299)
(1.54140,2.188844)
(1.57081,2.397227)
};
\addlegendentry{$T=383$ K}

\addplot[
    only marks,
    mark=*,
    mark size=1.5pt,
    mark options={
        line width=1.5pt
    },
    clr8!45,
] coordinates {
(1.00000,0.000000)
(1.06077,0.318900)
(1.10742,0.530547)
(1.15566,0.724084)
(1.19929,0.861042)
(1.24398,0.987548)
(1.28524,1.108225)
(1.31793,1.209076)
(1.35812,1.350267)
(1.39659,1.472431)
(1.43458,1.625800)
(1.47214,1.789919)
(1.50504,1.987391)
(1.53932,2.191808)
(1.57336,2.370705)
};
\addlegendentry{$T=393$ K}

\addplot[
    line width=1.5pt,
    clr1
] table[x expr={1 + \thisrow{ux_right_mean}}, y=Fx_right_sum, col sep=space]
{./figures/CarbonFiller/Carbon_Filler_00_F_and_u.dat};

\addplot[
    line width=1.5pt,
    clr2
] table[x expr={1 + \thisrow{ux_right_mean}}, y=Fx_right_sum, col sep=space]
{./figures/CarbonFiller/Carbon_Filler_01_F_and_u.dat};

\addplot[
    line width=1.5pt,
    clr3
] table[x expr={1 + \thisrow{ux_right_mean}}, y=Fx_right_sum, col sep=space]
{./figures/CarbonFiller/Carbon_Filler_02_F_and_u.dat};

\addplot[
    line width=1.5pt,
    clr4
] table[x expr={1 + \thisrow{ux_right_mean}}, y=Fx_right_sum, col sep=space]
{./figures/CarbonFiller/Carbon_Filler_03_F_and_u.dat};

\addplot[
    line width=1.5pt,
    clr5
] table[x expr={1 + \thisrow{ux_right_mean}}, y=Fx_right_sum, col sep=space]
{./figures/CarbonFiller/Carbon_Filler_04_F_and_u.dat};

\addplot[
    line width=1.5pt,
    clr9
] table[x expr={1 + \thisrow{ux_right_mean}}, y=Fx_right_sum, col sep=space]
{./figures/CarbonFiller/Carbon_Filler_05_F_and_u.dat};

\addplot[
    line width=1.5pt,
    clr7
] table[x expr={1 + \thisrow{ux_right_mean}}, y=Fx_right_sum, col sep=space]
{./figures/CarbonFiller/Carbon_Filler_06_F_and_u.dat};

\addplot[
    line width=1.5pt,
    clr8
] table[x expr={1 + \thisrow{ux_right_mean}}, y=Fx_right_sum, col sep=space]
{./figures/CarbonFiller/Carbon_Filler_07_F_and_u.dat};

\draw[gray, thin] (axis cs:1.55,2) rectangle (axis cs:1.58,3.8);

\end{axis}
\end{tikzpicture}
\hspace*{10pt}
\begin{tikzpicture}
\begin{axis}[
    width=0.25\textwidth,
    height=0.45\textwidth,
    xlabel={$F_{11}$ [-]},
    xmin=1.55, xmax=1.58,
    ymin=2.00, ymax=3.75,
    grid=major,
    major grid style={gray!40},
    axis line style={very thick},
    tick style={very thick},
    tick label style={font=\large},
    label style={font=\large},
    enlargelimits=0.05,
]

\addplot[
    only marks,
    mark=*,
    mark size=1.5pt,
    mark options={
        line width=1.5pt
    },
    clr1!80,
] coordinates {
(1.00000,0.000000)
(1.06605,0.450927)
(1.11608,0.638094)
(1.17084,0.836345)
(1.22446,1.003285)
(1.27520,1.144868)
(1.31610,1.271632)
(1.36402,1.435798)
(1.41088,1.665726)
(1.44961,1.928275)
(1.48409,2.297364)
(1.51142,2.638657)
(1.53723,2.999732)
(1.55437,3.310288)
(1.56927,3.647574)
};

\addplot[
    only marks,
    mark=*,
    mark size=1.5pt,
    mark options={
        line width=1.5pt
    },
    clr2!70,
] coordinates {
(1.00000,0.000000)
(1.07206,0.398578)
(1.12397,0.601937)
(1.19056,0.852403)
(1.23686,0.971640)
(1.29581,1.140401)
(1.34446,1.284534)
(1.39946,1.488493)
(1.44406,1.744582)
(1.47214,1.959235)
(1.49862,2.281025)
(1.52043,2.511789)
(1.54088,2.825084)
(1.55592,3.072414)
(1.57183,3.375458)
};

\addplot[
    only marks,
    mark=*,
    mark size=1.5pt,
    mark options={
        line width=1.5pt
    },
    clr3!70,
] coordinates {
(1.00000,0.000000)
(1.06455,0.351218)
(1.13816,0.617895)
(1.21525,0.849735)
(1.27205,1.014738)
(1.33427,1.207609)
(1.38272,1.364871)
(1.42278,1.520404)
(1.44906,1.689422)
(1.47487,1.871115)
(1.49969,2.101319)
(1.52464,2.323838)
(1.54140,2.570015)
(1.55798,2.794081)
(1.57183,3.018656)
};

\addplot[
    only marks,
    mark=*,
    mark size=1.5pt,
    mark options={
        line width=1.5pt
    },
    clr4!60,
] coordinates {
(1.00000,0.000000)
(1.05925,0.302547)
(1.11824,0.581138)
(1.17495,0.765233)
(1.22774,0.899099)
(1.26509,0.992175)
(1.31732,1.142058)
(1.36520,1.304146)
(1.42109,1.528480)
(1.46120,1.772867)
(1.48787,1.998354)
(1.51355,2.205599)
(1.53618,2.433870)
(1.55540,2.649906)
(1.57132,2.883669)
};

\addplot[
    only marks,
    mark=*,
    mark size=1.5pt,
    mark options={
        line width=1.5pt
    },
    clr5!60,
] coordinates {
(1.00000,0.000000)
(1.06152,0.285128)
(1.11104,0.480744)
(1.15913,0.622078)
(1.19393,0.738158)
(1.23815,0.848395)
(1.28148,0.951697)
(1.32340,1.062818)
(1.36048,1.177797)
(1.40003,1.297129)
(1.43794,1.442473)
(1.46723,1.565359)
(1.50290,1.771054)
(1.53671,1.992777)
(1.56927,2.246406)
};

\addplot[
    only marks,
    mark=*,
    mark size=1.5pt,
    mark options={
        line width=1.5pt
    },
    clr9!40,
] coordinates {
(1.00000,0.000000)
(1.07430,0.397746)
(1.13038,0.598524)
(1.18448,0.781625)
(1.22118,0.874767)
(1.26319,0.986624)
(1.29581,1.078572)
(1.33547,1.179858)
(1.37632,1.306539)
(1.40575,1.399503)
(1.44462,1.546719)
(1.47214,1.687119)
(1.50930,1.881512)
(1.53618,2.062994)
(1.57029,2.301634)
};

\addplot[
    only marks,
    mark=*,
    mark size=1.5pt,
    mark options={
        line width=1.5pt
    },
    clr7!50,
] coordinates {
(1.00000,0.000000)
(1.04399,0.238756)
(1.10160,0.509104)
(1.15635,0.746744)
(1.19527,0.863940)
(1.23880,0.998858)
(1.28960,1.125183)
(1.33547,1.273180)
(1.36755,1.373510)
(1.40175,1.492410)
(1.43794,1.652962)
(1.47650,1.832872)
(1.50717,2.002299)
(1.54140,2.188844)
(1.57081,2.397227)
};

\addplot[
    only marks,
    mark=*,
    mark size=1.5pt,
    mark options={
        line width=1.5pt
    },
    clr8!45,
] coordinates {
(1.00000,0.000000)
(1.06077,0.318900)
(1.10742,0.530547)
(1.15566,0.724084)
(1.19929,0.861042)
(1.24398,0.987548)
(1.28524,1.108225)
(1.31793,1.209076)
(1.35812,1.350267)
(1.39659,1.472431)
(1.43458,1.625800)
(1.47214,1.789919)
(1.50504,1.987391)
(1.53932,2.191808)
(1.57336,2.370705)
};

\addplot[
    line width=1.5pt,
    clr1
] table[x expr={1 + \thisrow{ux_right_mean}}, y=Fx_right_sum, col sep=space]
{./figures/CarbonFiller/Carbon_Filler_00_F_and_u.dat};

\addplot[
    line width=1.5pt,
    clr2
] table[x expr={1 + \thisrow{ux_right_mean}}, y=Fx_right_sum, col sep=space]
{./figures/CarbonFiller/Carbon_Filler_01_F_and_u.dat};

\addplot[
    line width=1.5pt,
    clr3
] table[x expr={1 + \thisrow{ux_right_mean}}, y=Fx_right_sum, col sep=space]
{./figures/CarbonFiller/Carbon_Filler_02_F_and_u.dat};

\addplot[
    line width=1.5pt,
    clr4
] table[x expr={1 + \thisrow{ux_right_mean}}, y=Fx_right_sum, col sep=space]
{./figures/CarbonFiller/Carbon_Filler_03_F_and_u.dat};

\addplot[
    line width=1.5pt,
    clr5
] table[x expr={1 + \thisrow{ux_right_mean}}, y=Fx_right_sum, col sep=space]
{./figures/CarbonFiller/Carbon_Filler_04_F_and_u.dat};

\addplot[
    line width=1.5pt,
    clr9
] table[x expr={1 + \thisrow{ux_right_mean}}, y=Fx_right_sum, col sep=space]
{./figures/CarbonFiller/Carbon_Filler_05_F_and_u.dat};

\addplot[
    line width=1.5pt,
    clr7
] table[x expr={1 + \thisrow{ux_right_mean}}, y=Fx_right_sum, col sep=space]
{./figures/CarbonFiller/Carbon_Filler_06_F_and_u.dat};

\addplot[
    line width=1.5pt,
    clr8
] table[x expr={1 + \thisrow{ux_right_mean}}, y=Fx_right_sum, col sep=space]
{./figures/CarbonFiller/Carbon_Filler_07_F_and_u.dat};

\end{axis}
\end{tikzpicture}
\caption{Comparison of the discovered model and the reference data for carbon-filled black rubber under incompressible uniaxial loading at different temperatures. The Piola stress $P_{11}$ is plotted as a function of the axial stretch $F_{11}$. The reference data are shown as discrete points, while the continuous curves represent the response of the discovered model. Left: Full stretch range. Right: Enlarged view, demonstrating that the discovered model is unable to explain the thermal stiffening effect in high temperature ranges.}
\label{fig:results_rubber}
\end{figure}

The identified constitutive response is shown in \cref{fig:results_rubber}.
First, it is noted that, although less pronounced than in the previous example, the stress at $F_{11}=1$ is non-zero for all temperatures.
This is consistent with the earlier discussion, as thermal expansion is not explicitly accounted for and the reference configuration is not temperature-dependent.
As before, we set the normalization factor to $n_T = 1$ and allow the network to implicitly identify the reference temperature.

In contrast to the findings reported in \cite{Fuhg2024}, the best-performing model in our study is not able to capture the temperature dependence in the range of $T = 363\,\mathrm{K}$ to $T = 393\,\mathrm{K}$.
Specifically, the model predicts a continued softening of the mechanical response with increasing temperature in this regime, rather than the observed stiffening behavior.
In \cite{Fuhg2024}, this behavior is attributed to thermoelastic inversion \cite{flory1953principles,Kirkinis2002}.
It remains unclear whether the inability of the present model to reproduce this effect is due to limitations of the network architecture or arises from the differing modeling assumptions, such as the treatment of reference configurations and compressibility.
Apart from this discrepancy, the model achieves a comparable level of accuracy to \cite{Fuhg2024} in the lower temperature range, indicating that the essential temperature-dependent behavior is captured in this regime.

\subsection{Synthetic data: Structural examples}
\label{sec:structural_example}

Up to this point, the investigations have primarily focused on homogeneous problems.
Moreover, the considered examples either addressed purely thermal behavior or mechanical responses parameterized by temperature, rather than a fully coupled interaction between thermal and mechanical fields.
In particular, essential coupling mechanisms such as thermal expansion, the influence of large deformations on heat conduction, or the generation of heat due to elastic deformation \cite{truesdell1992non} have not yet been taken into account.
Experimental evidence for the latter effect, even in metallic materials, is reported, for instance, in \cite{Rose2020}.

To address this limitation, we now investigate the discovery of a fully coupled thermomechanical model, as introduced in \ref{app:constitutive}, with material parameters adopted from the literature \cite{Ambati2015,Felder2022,Dittmann2020}.
In addition, a nonlinear contribution to the caloric part of the Helmholtz energy is employed to account for temperature-dependent heat capacity.

\cref{sec:structural_example_training} presents the specimen used for training and outlines the corresponding training procedure.
In \cref{sec:structural_example_testing}, the predictive capabilities of the discovered model are evaluated on an unseen boundary value problem.

\subsubsection{Training: Plate with an elliptic hole}
\label{sec:structural_example_training}

Training neural networks requires datasets with sufficiently high information content to enable an (almost) unique identification of the network parameters.
A common strategy in physics-based modeling, which we also employed in previous studies on inelasticity, is to consider specimens with highly non-homogeneous geometries in order to induce rich spatial variations in the physical fields.
In addition, complex loading paths are typically applied to further increase the diversity of the data.
While this approach is effective, it also increases the complexity of both the experimental setup and the data generation process.

In contrast, we adopt a training strategy based on simple geometries and loading paths, while incorporating multiple independent scenarios simultaneously.
From the perspective of automation and data generation, this approach offers a scalable alternative.

\begin{figure}[!t]
\centering
\begin{tikzpicture}
\pgfplotsset{
    colormap={redblue}{
        rgb255(0cm)=(255,0,0)
        rgb255(1cm)=(0,0,255)
    }
}
\usetikzlibrary{calc}

\node[anchor=center, inner sep=0] (img) at (0,0)
    {\includegraphics[width=0.8\textwidth]{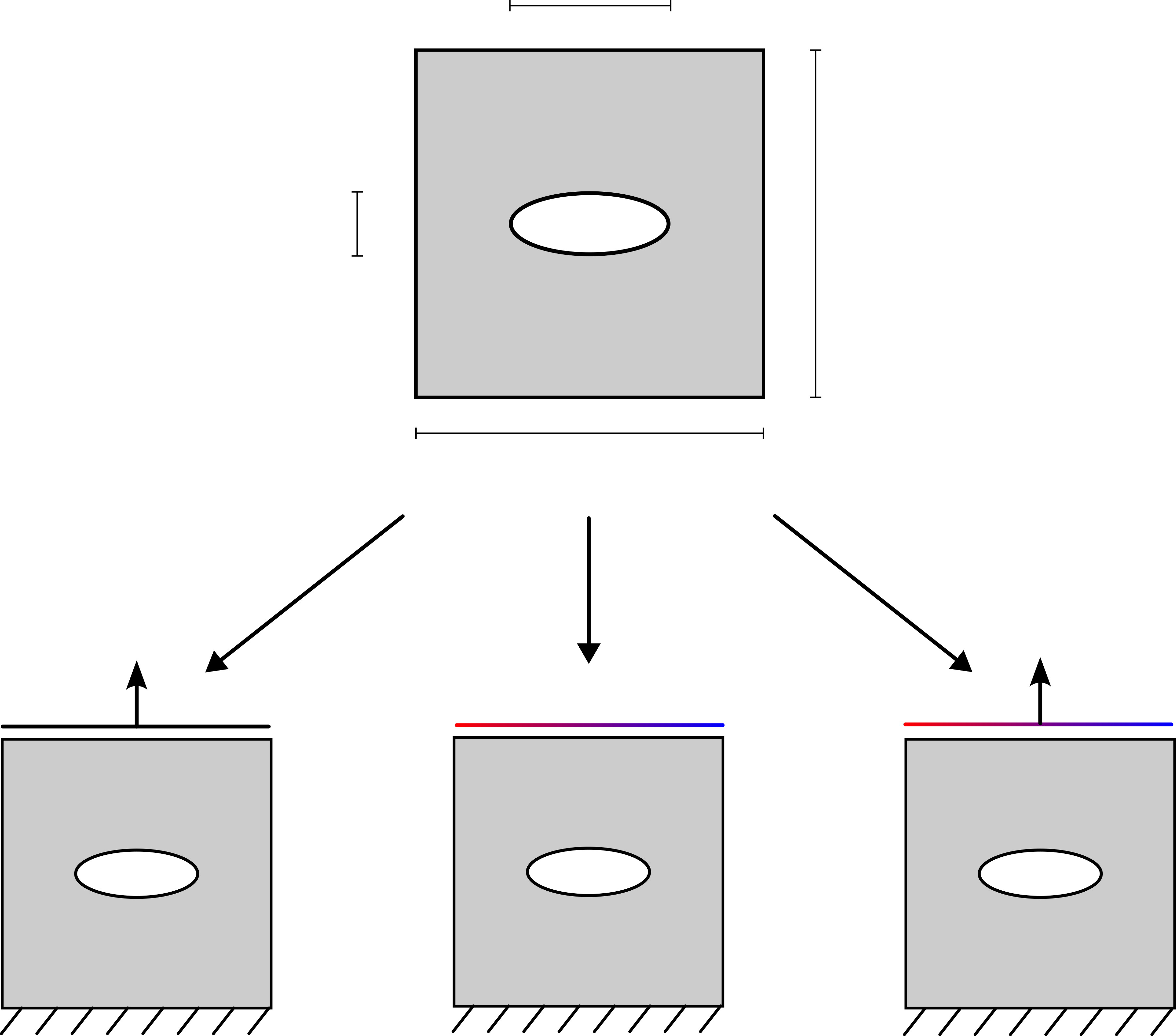}};

\node at (3.2,3.7) {$10$};
\node at (0.0,0.7) {$10$};
\node at (0.0,6.7) {$6$};
\node at (-3.2,3.7) {$3$};
\node at (3.0,1.0) {[mm]};

\node[rotate=38]  at (-3.6,-0.6) {Training \#1};
\node[rotate=90]  at (-0.5,-0.9) {Training \#2};
\node[rotate=-38] at ( 3.6,-0.6) {Training \#3};

\node at (-5.7,-5.7) {$T_{\mathrm{init}}=293.15\ \mathrm{K}$};
\node at ( 0.0,-5.7) {$T_{\mathrm{init}}=293.15\ \mathrm{K}$};
\node at ( 5.7,-5.7) {$T_{\mathrm{init}}=293.15\ \mathrm{K}$};

\node at (-5.7,-1.5) {$u_y(t)$};
\node at (0.0,-2.3) {$u_y=0$, $T(t)$};
\node at (5.7,-1.5) {$u_y(t)$};
\node at (6.4,-2.3) {$T(t)$};


\begin{axis}[
    at={($(img.south)+(-5.7cm,-0.5cm)$)},
    anchor=north,
    width=4.8cm,
    height=3.8cm,
    xlabel={Time $t$ [s]},
    ylabel={$u_y(t)$ [mm]},
    ylabel style={at={(axis description cs:0.15,0.5)}},
    grid=both,
    enlargelimits=0.1
]
\addplot[very thick] coordinates {
    (0,0) (10.0,2.0)
};
\end{axis}

\begin{axis}[
    at={($(img.south)+(0.0cm,-0.5cm)$)},
    anchor=north,
    width=4.8cm,
    height=3.8cm,
    xlabel={Time $t$ [s]},
    ylabel={$u_y(t)$ [mm]},
    xtick={0,7.5,15},
    ytick={0,1},
    ylabel style={at={(axis description cs:+0.15,0.5)}},
    grid=both,
    ymax=1
]
\addplot[very thick] coordinates {
    (0,0)
    (15,0)
};
\end{axis}

\begin{axis}[
    at={($(img.south)+(0.0cm,-0.5cm)$)},
    anchor=north,
    width=4.8cm,
    height=3.8cm,
    axis x line=none,
    axis y line*=right,
    ytick={293.15,310},
    ylabel={$T(t)$ [K]},
    ylabel style={at={(axis description cs:1.7,0.5)}},
    grid=both,
]
\addplot[
    mesh,
    no markers,
    very thick,
    point meta=x,
    colormap name=redblue,
] coordinates {
    (0,293.15)
    (7.5,310)
    (15,310)
};
\end{axis}


\begin{axis}[
    at={($(img.south)+(5.7cm,-0.5cm)$)},
    anchor=north,
    width=4.8cm,
    height=3.8cm,
    xlabel={Time $t$ [s]},
    ylabel={$u_y(t)$ [mm]},
    xtick={0,10,20},
    ytick={0,2},
    ylabel style={at={(axis description cs:+0.15,0.5)}},
    grid=both,
    ymax=3
]
\addplot[very thick] coordinates {
    (0,0)
    (10,2)
    (20,2)
};
\end{axis}

\begin{axis}[
    at={($(img.south)+(5.7cm,-0.5cm)$)},
    anchor=north,
    width=4.8cm,
    height=3.8cm,
    axis x line=none,
    axis y line*=right,
    ytick={293.15,310},
    ylabel={$T(t)$ [K]},
    ylabel style={at={(axis description cs:1.7,0.5)}},
    grid=both,
]
\addplot[
    mesh,
    no markers,
    very thick,
    point meta=x,
    colormap name=redblue,
] coordinates {
    (0,293.15)
    (10,310)
    (20,310)
};
\end{axis}

\end{tikzpicture}
\caption{Setup of the boundary value problem and training scenarios for the plate with an elliptic hole. Top: Geometry and dimensions of the specimen \cite{kehls2025autoencoder}, the thickness is $t=1$ mm. Middle: Three distinct training scenarios with different combinations of mechanical loading $u_y(t)$ and thermal loading $T(t)$ applied at the top boundary, while the bottom boundary is fixed and the initial temperature is $T_{\mathrm{init}}=293.15\,\mathrm{K}$. Bottom: Corresponding temporal evolution of the prescribed displacement and temperature for each training case. Solid black lines denote the prescribed displacement, whereas the temperature evolution is represented by colored lines transitioning from red to blue.}
\label{fig:sketch_PWEH}
\end{figure}
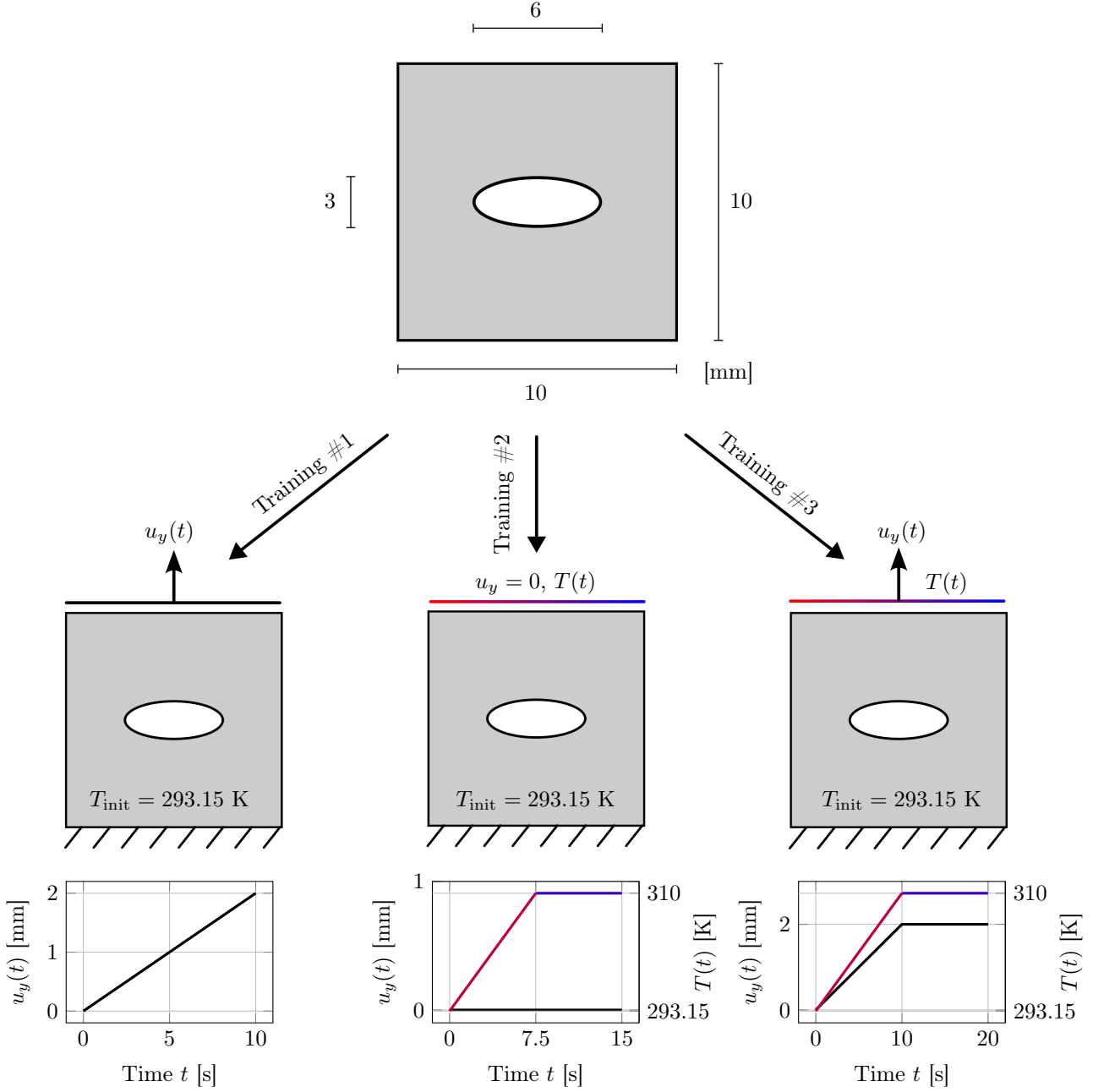

To this end, the specimen and the three loading scenarios used simultaneously for training are depicted in \cref{fig:sketch_PWEH}.
The first scenario is purely deformation-driven, which nevertheless induces temperature changes due to the intrinsic thermomechanical coupling.
The second scenario corresponds to purely thermal loading.
The third scenario considers a fully coupled loading involving both deformation and temperature.
In all cases involving temperature, a holding phase is introduced to facilitate the identification of transient effects.

The domain is discretized following \cite{kehls2025autoencoder} using $952$ hexahedral elements and $1647$ nodes.
Each scenario is simulated using $40$ time steps; however, the total duration differs between scenarios, resulting in different time increments.

\begin{figure}[!t]
\centering
\begin{tikzpicture}[baseline=(current bounding box.north)]
\begin{axis}[
    width=0.45\textwidth,
    height=0.40\textwidth,
    xlabel={Epochs},
    ylabel={Loss},
    grid=major,
    ymode=log,
    major grid style={gray!40},
    axis line style={very thick},
    tick style={very thick},
    tick label style={font=\large},
    label style={font=\large},
    legend style={
        at={(0.97,0.97)},
        anchor=north east,
        draw=black,
        fill=white,
        font=\large,
    },
    legend columns=2,
    legend cell align=left,
    enlargelimits=0.1
    ]

\addplot[
    ekA, solid, line width=1.5pt
] table[
    col sep=comma,
    x=epoch,
    y=loss
]{graphs/losses/reduced_loss_output_losses_PWEH/loss_reduced.csv};

\addlegendentry{$\mathcal{L}$}

\addplot[
    ekB, line width=1.5pt
] table[
    col sep=comma,
    x=epoch,
    y=loss_force_DC
]{graphs/losses/reduced_loss_output_losses_PWEH/loss_force_DC_reduced.csv};

\addlegendentry{$\lambda_{\mathrm{D}}\,\mathcal{L}_{\mathrm{D}}$}

\addplot[
    ekC, line width=1.5pt
] table[
    col sep=comma,
    x=epoch,
    y=loss_force_AC
]{graphs/losses/reduced_loss_output_losses_PWEH/loss_force_AC_reduced.csv};

\addlegendentry{$\lambda_{\mathrm{N}}\,\mathcal{L}_{\mathrm{N}}$}

\addplot[
    ekD, line width=1.5pt
] table[
    col sep=comma,
    x=epoch,
    y=Reg
]{graphs/losses/reduced_loss_output_losses_PWEH/Reg_reduced.csv};

\addlegendentry{$\lambda_{\mathrm{R}}\,\mathcal{L}_{\mathrm{R}}$}

\addplot[
    ekE, line width=1.5pt
] table[
    col sep=comma,
    x=epoch,
    y=Mismatch
]{graphs/losses/reduced_loss_output_losses_PWEH/Mismatch_reduced.csv};

\addlegendentry{$\lambda_{\mathrm{A}}\,\mathcal{L}_{\mathrm{aux}}$}

\end{axis}

\end{tikzpicture}
\hspace*{1cm}
\begin{tikzpicture}[baseline=(current bounding box.north)]
\begin{axis}[
width=0.45\textwidth,
height=0.40\textwidth,
major grid style={gray!40},
axis line style={very thick},
tick style={very thick},
tick label style={font=\large},
label style={font=\large},
ymajorgrids=true,
x tick label style={rotate=30, anchor=east},
ybar,
bar width=12pt,
enlarge x limits=0.15,
ylabel={Relative activity $\tilde{A}$ [-]},
symbolic x coords={Em,Ef,Es,Di,Aux},
xtick={Em,Ef,Es,Di,Aux},
xticklabels={
    $\mathrm{FICNN}_{\tens{F},s}$,
    $\mathrm{FICNN}_{\tens{F}}$,
    $\mathrm{FICNN}_{s}$,
    $\mathrm{PICNN}_{\tens{g}}$,
    $\mathrm{MLP}_{s}$
},
nodes near coords,
nodes near coords style={
    /pgf/number format/.cd,
    fixed,
    precision=3
},
nodes near coords align={vertical},
]
\addplot[draw=black,fill=ekB!50!ekC,line width=1.0pt,bar shift=0pt] coordinates {(Em,2.831332234507e-01)};
\addplot[draw=black,fill=ekB!50!ekC,line width=1.0pt,bar shift=0pt] coordinates {(Ef,9.455100150320e-03)};
\addplot[draw=black,fill=ekB!50!ekC,line width=1.0pt,bar shift=0pt] coordinates {(Es,7.429073881072e-02)};
\addplot[draw=black,fill=ekB!50!ekC,line width=1.0pt,bar shift=0pt] coordinates {(Di,3.509693503708e-01)};
\addplot[draw=black,fill=ekE,line width=1.0pt,bar shift=0pt]        coordinates {(Aux,2.821515872175e-01)};
\end{axis}
\end{tikzpicture}
\caption{Left: Training history of the physics-based neural network for the plate with an elliptic hole problem. Shown are the total loss as well as its individual contributions, including the physics-based residual losses, the loss of the auxiliary network, and the regularization term, over the course of training iterations.
Right: Relative activity according to \cref{eq:activity} of the individual subnetworks for the plate with an elliptic hole problem. The diagram shows the normalized contributions of internal energy, dissipation, and auxiliary MLP to the overall activity.}
\label{fig:loss_PWEH}
\end{figure}
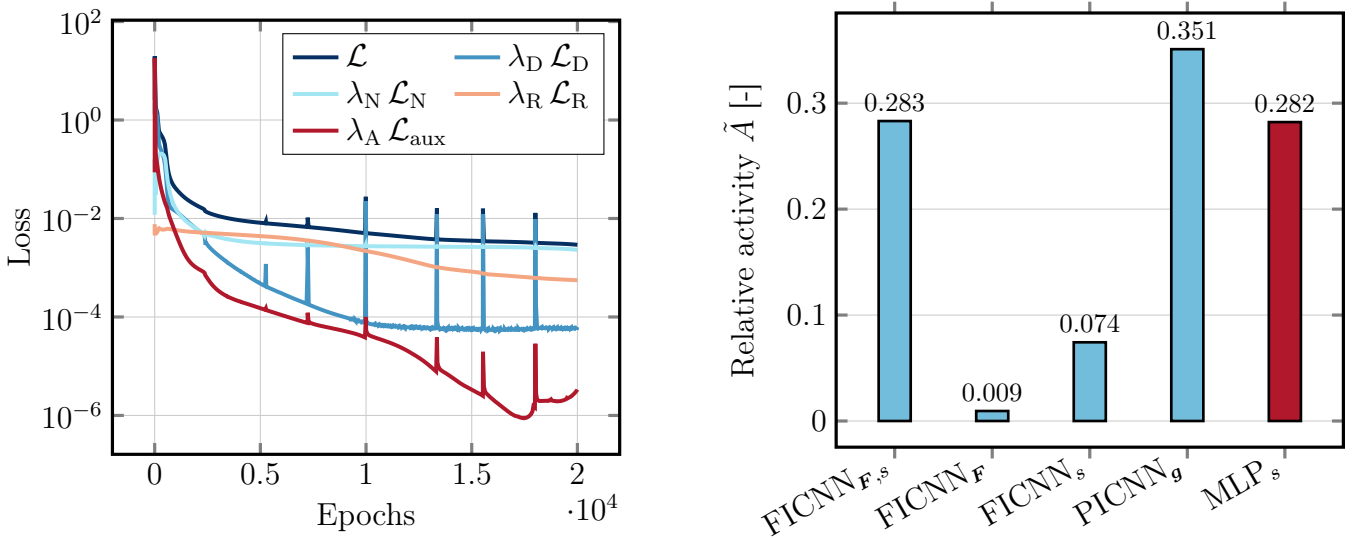

The network is trained for $20{,}000$ epochs with a weighting parameter of $\lambda_{\mathrm{A}} = 10^1$.
The evolution of the total loss and all individual loss components is shown in \cref{fig:loss_PWEH} (left).
Initially, all loss terms are of comparable magnitude.
As training progresses, the losses decrease steadily, with the entropy-related loss $\mathcal{L}_{\mathrm{aux}}$ exhibiting the most pronounced reduction.

The corresponding relative activity of the subnetworks for the best-performing model is shown in \cref{fig:loss_PWEH} (right).
From a constitutive perspective, the coupled subnetwork $\mathrm{FICNN}_{\tens{F},s}$ dominates the internal energy, whereas the purely mechanical subnetwork $\mathrm{FICNN}_{\tens{F}}$ contributes only marginally.
Furthermore, the dissipation network exhibits significant activity, which can again be attributed to the parametrization induced by Fourier-type heat conduction.

\subsubsection{Testing: Spring-like specimen}
\label{sec:structural_example_testing}

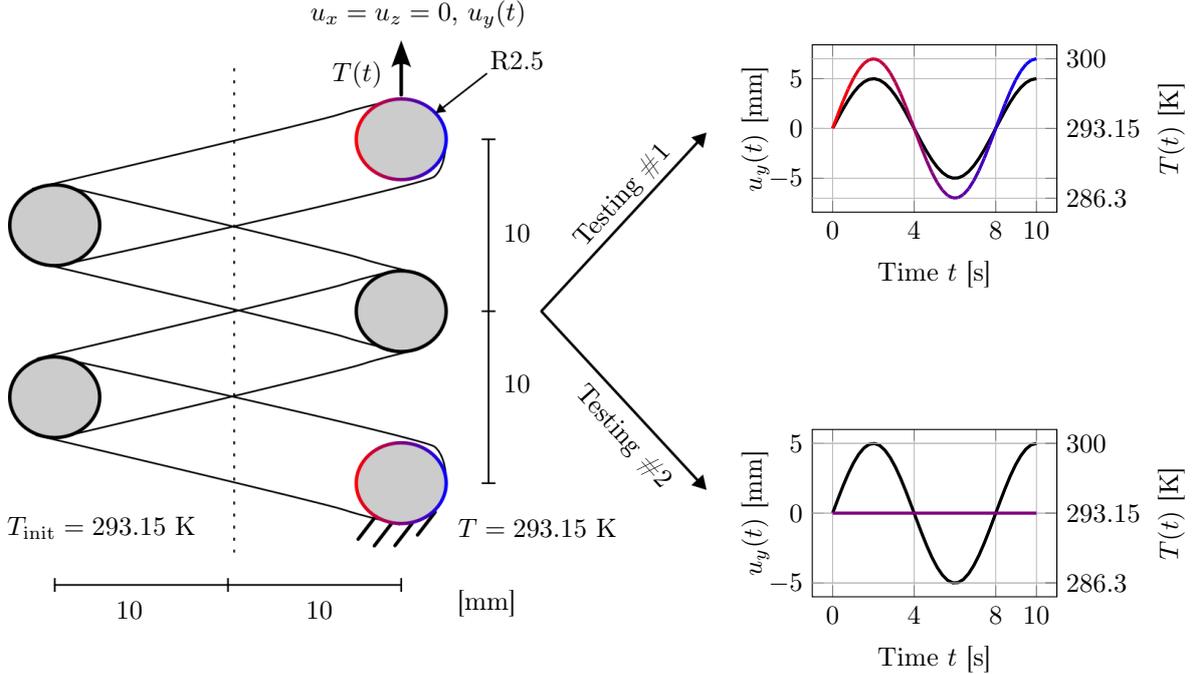
\begin{figure}[!ht]
\centering
\begin{tikzpicture}
\pgfplotsset{
    colormap={redblue}{
        rgb255(0cm)=(255,0,0)
        rgb255(1cm)=(0,0,255)
    }
}
\usetikzlibrary{calc}

\node[anchor=center, inner sep=0] (img) at (0,0)
    {\includegraphics[width=0.5\textwidth]{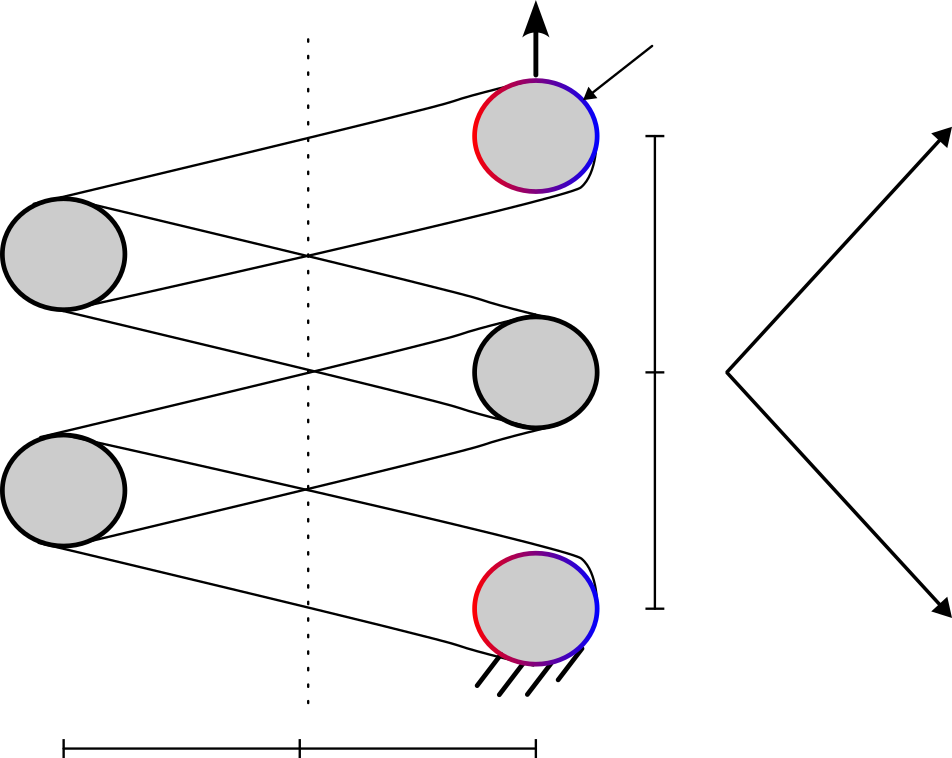}};

\node at (2.1,1.1) {$10$};
\node at (2.1,-0.9) {$10$};
\node at (-3.0,-3.9) {$10$};
\node at (-0.5,-3.9) {$10$};
\node at (2.1,3.4) {R$2.5$};
\node at (1.7,-3.8) {[mm]};

\node at (0.8,4.0) {$u_x=u_z=0$, $u_y(t)$};
\node at (0.0,3.2) {$T(t)$};
\node[anchor=west] at (1.2,-2.8) {$T=293.15$ K};

\node[anchor=east] at (-2.0,-2.8) {$T_{\mathrm{init}}=293.15$ K};

\node[rotate=48] at ($(img.east)+(-1.1cm,+1.6cm)$) {Testing \#1};
\node[rotate=-48] at ($(img.east)+(-1.1cm,-1.6cm)$) {Testing \#2};

\begin{axis}[
    at={($(img.east)+(3.0cm,+3.6cm)$)},
    anchor=north,
    width=4.8cm,
    height=3.8cm,
    xlabel={Time $t$ [s]},
    ylabel={$u_y(t)$ [mm]},
    xtick={0,4,8,10},
    ytick={-5,0,5},
    ylabel style={at={(axis description cs:+0.15,0.5)}},
    grid=both,
    enlargelimits=0.1,
    ymax=7,ymin=-7
]
\addplot[
    very thick,
    domain=0:10,
    samples=40
]
{5*sin(deg(pi/4 * x))};
\end{axis}

\begin{axis}[
    at={($(img.east)+(3.0cm,3.6cm)$)},
    anchor=north,
    width=4.8cm,
    height=3.8cm,
    axis x line=none,
    axis y line*=right,
    ytick={286.3,293.15,300},
    ylabel={$T(t)$ [K]},
    ylabel style={at={(axis description cs:1.85,0.5)}},
    grid=both,
    ymin=286.3, ymax=300,
    enlargelimits=0.1
]
\addplot[
    mesh,
    no markers,
    very thick,
    point meta=x,
    colormap name=redblue,
    domain=0:10,
    samples=40,
]
{293.15+(300-293.15)*sin(deg(pi/4 * x))};
\end{axis}

\begin{axis}[
    at={($(img.east)+(3.0cm,-1.5cm)$)},
    anchor=north,
    width=4.8cm,
    height=3.8cm,
    xlabel={Time $t$ [s]},
    ylabel={$u_y(t)$ [mm]},
    xtick={0,4,8,10},
    ytick={-5,0,5},
    ylabel style={at={(axis description cs:+0.15,0.5)}},
    grid=both,
    enlargelimits=0.1
]
\addplot[
    very thick,
    domain=0:10,
    samples=40
]
{5*sin(deg(pi/4 * x))};
\end{axis}

\begin{axis}[
    at={($(img.east)+(3.0cm,-1.5cm)$)},
    anchor=north,
    width=4.8cm,
    height=3.8cm,
    axis x line=none,
    axis y line*=right,
    ytick={286.3,293.15,300},
    ylabel={$T(t)$ [K]},
    ylabel style={at={(axis description cs:1.85,0.5)}},
    grid=both,
    ymin=286.3, ymax=300,
    enlargelimits=0.1
]
\addplot[
    mesh,
    no markers,
    very thick,
    point meta=x,
    colormap name=redblue,
    domain=0:10,
    samples=40,
]
coordinates {
    (0,293.15)
    (10,293.15)
};
\end{axis}

\end{tikzpicture}
\caption{Boundary value problem and testing scenarios for the spring-like structure. Left: Geometry and boundary conditions, including prescribed displacement $u_y(t)$ and temperature $T(t)$ at the top boundary, while the bottom boundary is fixed and maintained at constant temperature $T=293.15\,\mathrm{K}$. The initial temperature is set to $T_{\mathrm{init}}=293.15\,\mathrm{K}$. Right: Two testing scenarios used for model evaluation. In Testing~\#1, coupled thermo-mechanical loading is applied via time-dependent displacement and temperature. In Testing~\#2, only mechanical loading is applied while the temperature is kept constant, such that thermal effects arise solely from the deformation. Solid black lines denote the prescribed displacement, whereas the temperature evolution is represented by colored lines transitioning from red to blue.}
\label{fig:sketch_spring}
\end{figure}

It remains to assess the predictive capability of the discovered neural network with respect to nodal reactions for an unseen boundary value problem.
To this end, we consider a spring-like specimen, depicted in \cref{fig:sketch_spring}, which is evaluated under two different testing scenarios.
In the first scenario, both sinusoidal deformation and temperature loading are applied at the top clamping of the specimen.
In contrast, the second scenario maintains a constant temperature at both the top and bottom boundaries, equal to the initial temperature throughout the domain.
Thus, any temperature evolution arises solely from deformation-induced effects.

The geometry is discretized using $3640$ hexahedral elements and $4716$ nodes.
Both testing scenarios are simulated using $40$ time steps.

It should be noted that the auxiliary MLP for entropy prediction is not employed during testing, as it primarily serves to accelerate and stabilize training.
Instead, all evaluations are performed using the Newton--Raphson scheme.

\begin{figure}[!ht]
\centering
\vspace*{-2.5cm}
\begin{tikzpicture}
\node[anchor=center, inner sep=0] (img) at (0,0)
    {\includegraphics[height=\textheight]{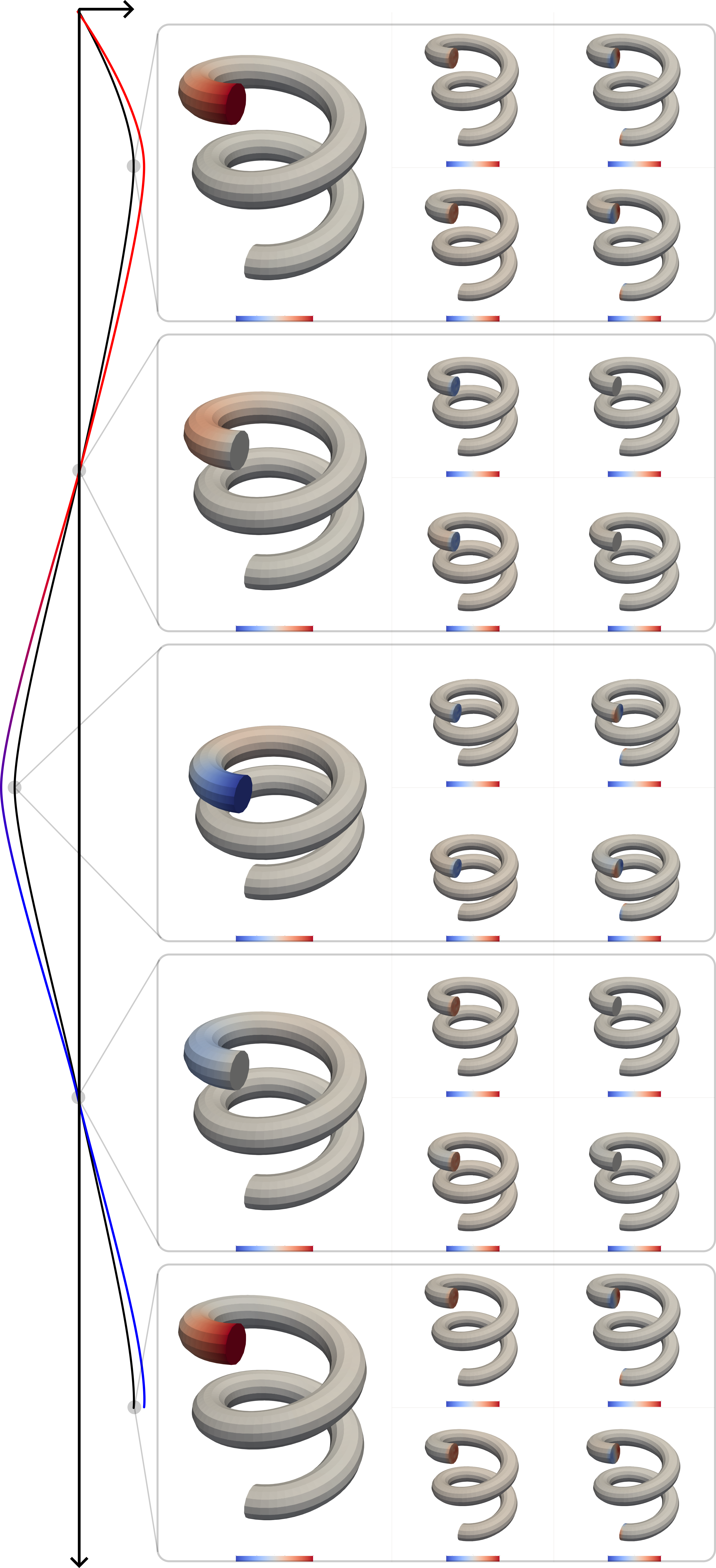}};

\node[rotate=-90] at (-4.6,-11) {Time};

\begin{scope}
\node at (-2.8,11.3) {$T$};
\node at (-1.35,7.35) {\scriptsize [K]};
\node[anchor=east] at (-1.9+0.07,7.2) {\scriptsize $286.3$};
\node[anchor=west] at (-0.8+0.07,7.2) {\scriptsize $300$};

\node[anchor=east] at (1.+0.07,11.3) {$Q_{\mathrm{ref}}$};
\node[rotate=90] at (0.4+0.07,10.65) {\scriptsize [mW]};
\node at (0.95+0.07,9.5) {\scriptsize $-13$};
\node at (2.35+0.07,9.5) {\scriptsize $12$};

\node[anchor=east] at (3.5+0.07,11.3) {$F_{y,\mathrm{ref}}$};
\node[rotate=90] at (2.95+0.07,10.75) {\scriptsize [kN]};
\node at (3.3+0.07,9.5) {\scriptsize $-0.45$};
\node at (4.9+0.07,9.5) {\scriptsize $0.45$};

\node[anchor=east] at (0.9+0.07,8.9) {$Q$};
\node[rotate=90] at (0.4+0.07,8.3) {\scriptsize [mW]};
\node at (0.95+0.07,7.15) {\scriptsize $-13$};
\node at (2.35+0.07,7.15) {\scriptsize $12$};

\node[anchor=east] at (3.4+0.07,8.9) {$F_y$};
\node[rotate=90] at (2.95+0.07,8.4) {\scriptsize [kN]};
\node at (3.3+0.07,7.15) {\scriptsize $-0.45$};
\node at (4.9+0.07,7.15) {\scriptsize $0.45$};
\end{scope}

\begin{scope}[shift={(0,-4.72)}]
\node[anchor=east] at (-1.9+0.07,7.2) {\scriptsize $286.3$};
\node[anchor=west] at (-0.8+0.07,7.2) {\scriptsize $300$};

\node at (0.95+0.07,9.5) {\scriptsize $-13$};
\node at (2.35+0.07,9.5) {\scriptsize $12$};

\node at (3.3+0.07,9.5) {\scriptsize $-0.45$};
\node at (4.9+0.07,9.5) {\scriptsize $0.45$};

\node at (0.95+0.07,7.15) {\scriptsize $-13$};
\node at (2.35+0.07,7.15) {\scriptsize $12$};

\node at (3.3+0.07,7.15) {\scriptsize $-0.45$};
\node at (4.9+0.07,7.15) {\scriptsize $0.45$};
\end{scope}

\begin{scope}[shift={(0,-9.46)}]
\node[anchor=east] at (-1.9+0.07,7.2) {\scriptsize $286.3$};
\node[anchor=west] at (-0.8+0.07,7.2) {\scriptsize $300$};

\node at (0.95+0.07,9.5) {\scriptsize $-13$};
\node at (2.35+0.07,9.5) {\scriptsize $12$};

\node at (3.3+0.07,9.5) {\scriptsize $-0.45$};
\node at (4.9+0.07,9.5) {\scriptsize $0.45$};

\node at (0.95+0.07,7.15) {\scriptsize $-13$};
\node at (2.35+0.07,7.15) {\scriptsize $12$};

\node at (3.3+0.07,7.15) {\scriptsize $-0.45$};
\node at (4.9+0.07,7.15) {\scriptsize $0.45$};
\end{scope}

\begin{scope}[shift={(0,-14.2)}]
\node[anchor=east] at (-1.9+0.07,7.2) {\scriptsize $286.3$};
\node[anchor=west] at (-0.8+0.07,7.2) {\scriptsize $300$};

\node at (0.95+0.07,9.5) {\scriptsize $-13$};
\node at (2.35+0.07,9.5) {\scriptsize $12$};

\node at (3.3+0.07,9.5) {\scriptsize $-0.45$};
\node at (4.9+0.07,9.5) {\scriptsize $0.45$};

\node at (0.95+0.07,7.15) {\scriptsize $-13$};
\node at (2.35+0.07,7.15) {\scriptsize $12$};

\node at (3.3+0.07,7.15) {\scriptsize $-0.45$};
\node at (4.9+0.07,7.15) {\scriptsize $0.45$};
\end{scope}

\begin{scope}[shift={(0,-18.93)}]
\node[anchor=east] at (-1.9+0.07,7.2) {\scriptsize $286.3$};
\node[anchor=west] at (-0.8+0.07,7.2) {\scriptsize $300$};

\node at (0.95+0.07,9.5) {\scriptsize $-13$};
\node at (2.35+0.07,9.5) {\scriptsize $12$};

\node at (3.3+0.07,9.5) {\scriptsize $-0.45$};
\node at (4.9+0.07,9.5) {\scriptsize $0.45$};

\node at (0.95+0.07,7.15) {\scriptsize $-13$};
\node at (2.35+0.07,7.15) {\scriptsize $12$};

\node at (3.3+0.07,7.15) {\scriptsize $-0.45$};
\node at (4.9+0.07,7.15) {\scriptsize $0.45$};
\end{scope}


\end{tikzpicture}
\caption{Results for Testing~\#1 of the spring-like structure under coupled thermo-mechanical loading. Shown are five representative time steps during the loading process. For each snapshot, the temperature field $T$, the heat flow $Q$, and the forces $F_y$ are depicted. The reference solution $(\bullet)_{\mathrm{ref}}$ is compared against the predictions of the discovered model. The results demonstrate that the learned model accurately captures the fully coupled thermo-mechanical response.}
\label{fig:results_spring_testing01}
\end{figure}

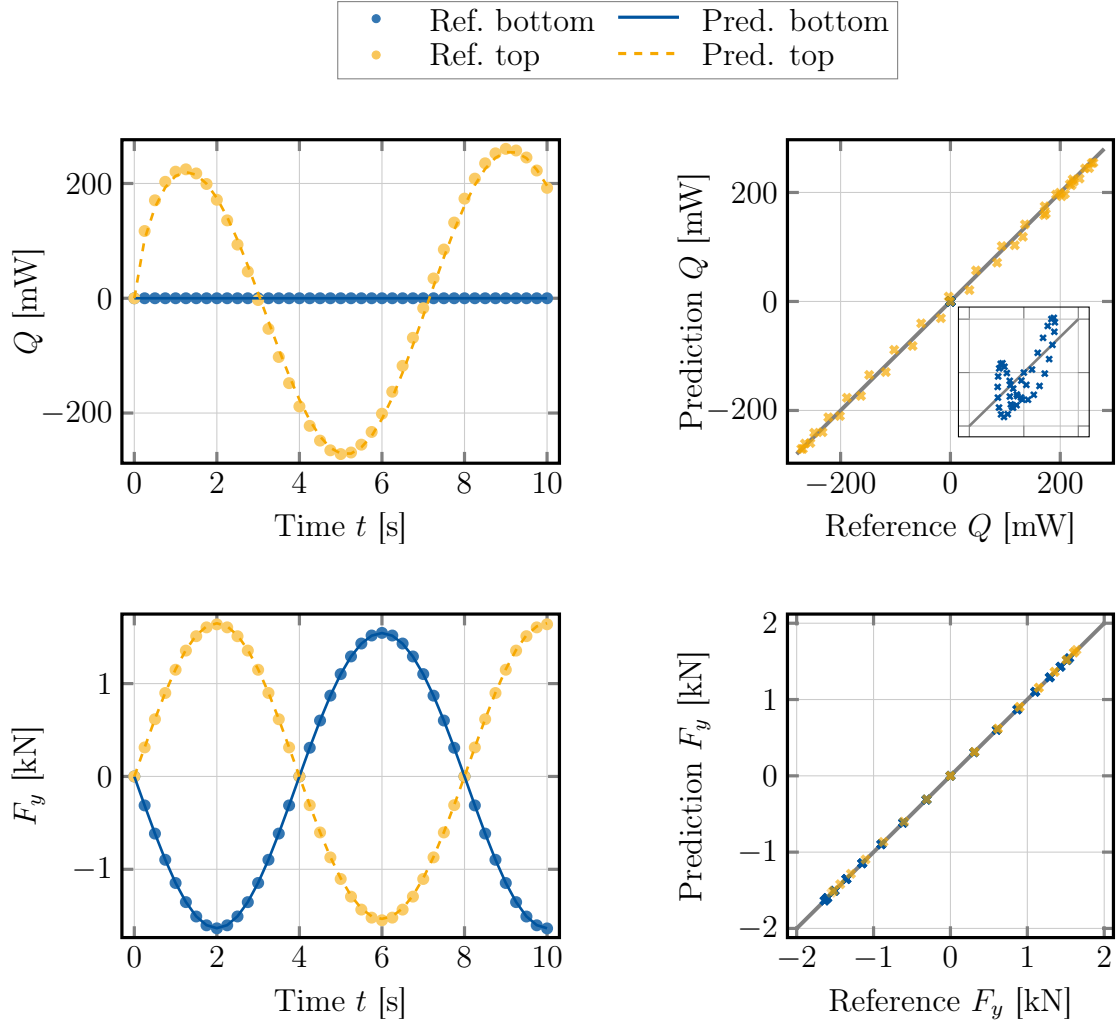
\begin{figure}[!ht]
\centering
\begin{tikzpicture}

  \begin{groupplot}[
    group style={
      group size=2 by 2,
      horizontal sep=3cm,
      vertical sep=2cm
    },
    width=0.40\textwidth,
    height=0.32\textwidth,
    grid=major,
    major grid style={gray!40},
    axis line style={very thick},
    tick style={very thick},
    tick label style={font=\large},
    label style={font=\large},
    enlargelimits=0.03,
    clip mode=individual
  ]

    \nextgroupplot[
      xlabel={Time $t$ [s]},
      ylabel={$Q$ [mW]},
    ]

    \addplot[
      only marks,
      mark=*,
      mark size=1.5pt,
      mark options={line width=1.5pt},
      clr1!80,
    ] table[x=t, y=FT_0, col sep=comma]
    {./graphs/Spring/Spring02_results.csv};

    \addplot[
      only marks,
      mark=*,
      mark size=1.5pt,
      mark options={line width=1.5pt},
      clr4!60,
    ] table[x=t, y=FT_1, col sep=comma]
    {./graphs/Spring/Spring02_results.csv};

    \addplot[
      line width=1.0pt,
      clr1
    ] table[x=t, y=FT_NN_Newton_0, col sep=comma]
    {./graphs/Spring/Spring02_results.csv};

    \addplot[
      line width=1.0pt,
      clr4,
      dashed
    ] table[x=t, y=FT_NN_Newton_1, col sep=comma]
    {./graphs/Spring/Spring02_results.csv};

    \nextgroupplot[
      name=parityQ,
      width=0.32\textwidth,
      xlabel={Reference $Q$ [mW]},
      ylabel={Prediction $Q$ [mW]},
    ]

    \addplot[
      gray,
      line width=1.5pt,
      domain=-280:280,
      samples=2
    ] {x};

    \addplot[
      only marks,
      mark=x,
      mark size=2pt,
      mark options={line width=1.5pt},
      clr1
    ] table[x=FT_0, y=FT_NN_Newton_0, col sep=comma]
    {./graphs/Spring/Spring02_results.csv};

    \addplot[
      only marks,
      mark=x,
      mark size=2pt,
      opacity=0.7,
      mark options={line width=1.5pt},
      clr4
    ] table[x=FT_1, y=FT_NN_Newton_1, col sep=comma]
    {./graphs/Spring/Spring02_results.csv};

    \nextgroupplot[
      xlabel={Time $t$ [s]},
      ylabel={$F_y$ [kN]},
    ]

    \addplot[
      only marks,
      mark=*,
      mark size=1.5pt,
      mark options={line width=1.5pt},
      clr1!80,
    ] table[x=t, y=Fy_0, col sep=comma]
    {./graphs/Spring/Spring02_results.csv};

    \addplot[
      only marks,
      mark=*,
      mark size=1.5pt,
      mark options={line width=1.5pt},
      clr4!60,
    ] table[x=t, y=Fy_1, col sep=comma]
    {./graphs/Spring/Spring02_results.csv};

    \addplot[
      line width=1.0pt,
      clr1
    ] table[x=t, y=Fy_NN_Newton_0, col sep=comma]
    {./graphs/Spring/Spring02_results.csv};

    \addplot[
      line width=1.0pt,
      clr4,
      dashed
    ] table[x=t, y=Fy_NN_Newton_1, col sep=comma]
    {./graphs/Spring/Spring02_results.csv};

    \nextgroupplot[
      width=0.32\textwidth,
      xlabel={Reference $F_y$ [kN]},
      ylabel={Prediction $F_y$ [kN]},
    ]

    \addplot[
      gray,
      line width=1.5pt,
      domain=-2:2,
      samples=2
    ] {x};

    \addplot[
      only marks,
      mark=x,
      mark size=2pt,
      mark options={line width=1.5pt},
      clr1
    ] table[x=Fy_0, y=Fy_NN_Newton_0, col sep=comma]
    {./graphs/Spring/Spring02_results.csv};

    \addplot[
      only marks,
      mark=x,
      mark size=2pt,
      opacity=0.4,
      mark options={line width=1.5pt},
      clr4
    ] table[x=Fy_1, y=Fy_NN_Newton_1, col sep=comma]
    {./graphs/Spring/Spring02_results.csv};

  \end{groupplot}

\begin{axis}[
  at={($(parityQ.south west)+(2.25cm,0.35cm)$)},
  anchor=south west,
  width=0.18\textwidth,
  height=0.18\textwidth,
  xticklabels=\empty,
  yticklabels=\empty,
  scaled ticks=false,
  grid=both,
]

  \addplot[
    gray,
    line width=1pt,
    domain=-0.001:0.001,
    samples=2
  ] {x};

  \addplot[
    only marks,
    mark=x,
    mark size=1.5pt,
    mark options={line width=1pt},
    clr1
  ] table[x=FT_0, y=FT_NN_Newton_0, col sep=comma]
  {./graphs/Spring/Spring02_results.csv};

  \end{axis}

  \coordinate (legendtop) at ($(group c1r1.north west)!0.5!(group c2r1.north east)+(0,8mm)$);

\def\legw{8mm}

\matrix[
  matrix of nodes,
  anchor=south,
  at={(legendtop)},
  draw=gray,
  inner sep=3pt,
  row sep=1.5mm,
  column sep=3mm,
  nodes={
    anchor=west,
    inner sep=0pt,
    outer sep=0pt,
    font=\large
  }
]{
\makebox[\legw][c]{\tikz[baseline=-0.6ex]\filldraw[clr1!80] (0,0) circle (1.5pt);} &
{Ref.\ bottom} &
\makebox[\legw][c]{\tikz[baseline=-0.6ex]\draw[clr1, line width=1.2pt] (0,0) -- (\legw,0);} &
{Pred. bottom} \\

\makebox[\legw][c]{\tikz[baseline=-0.6ex]\filldraw[clr4!65] (0,0) circle (1.5pt);} &
{Ref.\ top} &
\makebox[\legw][c]{\tikz[baseline=-0.6ex]\draw[clr4, dashed, line width=1.2pt] (0,0) -- (\legw,0);} &
{Pred. top} \\
  };

\end{tikzpicture}
\caption{Quantitative comparison of the predicted and reference responses for Testing~\#1 of the spring-like structure under coupled thermo-mechanical loading. Left: Temporal evolution of the reaction heat flow $Q$ (top) and the reaction forces $F_y$ (bottom) at the top and bottom boundaries. Right: Corresponding parity plots comparing predictions and reference values. The results show good agreement for both mechanical and thermal quantities, with only minor deviations in the heat flow at the bottom boundary.}
\label{fig:results_spring_testing01_curves}
\end{figure}

\paragraph{Testing \#1}
The results at five representative snapshots during loading are shown in \cref{fig:results_spring_testing01}.
The spatial distribution of the nodal forces in vertical direction, comparing the reference solution $F_{y,\mathrm{ref}}$ and the prediction $F_y$, as well as the heat flows $Q_{\mathrm{ref}}$ and $Q$, are in good agreement across all snapshots.
Note that the forces $F_y$ correspond to the surface integrals of the tractions $\ten{t}$, while the heat flow corresponds to the surface integral of the heat flux $q$.
This agreement is further confirmed quantitatively in \cref{fig:results_spring_testing01_curves}.
The corresponding parity plots indicate good agreement for both the reaction forces at the top and bottom boundaries and the reaction heat flow.
Only minor deviations in the heat flow at the bottom boundary are observed, which are considered negligible.

These results indicate that the discovered model is capable of accurately predicting the thermomechanical response for an unseen boundary value problem with combined loading conditions.

\begin{figure}[!ht]
\centering
\vspace*{-2cm}
\begin{tikzpicture}
\node[anchor=center, inner sep=0] (img) at (0,0)
    {\includegraphics[height=\textheight]{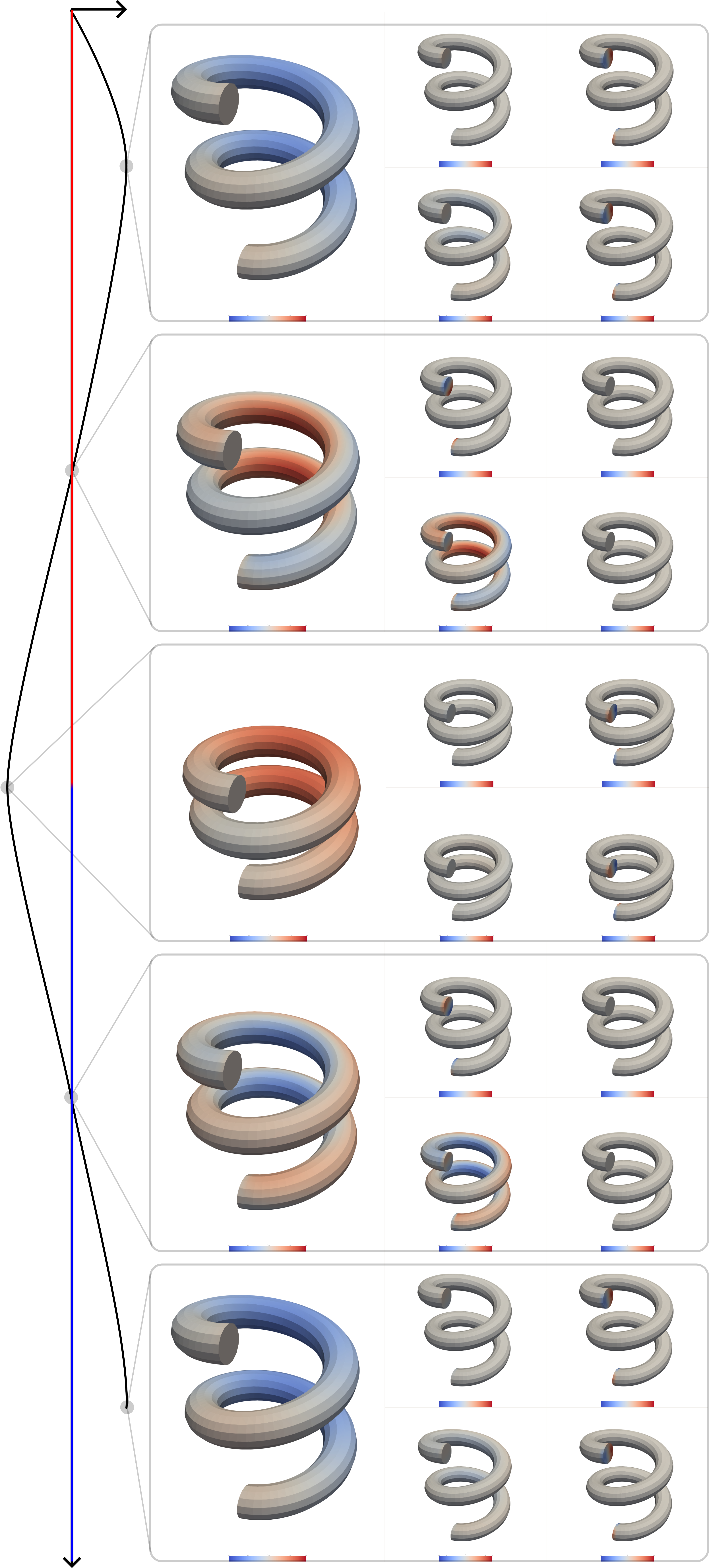}};

\node[rotate=-90] at (-4.6,-11) {Time};

\begin{scope}
\node at (-2.8,11.3) {$T$};
\node at (-1.35,7.35) {\scriptsize [K]};
\node[anchor=east] at (-1.9,7.2) {\scriptsize $293.14$};
\node[anchor=west] at (-0.78,7.2) {\scriptsize $293.15$};

\node[anchor=east] at (1.,11.3) {$Q_{\mathrm{ref}}$};
\node[rotate=90] at (0.4,10.65) {\scriptsize [mW]};
\node at (0.95,9.5) {\scriptsize $-2.4$};
\node at (2.35,9.5) {\scriptsize $2.4$};
\node at (2.5,9.8) {\scriptsize $\cdot 10^{-4}$};

\node[anchor=east] at (3.5,11.3) {$F_{y,\mathrm{ref}}$};
\node[rotate=90] at (2.95,10.75) {\scriptsize [kN]};
\node at (3.3,9.5) {\scriptsize $-0.45$};
\node at (4.9,9.5) {\scriptsize $0.45$};

\node[anchor=east] at (0.9,8.9) {$Q$};
\node[rotate=90] at (0.4,8.3) {\scriptsize [mW]};
\node at (0.95,7.15) {\scriptsize $-2.8$};
\node at (2.35,7.15) {\scriptsize $2.7$};
\node at (2.5,7.45) {\scriptsize $\cdot 10^{-4}$};

\node[anchor=east] at (3.4,8.9) {$F_y$};
\node[rotate=90] at (2.95,8.4) {\scriptsize [kN]};
\node at (3.3,7.15) {\scriptsize $-0.45$};
\node at (4.9,7.15) {\scriptsize $0.45$};
\end{scope}

\begin{scope}[shift={(0,-4.72)}]
\node[anchor=east] at (-1.9,7.2) {\scriptsize $293.14$};
\node[anchor=west] at (-0.78,7.2) {\scriptsize $293.15$};

\node at (0.95,9.5) {\scriptsize $-2.4$};
\node at (2.35,9.5) {\scriptsize $2.4$};
\node at (2.5,9.8) {\scriptsize $\cdot 10^{-4}$};

\node at (3.3,9.5) {\scriptsize $-0.45$};
\node at (4.9,9.5) {\scriptsize $0.45$};

\node at (0.95,7.15) {\scriptsize $-2.8$};
\node at (2.35,7.15) {\scriptsize $2.7$};
\node at (2.5,7.45) {\scriptsize $\cdot 10^{-4}$};

\node at (3.3,7.15) {\scriptsize $-0.45$};
\node at (4.9,7.15) {\scriptsize $0.45$};
\end{scope}

\begin{scope}[shift={(0,-9.46)}]
\node[anchor=east] at (-1.9,7.2) {\scriptsize $293.14$};
\node[anchor=west] at (-0.78,7.2) {\scriptsize $293.15$};

\node at (0.95,9.5) {\scriptsize $-2.4$};
\node at (2.35,9.5) {\scriptsize $2.4$};
\node at (2.5,9.8) {\scriptsize $\cdot 10^{-4}$};

\node at (3.3,9.5) {\scriptsize $-0.45$};
\node at (4.9,9.5) {\scriptsize $0.45$};

\node at (0.95,7.15) {\scriptsize $-2.8$};
\node at (2.35,7.15) {\scriptsize $2.7$};
\node at (2.5,7.45) {\scriptsize $\cdot 10^{-4}$};

\node at (3.3,7.15) {\scriptsize $-0.45$};
\node at (4.9,7.15) {\scriptsize $0.45$};
\end{scope}

\begin{scope}[shift={(0,-14.2)}]
\node[anchor=east] at (-1.9,7.2) {\scriptsize $293.14$};
\node[anchor=west] at (-0.78,7.2) {\scriptsize $293.15$};

\node at (0.95,9.5) {\scriptsize $-2.4$};
\node at (2.35,9.5) {\scriptsize $2.4$};
\node at (2.5,9.8) {\scriptsize $\cdot 10^{-4}$};

\node at (3.3,9.5) {\scriptsize $-0.45$};
\node at (4.9,9.5) {\scriptsize $0.45$};

\node at (0.95,7.15) {\scriptsize $-2.8$};
\node at (2.35,7.15) {\scriptsize $2.7$};
\node at (2.5,7.45) {\scriptsize $\cdot 10^{-4}$};

\node at (3.3,7.15) {\scriptsize $-0.45$};
\node at (4.9,7.15) {\scriptsize $0.45$};
\end{scope}

\begin{scope}[shift={(0,-18.93)}]
\node[anchor=east] at (-1.9,7.2) {\scriptsize $293.14$};
\node[anchor=west] at (-0.78,7.2) {\scriptsize $293.15$};

\node at (0.95,9.5) {\scriptsize $-2.4$};
\node at (2.35,9.5) {\scriptsize $2.4$};
\node at (2.5,9.8) {\scriptsize $\cdot 10^{-4}$};

\node at (3.3,9.5) {\scriptsize $-0.45$};
\node at (4.9,9.5) {\scriptsize $0.45$};

\node at (0.95,7.15) {\scriptsize $-2.8$};
\node at (2.35,7.15) {\scriptsize $2.7$};
\node at (2.5,7.45) {\scriptsize $\cdot 10^{-4}$};

\node at (3.3,7.15) {\scriptsize $-0.45$};
\node at (4.9,7.15) {\scriptsize $0.45$};
\end{scope}


\end{tikzpicture}
\caption{Results for Testing~\#2 of the spring-like structure under purely mechanical loading with constant temperature boundary conditions. Shown are five representative time steps during the loading process. For each snapshot, the temperature field $T$, the heat flow $Q$, and the forces $F_y$ are depicted. The reference solution $(\bullet)_{\mathrm{ref}}$ is compared against the predictions of the discovered model. While the forces are accurately reproduced, noticeable discrepancies are observed in the heat flow, including non-physical fluxes within the interior of the domain.}
\label{fig:results_spring_testing02}
\end{figure}

\paragraph{Testing \#2}
While the first scenario demonstrates strong predictive performance, the second scenario specifically targets the ability of the model to capture deformation-induced self-heating effects.
The results for five representative snapshots are shown in \cref{fig:results_spring_testing02}.
As expected, the temperature decreases under tensile loading and increases under compressive loading, reflecting thermoelastic coupling.
This cyclic behavior is consistently observed due to the sinusoidal deformation.

For the nodal forces, the predicted distributions closely match the reference solution across all time steps.
However, in contrast to the first scenario, noticeable discrepancies are observed in the heat flow.
In particular, for the second and fourth snapshots, non-zero heat flows appear within the interior of the domain, which is inconsistent with the absence of thermal Dirichlet boundary conditions.

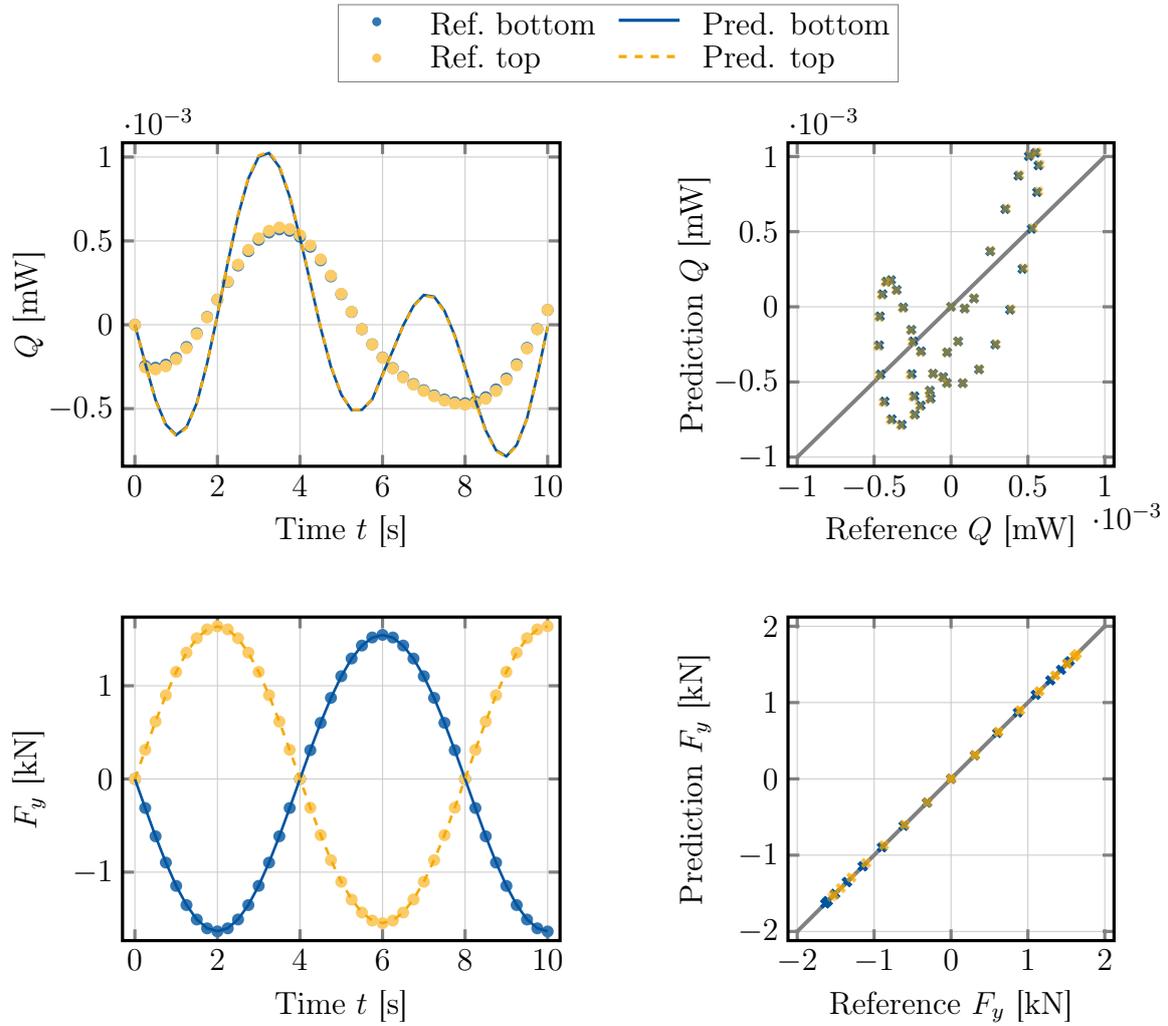
\begin{figure}[!ht]
\centering
\begin{tikzpicture}

  \begin{groupplot}[
    group style={
      group size=2 by 2,
      horizontal sep=3cm,
      vertical sep=2cm
    },
    width=0.40\textwidth,
    height=0.32\textwidth,
    grid=major,
    major grid style={gray!40},
    axis line style={very thick},
    tick style={very thick},
    tick label style={font=\large},
    label style={font=\large},
    enlargelimits=0.03,
    clip mode=individual
  ]

    \nextgroupplot[
      xlabel={Time $t$ [s]},
      ylabel={$Q$ [mW]},
    ]

    \addplot[
      only marks,
      mark=*,
      mark size=1.5pt,
      mark options={line width=1.5pt},
      clr1!80,
    ] table[x=t, y=FT_0, col sep=comma]
    {./graphs/Spring/Spring01_results.csv};

    \addplot[
      only marks,
      mark=*,
      mark size=1.5pt,
      mark options={line width=1.5pt},
      clr4!60,
    ] table[x=t, y=FT_1, col sep=comma]
    {./graphs/Spring/Spring01_results.csv};

    \addplot[
      line width=1.0pt,
      clr1
    ] table[x=t, y=FT_NN_Newton_0, col sep=comma]
    {./graphs/Spring/Spring01_results.csv};

    \addplot[
      line width=1.0pt,
      clr4,
      dashed
    ] table[x=t, y=FT_NN_Newton_1, col sep=comma]
    {./graphs/Spring/Spring01_results.csv};

    \nextgroupplot[
      width=0.32\textwidth,
      xlabel={Reference $Q$ [mW]},
      ylabel={Prediction $Q$ [mW]},
      scaled x ticks=true,
      x tick scale label style={
        at={(axis description cs:1.15,0.05)},
        anchor=north east
      },
    ]

    \addplot[
      gray,
      line width=1.5pt,
      domain=-0.001:0.001,
      samples=2
    ] {x};

    \addplot[
      only marks,
      mark=x,
      mark size=2pt,
      mark options={line width=1.5pt},
      clr1
    ] table[x=FT_0, y=FT_NN_Newton_0, col sep=comma]
    {./graphs/Spring/Spring01_results.csv};

    \addplot[
      only marks,
      mark=x,
      mark size=2pt,
      opacity=0.5,
      mark options={line width=1.5pt},
      clr4
    ] table[x=FT_1, y=FT_NN_Newton_1, col sep=comma]
    {./graphs/Spring/Spring01_results.csv};

    \nextgroupplot[
      xlabel={Time $t$ [s]},
      ylabel={$F_y$ [kN]},
    ]

    \addplot[
      only marks,
      mark=*,
      mark size=1.5pt,
      mark options={line width=1.5pt},
      clr1!80,
    ] table[x=t, y=Fy_0, col sep=comma]
    {./graphs/Spring/Spring01_results.csv};

    \addplot[
      only marks,
      mark=*,
      mark size=1.5pt,
      mark options={line width=1.5pt},
      clr4!60,
    ] table[x=t, y=Fy_1, col sep=comma]
    {./graphs/Spring/Spring01_results.csv};

    \addplot[
      line width=1.0pt,
      clr1
    ] table[x=t, y=Fy_NN_Newton_0, col sep=comma]
    {./graphs/Spring/Spring01_results.csv};

    \addplot[
      line width=1.0pt,
      clr4,
      dashed
    ] table[x=t, y=Fy_NN_Newton_1, col sep=comma]
    {./graphs/Spring/Spring01_results.csv};

    \nextgroupplot[
      width=0.32\textwidth,
      xlabel={Reference $F_y$ [kN]},
      ylabel={Prediction $F_y$ [kN]},
    ]

    \addplot[
      gray,
      line width=1.5pt,
      domain=-2:2,
      samples=2
    ] {x};

    \addplot[
      only marks,
      mark=x,
      mark size=2pt,
      mark options={line width=1.5pt},
      clr1
    ] table[x=Fy_0, y=Fy_NN_Newton_0, col sep=comma]
    {./graphs/Spring/Spring01_results.csv};

    \addplot[
      only marks,
      mark=x,
      mark size=2pt,
      opacity=0.5,
      mark options={line width=1.5pt},
      clr4
    ] table[x=Fy_1, y=Fy_NN_Newton_1, col sep=comma]
    {./graphs/Spring/Spring01_results.csv};

  \end{groupplot}

  \coordinate (legendtop) at ($(group c1r1.north west)!0.5!(group c2r1.north east)+(0,8mm)$);

\def\legw{8mm}

\matrix[
  matrix of nodes,
  anchor=south,
  at={(legendtop)},
  draw=gray,
  inner sep=3pt,
  row sep=1.5mm,
  column sep=3mm,
  nodes={
    anchor=west,
    inner sep=0pt,
    outer sep=0pt,
    font=\large
  }
]{
\makebox[\legw][c]{\tikz[baseline=-0.6ex]\filldraw[clr1!80] (0,0) circle (1.5pt);} &
{Ref.\ bottom} &
\makebox[\legw][c]{\tikz[baseline=-0.6ex]\draw[clr1, line width=1.2pt] (0,0) -- (\legw,0);} &
{Pred. bottom} \\

\makebox[\legw][c]{\tikz[baseline=-0.6ex]\filldraw[clr4!65] (0,0) circle (1.5pt);} &
{Ref.\ top} &
\makebox[\legw][c]{\tikz[baseline=-0.6ex]\draw[clr4, dashed, line width=1.2pt] (0,0) -- (\legw,0);} &
{Pred. top} \\
  };

\end{tikzpicture}
\caption{Quantitative comparison of the predicted and reference responses for Testing~\#2 of the spring-like structure under purely mechanical loading with constant temperature boundary conditions. Left: Temporal evolution of the reaction heat flow $Q$ (top) and the reaction forces $F_y$ (bottom) at the top and bottom boundaries. Right: Corresponding parity plots comparing predictions and reference values. While the reaction forces are accurately captured, significant deviations are observed in the heat flow, both in magnitude and temporal evolution, including spurious oscillations and increased scatter in the parity plot.}
\label{fig:results_spring_testing02_curves}
\end{figure}

The quantitative comparison in \cref{fig:results_spring_testing02_curves} confirms these observations.
While the reaction forces at both boundaries are accurately captured, the predicted heat flow deviates significantly from the reference solution.
These deviations are evident not only in magnitude but also in the qualitative temporal evolution, where the predicted response exhibits higher-frequency oscillations.

In summary, the model accurately captures the mechanical response across all scenarios.
The thermal response is also well predicted when a pronounced temperature field is present.
However, the model fails to accurately reproduce deformation-induced heating effects.
This limitation is likely due to the significantly smaller magnitude of the temperature changes associated with this phenomenon, which are on the order of $10^{-2}\,\mathrm{K}$, and thus are difficult to resolve during training.

Whether alternative training strategies—such as selectively constraining parts of the network or focusing specific subnetworks on deformation-induced heating—can overcome this limitation remains an open question and is left for future work.

\section{Discussion}
\label{sec:discussion}

\paragraph{Physics-based neural networks learn entropy-based thermomechanics}
The central result of this work is that physics-based neural networks can successfully learn constitutive behavior in fully coupled thermomechanics when formulated in terms of internal energy and dissipation. While such entropy-based formulations are well-established from a thermodynamic perspective \cite{ColemanGurtin1967}, they have so far found limited use in constitutive modeling, partly due to the difficulty of constructing intuitive and expressive models directly in terms of entropy and internal energy. 
The proposed framework overcomes this limitation by enabling not only the identification of model parameters, but the discovery of suitable constitutive representations themselves. In this sense, the neural network does not rely on a predefined model structure, but is able to approximate constitutive relations that may not be available in closed form a priori.
Across all examples, the discovered models reproduce the governing mechanical and thermal responses while satisfying thermodynamic admissibility by construction. This is particularly important because the present framework is formulated in terms of deformation and entropy, whereas temperature remains the experimentally accessible quantity. The results therefore show that the entropy-based formulation is not merely a theoretical alternative, but a practically viable route for constitutive discovery in thermomechanics. This is confirmed consistently across the purely thermal setting (\cref{sec:heat_diffusion}), the temperature-dependent mechanical examples (\cref{sec:porcine_tissue,sec:carbon_filler}), and the fully coupled structural problem (\cref{sec:structural_example}). In this sense, the network does not only fit data, but learns a constitutive structure that is consistent with the energetic and dissipative principles of coupled thermomechanics.

\paragraph{The internal-energy formulation enables robust, consistent, and architecture-embedded thermodynamics}
A key conceptual advantage of the proposed framework lies in the choice of internal energy as primary energetic potential. In contrast to Helmholtz-based formulations, this avoids the need to impose mixed convexity--concavity conditions with respect to deformation and temperature. Instead, the model is constructed from the convex-like internal energy and a convex dissipation potential, which can be represented naturally through input-convex neural networks. Moreover, thermodynamic consistency is not enforced a posteriori but embedded directly into the architecture through invariant-based representations and zero-anchored convex constructions. Objectivity, isotropy, monotonicity in entropy, and convexity of the energetic and dissipative parts are therefore guaranteed by design. The numerical examples demonstrate that this combination is not only thermodynamically sound but also computationally robust, as stable training is achieved across all considered regimes (\cref{sec:heat_diffusion,sec:porcine_tissue,sec:carbon_filler,sec:structural_example}). This supports the view that constitutive learning should be formulated as structured model discovery rather than unconstrained regression \cite{Linka2021,Linka2023,Klein2022,Linden2023}.

\paragraph{Multiphysics discovery requires comprehensive loading paths and benefits from structured training strategies}
The examples show clearly that constitutive discovery in thermomechanics cannot be judged from isolated observations, but requires testing procedures that span the full loading history, including transient, coupled, and path-dependent effects. This is particularly evident in the heat conduction example, where resolving the temporal evolution is essential to identify dissipation (\cref{fig:results_heat_diffusion}). In the structural setting, multiple training scenarios are deliberately combined to activate deformation-driven, thermal, and fully coupled mechanisms (\cref{fig:sketch_PWEH}). This strategy proves sufficient to identify a model that generalizes to a new geometry and unseen loading conditions (\cref{fig:results_spring_testing01,fig:results_spring_testing01_curves}). These results indicate that, for multiphysics discovery, diversity and complementarity of loading paths are equally important to the complexity of individual experiments, as they allow the network to disentangle energetic storage, dissipation, and coupling effects over time, cf. \cite{wang2023respecting}.

\paragraph{Interpretability and efficiency: subnetworks learn physically meaningful roles while auxiliary models accelerate training}
A notable feature of the framework is that the learned subnetworks acquire physically interpretable roles. In the heat conduction example, the relative activity analysis shows that the dissipation network dominates, while purely mechanical energetic contributions remain inactive, as expected (\cref{fig:loss_heat_diffusion}). In contrast, all energetic branches become relevant in temperature-dependent mechanical problems, and the coupled energetic contribution is most active in the fully coupled structural case (\cref{fig:loss_PWEH}). This provides a first layer of interpretability by linking network components to physical mechanisms. At the same time, the auxiliary entropy network significantly accelerates training without degrading the final constitutive response, similar to its usage in case of inelasticity \cite{Asad2023}. As demonstrated in the thermal example, predictions obtained with the auxiliary network closely match those based on the reinstated Newton solver (\cref{fig:results_heat_diffusion}), indicating that the auxiliary model serves as an efficient surrogate during training while preserving the thermodynamic structure during inference.

\paragraph{Data quality and scale separation fundamentally limit thermomechanical discovery}
The porcine tissue and rubber examples highlight that constitutive discovery is inherently limited by the consistency and completeness of the available data. Missing information, such as thermal expansion or consistent rest configurations, restricts identifiability and prevents a fully consistent thermomechanical interpretation (\cref{fig:results_porcine_tissue,fig:results_rubber}). While qualitative trends, such as temperature-dependent softening, are still captured, these examples demonstrate that data must be compatible with the underlying thermodynamic framework. In addition, the structural test reveals that weak coupling effects, such as deformation-induced temperature changes, remain difficult to identify (\cref{fig:results_spring_testing02,fig:results_spring_testing02_curves}). Since these effects are orders of magnitude smaller than the dominant thermal contributions, they are overshadowed during training \cite{tarantola2005inverse}. This indicates that small-scale multiphysics interactions require targeted excitation or adapted training strategies to become identifiable. In this context, it remains an open question whether enriched data generation strategies, such as multiscale simulations as used in neural networks for magneto-elasticity \cite{roth2025datadriven}, all-at-once approaches \cite{Roemer2025}, or tailored dual-stage approaches combining data-driven identification with physics-augmented neural networks \cite{Linden2025}, can improve identifiability and robustness in thermomechanical discovery. Overall, the results show that the success of thermomechanical discovery is inseparable from the design of the data and the relative scale of the underlying physical effects.

\section*{Conclusion}

A physics-based neural network framework for the discovery of constitutive models in fully coupled thermomechanics has been presented. 
The approach is formulated in terms of the deformation gradient and entropy that uses the internal energy and a dissipation potential. This enables a thermodynamically consistent treatment of thermomechanical processes while retaining temperature as the experimentally observable variable.
Thermodynamic admissibility is ensured by construction through convex energetic and dissipative potentials and invariant-based representations embedded in the network architecture.

The numerical investigations demonstrate that the proposed framework captures transient thermal behavior, temperature-dependent mechanical response, and fully coupled thermomechanical effects within a unified setting. In particular, the learned models generalize to unseen boundary value problems and yield accurate predictions of structural reactions, indicating that the approach identifies constitutive behavior beyond pointwise data fitting.

At the same time, the study highlights inherent limitations of data-driven thermomechanical discovery. The quality of the identified models depends critically on the consistency and completeness of the available data, and weak coupling effects may not be resolved if they are not sufficiently represented in the training scenarios.

Overall, the results indicate that internal-energy-dissipation-based neural networks provide a promising and structurally consistent framework for constitutive discovery in thermomechanics. Future works may focus on improving the identification of small-scale coupling effects, extending the approach to inelastic materials, and designing data generation strategies tailored to multiphysics settings.

\subsection*{Acknowledgements}   
\noindent
We thank Karl A. Kalina for some enlightening discussions on the topic of physics-constrained neural networks.
This work was supported 
by the 
by the DFG TRR 280 417002380,  
by the Stanford Bio-X Snack Grant 2025, and
by the NSF CMMI Award 2320933.
Further,
Paul Steinmann and Ellen Kuhl acknowledge support from the European Research Council (ERC) under the Horizon Europe program (Grant -Nos. 101052785 and 101141626, projects: SoftFrac and DISCOVER). Funded by the European Union. Views and opinions expressed are however those of the author(s) only and do not necessarily reflect those of the European Union or the European Research Council Executive Agency. Neither the European Union nor the granting authority can be held responsible for them.

\includegraphics[width=4cm]{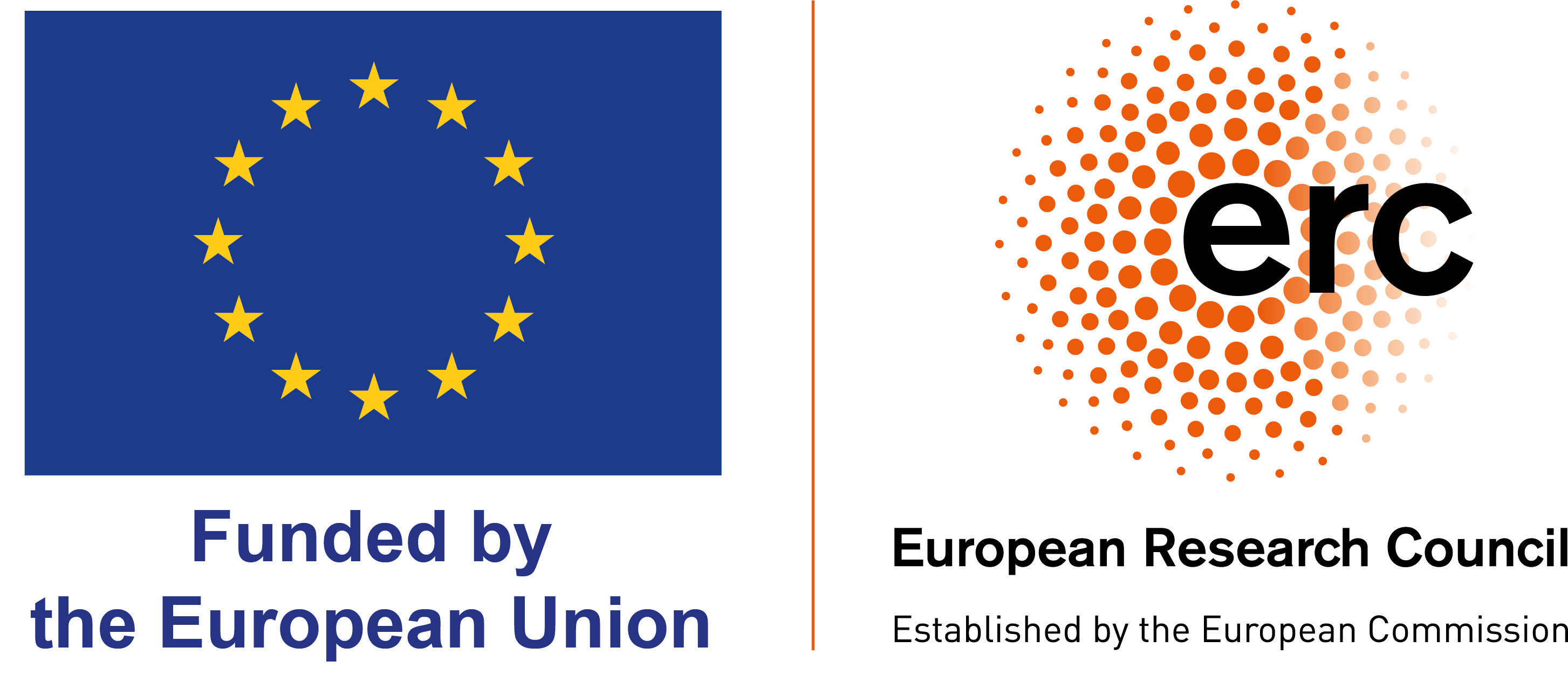}
\subsection*{CRediT authorship contribution statement}
\noindent 
HH: Conceptualization, Methodology, Software, Validation, Investigation, Formal Analysis, Data Curation, Writing - Original Draft, Writing - Review \& Editing, Funding acquisition \\
PS: Writing - Review \& Editing, Funding acquisition \\
EK: Writing - Review \& Editing, Supervision, Funding acquisition
\subsection*{Data availability}
\noindent
Our source code and examples are available at \href{https://doi.org/10.5281/zenodo.19248596}{https://doi.org/10.5281/zenodo.19248596}.
\subsection*{Statement of AI-assisted tools usage}
\noindent
The authors acknowledge the use of OpenAI’s ChatGPT, an AI language model, for assistance in generating and refining
text. The authors reviewed, edited, and take full responsibility for the content and conclusions of this work.
\FloatBarrier
\appendix
\section{Hyperparameters and neural networks architectures}
\label{app:hyperparameters}
The hyperparameters for training are listed in \cref{tab:training}, while the architectures for the individual neural networks are given in \cref{tab:architectures}.
\begin{table}[h]
\centering
\caption{Training hyperparameters.}
\label{tab:training}
\begin{tabular}{lll}
\toprule
\textbf{Category} & \textbf{Parameter} & \textbf{Value} \\
\midrule

\multirow{4}{*}{Optimization}
 & Number of epochs & - \\
 & Learning rate & $1\times 10^{-3}$ \\
 & Gradient clipping & disabled \\
 & Clip norm & $1.0$ \\
& Optimizer & ADAM \\

\midrule
\multirow{4}{*}{Loss weighting}
 & $\lambda_{\mathrm{D}}$ (Dirichlet force) & $1$ \\
 & $\lambda_{\mathrm{N}}$ (Active force) & $1$ \\
 & $\lambda_{\mathrm{R}}$ (regularization) & $1$ \\
 & $\lambda_{\mathrm{A}}$ (auxiliary mismatch) & - \\

\midrule
\multirow{2}{*}{Regularization}
 & Energy networks ($L^1$) & $10^{-5}$ \\
 & Dissipation networks ($L^1$) & $10^{-5}$ \\

\midrule
\multirow{2}{*}{Constraints}
 & Energy kernels & $w \geq 10^{-7}$ \\
 & Dissipation kernels & $w \geq 0$ \\

\midrule
Numerics
 & Precision & float64 \\

\bottomrule
\end{tabular}
\end{table}

\begin{table}[h]
\centering
\caption{Neural network architectures.}
\label{tab:architectures}
\resizebox{\textwidth}{!}{
\begin{tabular}{llllll}
\toprule
\textbf{Network} & \textbf{Layers} & \textbf{Activations} & \textbf{Output} & \textbf{Output act.} & \textbf{Output bias} \\
\midrule

FICNN$_{\tens{F},s}$ (joint energy)
& [12, 12]
& exp, exp
& 1
& softplus 
& Yes \\

FICNN$_{\tens{F}}$ (energy features)
& [12, 12, 12]
& exp, softplus, softplus
& 12
& identity 
& Yes \\

FICNN$_s$ (entropy features)
& [12, 12, 12]
& exp, exp, exp
& 12
& identity 
& Yes \\

PICNN$_{\tens{g}}$ (dissipation potential)
& [12, 12, 12]
& softplus, softplus, softplus
& 1
& ReLU 
& No \\

PICNN (coupling branch)
& [6, 6, 6]
& GELU, GELU, GELU
& --
& -- 
& -- \\

MLP$_s$ (auxiliary entropy)
& [12, 12]
& GELU, GELU
& 1
& identity 
& Yes \\

\bottomrule
\end{tabular}
}
\end{table}
\section{Constitutive Material laws}
\label{app:constitutive}

The training data are generated using two different constitutive models. In both cases, the formulation is based on a Helmholtz energy $\psi(\ten{F},T)$ and a dissipation (conduction) potential $\chi(\ten{g};\ten{F},T)$ defined in the reference configuration\footnote{We use the notation of $\chi$ here to emphasize that the potential is parameterized by the temperature instead of the entropy.}.

\paragraph{Thermal model}

For the purely thermal study, a model is considered with a passive mechanical response. The Helmholtz energy is defined as
\begin{equation}
\psi(\ten{F}, T)
=
\psi_{\mathrm{mech}}(\ten{C})
+
\psi_{\mathrm{th}}(T),
\end{equation}
with the mechanical contribution
\begin{equation}
\psi_{\mathrm{mech}}
=
a\, I_1 + b\, I_2 + c\, I_3 - \frac{d}{2} \ln(I_3),
\end{equation}
where $I_1 = \mathrm{tr}(\ten{C})$, $I_2 = \mathrm{tr}(\mathrm{cof}\,\ten{C})$, and $I_3 = \det(\ten{C})$, and
\begin{equation}
d = 2a + 4b + 2c.
\end{equation}
The thermal contribution is given by
\begin{equation}
\psi_{\mathrm{th}}(T)
=
c_{T0}
\left[
[ T - T_{0} ] - T \ln\left(\frac{T}{T_{0}}\right)
\right].
\end{equation}
The dissipation potential is defined as
\begin{equation}
\chi(\ten{g}; T, \ten{C})
=
\frac{\lambda_T}{2}\, T \, \ten{g} \cdot \mathrm{cof}(\ten{C}) \cdot \ten{g}.
\end{equation}
The material parameters are
\begin{equation}
a = 1~\mathrm{GPa}, \quad 
b = 1~\mathrm{GPa}, \quad 
c = 1~\mathrm{GPa},
\end{equation}
\begin{equation}
\lambda_T = 30.2~\mathrm{mW\,mm^{-1}K^{-1}}, \quad 
c_{T0} = 15.0~\mathrm{mJ\,mm^{-3}K^{-1}}, \quad 
T_{0} = 293.15~\mathrm{K}.
\end{equation}
\paragraph{Fully coupled thermomechanical model}

For the fully coupled analysis, a thermomechanical model is employed. The Helmholtz energy is decomposed as
\begin{equation}
\psi(\ten{F}, T)
=
\psi_{\mathrm{mech}}(\ten{C})
+
\psi_{\mathrm{th}}(T)
+
\psi_{\mathrm{cpl}}(\ten{C}, T).
\end{equation}
The mechanical contribution reads
\begin{equation}
\psi_{\mathrm{mech}}
=
\frac{\mu}{2}
\left[I_1 - 3 - \ln(I_3)\right]
+
\frac{\lambda}{4}
\left[I_3 - 1 - \ln(I_3)\right],
\end{equation}
where $\lambda$ and $\mu$ denote the Lamé parameters.
The thermal contribution is given by
\begin{equation}
\psi_{\mathrm{th}}(T)
=
c_{T0}
\left[
y + y^2 - T \ln\left(\frac{y}{y_0}\right)
\right],
\end{equation}
with
\begin{equation}
y = \frac{-1 + \sqrt{1 + 8T}}{4}, 
\qquad
y_0 = \frac{-1 + \sqrt{1 + 8T_{0}}}{4}.
\end{equation}

The thermo-mechanical coupling term is defined as
\begin{equation}
\psi_{\mathrm{cpl}}
=
\frac{3}{2}\, \kappa\, \alpha_0 \, [T - T_0]\, \ln(I_3),
\end{equation}
where $\kappa = \lambda + \tfrac{2}{3}\mu$ is the bulk modulus.
The dissipation potential is again given by
\begin{equation}
\chi(\ten{g}; T, \ten{C})
=
\frac{\lambda_T}{2}\, T \, \ten{g} \cdot \mathrm{cof}(\ten{C}) \cdot \ten{g}.
\end{equation}
The material parameters are taken from the literature \cite{Felder2022,Ambati2015,Dittmann2020}
\begin{equation}
\lambda = 101.160~\mathrm{GPa}, \qquad
\mu = 73.255~\mathrm{GPa}, \qquad
\alpha_0 = 1.1 \times 10^{-5}~\mathrm{K^{-1}},
\end{equation}
\begin{equation}
\lambda_T = 50.2~\mathrm{mW\,mm^{-1}K^{-1}}, \qquad
c_{T0} = 3.59~\mathrm{mJ\,mm^{-3}K^{-1}}, \qquad
T_{0} = 293.15~\mathrm{K}.
\end{equation}
\section{Balance of energy in terms of Helmholtz energy}
\label{app:balance_energy_psi}

\paragraph{State law}
Choosing the Helmholtz energy $\psi(\ten{F},T)$ as the energetic potential leads to the following alternative form of the state laws compared to \cref{eq:reduced_state_laws}
\begin{equation}
  \ten{P} = \frac{\partial\psi}{\partial\ten{F}}, 
  \qquad 
  s = -\frac{\partial \psi}{\partial T}.
  \label{eq:state_law_psi}
\end{equation}

\paragraph{Governing equation}
Inserting the state laws in \cref{eq:state_law_psi} into the energy balance \eqref{eq:balance_energy} yields the governing equation for heat diffusion
\begin{equation}
  -T\,\frac{\partial^2\psi}{\partial T^2}\,\dot{T}
  =
  -\mathrm{Div}\,\ten{q} + r
  + T\,\frac{\partial^2\psi}{\partial\ten{F}\partial T} : \dot{\ten{F}} .
\end{equation}

For implementation purposes, it is worth noting that the mixed derivative with respect to $\ten{F}$ can be expressed in terms of the right Cauchy--Green tensor $\ten{C}=\ten{F}^\mathrm{T}\ten{F}$ as
\begin{equation}
  \frac{\partial^2\psi}{\partial\ten{F}\partial T}
  =
  2\,\ten{F}\,\frac{\partial^2\psi}{\partial\ten{C}\partial T}.
\end{equation}

The weak form presented in \cref{sec:weak_forms} can be adapted accordingly.

\bibliographystyle{unsrt}  
\bibliography{references}
\end{document}